\documentclass[longauth]{aa}

\usepackage{euclid}

\usepackage{natbib}
\bibpunct{(}{)}{;}{a}{}{,} 

\usepackage{soul}
\usepackage{float,lscape}
\usepackage{graphicx}
\usepackage{subcaption}
\usepackage{txfonts}
\usepackage{amsmath}
\usepackage{physics} 	
\usepackage{color} 		
\usepackage{verbatim} 	
\usepackage{threeparttable} 
\usepackage[colorlinks,linkcolor=blue,urlcolor=blue,citecolor=blue]{hyperref} 
\usepackage{float,lscape}
\usepackage{geometry}
\usepackage{pdflscape}
\usepackage[separate-uncertainty=true, list-units=repeat]{siunitx}

\graphicspath{{./}{figures/}}

\begin{document}


\definecolor{blue}{rgb}{0,0.3,0.7}
\definecolor{red}{rgb}{0.85,0.08,0.05}
\definecolor{purple}{rgb}{0.65,0.00,0.35}
\definecolor{green}{rgb}{0.35,0.45,0.25}
\definecolor{orange}{rgb}{1.0,0.5,0.15}
\definecolor{yellow}{rgb}{0.9,0.65,0.15}
\definecolor{brown}{rgb}{0.7,0.25,0.0}
\definecolor{cyan}{rgb}{0.1,0.66,0.66}
\definecolor{pink}{rgb}{0.858, 0.688, 0.688}
\definecolor{violet}{rgb}{0.127, 0.2, 0.255}

\newcommand{\mum}{\,\si{\micron}\xspace}
\newcommand{\vect}[1]{\boldsymbol{#1}}
\newcommand{\zphot}{$z_{\rm phot}$}
\newcommand{\zspec}{$z_{\rm spec}$}
\newcommand{\sigmaLogMs}{\sigma_{\text{Log}M_*}}
\newcommand{\mstar}{M_{\star}}
\def\oiii{[\ion{O}{III}]}
\newcommand{\SEpp}{\texttt{SE++}}
\newcommand{\hotcold}{\texttt{hot+cold}}
\newcommand{\psfex}{\texttt{PSFEx}\xspace}
\newcommand{\lephare}{\texttt{LePHARE}\xspace}
\newcommand{\cigale}{\texttt{CIGALE}\xspace}
\newcommand{\warnfl}{\texttt{warn\_flag}\xspace}

\newcommand{\hst}{\textit{HST}}
\newcommand{\JWST}{\textit{JWST}}
\newcommand{\HSC}{HSC}
\newcommand{\UVISTA}{UltraVISTA}
\newcommand{\SCUBA}{SCUBA-2}
\newcommand{\CFHT}{CFHT}
\newcommand{\DES}{DES}
\newcommand{\Subaru}{\textit{Subaru}}
\newcommand{\AGEL}{AGEL}
\newcommand{\ALMA}{ALMA}
\newcommand{\NIRCAM}{NIRCam}
\newcommand{\MIRI}{MIRI}
\newcommand{\SPIRE}{SPIRE}
\newcommand{\SPITZER}{\textit{Spitzer}}
\newcommand{\PACS}{PACS}

\DeclareSIUnit{\Msun}{M_\odot}
\DeclareSIUnit{\year}{yr}
\DeclareSIUnit{\pc}{pc}
\DeclareSIUnit{\mag}{mag}
\DeclareSIUnit{\mas}{mas}
\DeclareSIUnit{\dex}{dex}
\DeclareSIUnit{\jansky}{Jy}
\DeclareSIUnit{\kpc}{kpc}

\title{COSMOS2025: The COSMOS-Web galaxy catalog of photometry, morphology, redshifts, and physical parameters from \JWST, \hst,   and ground-based imaging}
\subtitle{}  

\author{
Marko Shuntov\inst{\ref{DAWN},\ref{NBI},\ref{UG}}\fnmsep\thanks{\email{marko.shuntov@nbi.ku.dk}}
\and Hollis~B.~Akins\inst{\ref{UAT}} \and
Louise Paquereau\inst{\ref{IAP}} \and
Caitlin M.~Casey\inst{\ref{ucsb}, \ref{UAT}, \ref{DAWN}} \and
Olivier~Ilbert \inst{\ref{LAM}} \and%
Rafael~C.~Arango-Toro\inst{\ref{LAM}} \and
Henry~Joy~McCracken\inst{\ref{IAP}} \and
Maximilien~Franco\inst{\ref{CEA},\ref{UAT}} \and
Santosh~Harish\inst{\ref{Rochester}} \and
Jeyhan~S.~Kartaltepe\inst{\ref{Rochester}} \and
Anton~M.~Koekemoer\inst{\ref{STScI}} \and
Lilan~Yang\inst{\ref{Rochester}} \and
Marc~Huertas-Company\inst{\ref{IAC}, \ref{LERMA}, \ref{Paris-Cite}, \ref{laLaguna}} \and
Edward M.~Berman\inst{\ref{NorthEastern}} \and
Jacqueline E. McCleary\inst{\ref{NorthEastern}} \and
Sune Toft\inst{\ref{DAWN},\ref{NBI}} \and
Raphaël Gavazzi\inst{\ref{LAM}} \and
Mark J. Achenbach\inst{\ref{Honolulu2}} \and
Emmanuel Bertin\inst{\ref{CEA}} \and
Malte Brinch\inst{\ref{DAWN},\ref{DTU}}
Jackie Champagne\inst{\ref{Arizona}} \and
Nima Chartab\inst{\ref{Pasadena}} \and
Nicole E. Drakos\inst{\ref{HawaiiHilo}} \and
Eiichi Egami\inst{\ref{Arizona}} \and
Ryan Endsley\inst{\ref{UAT}} \and
Andreas L. Faisst\inst{\ref{Pasadena}} \and
Xiaohui Fan\inst{\ref{Arizona}} \and
Carter Flayhart\inst{\ref{Rochester}} \and
William~G.~Hartley\inst{\ref{UG}} \and
Hossein Hatamnia\inst{\ref{Riverside}} \and
Ghassem Gozaliasl\inst{\ref{Aalto},\ref{Helsinki}} \and
Fabrizio Gentile\inst{\ref{CEA},\ref{Bologna}} \and
Iris Jermann\inst{\ref{DAWN},\ref{DTU}} \and
Shuowen Jin\inst{\ref{DAWN},\ref{DTU}} \and
Koki Kakiichi\inst{\ref{DAWN},\ref{NBI}} \and
Ali Ahmad Khostovan\inst{\ref{Kentucky},\ref{Rochester}}
Martin Kümmel\inst{\ref{Munich}} \and
Clotilde Laigle\inst{\ref{IAP}} \and
Ronaldo Laishram\inst{\ref{NAOJapan}} \and
Erini Lambrides\inst{\ref{Goddard}}
Daizhong Liu\inst{\ref{PurpleMObs}} \and
Jianwei Lyu\inst{\ref{Arizona}} \and
Georgios Magdis\inst{\ref{DAWN},\ref{DTU}} \and
Bahram Mobasher\inst{\ref{Riverside}} \and
Thibaud Moutard\inst{\ref{ESA}} \and
Alvio Renzini\inst{\ref{Padova}} \and
Brant E. Robertson\inst{\ref{SCruz}} \and
Marc Schefer\inst{\ref{UG}} \and
Diana Scognamiglio\inst{\ref{JPL}} \and
Nick Scoville\inst{\ref{Caltech}} \and
Zahra Sattari\inst{\ref{Riverside},\ref{Pasadena}} \and
David B. Sanders\inst{\ref{Honolulu}} \and
Sina Taamoli\inst{\ref{Riverside}} \and
Benny Trakhtenbrot\inst{\ref{TelAviv},\ref{MPIEP},\ref{ECO}} \and
Francesco Valentino\inst{\ref{DAWN},\ref{DTU}} \and
Feige Wang\inst{\ref{UMichigan},\ref{Arizona}} \and
John R. Weaver\inst{\ref{Amherst}} \and
Jinyl Yang\inst{\ref{UMichigan}}
}

\institute{
Cosmic Dawn Center (DAWN), Denmark \label{DAWN}
\and
Niels Bohr Institute, University of Copenhagen, Jagtvej 128, 2200 Copenhagen, Denmark \label{NBI}
\and
University of Geneva, 24 rue du Général-Dufour, 1211 Genève 4, Switzerland \label{UG}
\and
The University of Texas at Austin, 2515 Speedway Blvd Stop C1400, Austin, TX 78712, USA\label{UAT}
\and
Institut d'Astrophysique de Paris, UMR 7095, CNRS, and Sorbonne Universit\'e, 98 bis boulevard Arago, 75014 Paris, France \label{IAP}
\and
Department of Physics, University of California, Santa Barbara, Santa Barbara, CA 93106 USA\label{ucsb}
\and
Aix Marseille Univ, CNRS, LAM, Laboratoire d'Astrophysique de Marseille, Marseille, France  \label{LAM}
\and
Université Paris-Saclay, Université Paris Cité, CEA, CNRS, AIM, 91191 Gif-sur-Yvette, France \label{CEA}
\and
Laboratory for Multiwavelength Astrophysics, School of Physics and Astronomy, Rochester Institute of Technology, 84 Lomb Memorial Drive, Rochester, NY 14623, USA \label{Rochester}
\and
Space Telescope Science Institute, 3700 San Martin Drive, Baltimore, MD 21218, USA \label{STScI}
\and
Instituto de Astrofísica de Canarias (IAC), La Laguna, E-38205, Spain \label{IAC}
\and
Observatoire de Paris, LERMA, PSL University, 61 avenue de l’Observatoire, F-75014 Paris, France \label{LERMA}
\and
Université Paris-Cité, 5 Rue Thomas Mann, 75014 Paris, France \label{Paris-Cite}
\and
Universidad de La Laguna, Avda. Astrofísico Fco. Sanchez, La Laguna, Tenerife, Spain \label{laLaguna}
\and
Department of Physics, Northeastern University, 360 Huntington Ave, Boston, MA \label{NorthEastern}
\and
Department of Physics and Astronomy, University of Hawaii at Manoa, 2505 Correa Rd, Honolulu, HI 96822, USA \label{Honolulu2}
\and
DTU-Space, Technical University of Denmark, Elektrovej 327, 2800 Kgs. Lyngby, Denmark \label{DTU}
\and
Steward Observatory, University of Arizona, 933 N. Cherry Ave., Tucson, AZ 85719, USA \label{Arizona}
\and
Caltech/IPAC, MS 314-6, 1200 E. California Blvd. Pasadena, CA 91125, USA \label{Pasadena}
\and
Department of Physics, University of Hawaii, Hilo, 200 W Kawili St, Hilo, HI 96720, USA \label{HawaiiHilo}
\and
Department of Physics and Astronomy, University of California, Riverside, 900 University Avenue, Riverside, CA 92521, USA \label{Riverside}
\and
Department of Computer Science, Aalto University, P.O. Box 15400, FI-00076 Espoo, Finland \label{Aalto}
\and
Department of Physics, University of, P.O. Box 64, FI-00014 Helsinki, Finland \label{Helsinki}
\and
University of Bologna - Department of Physics and Astronomy “Augusto Righi” (DIFA), Via Gobetti 93/2, I-40129 Bologna, Italy \label{Bologna}
\and
Department of Physics and Astronomy, University of Kentucky, 505 Rose Street, Lexington, KY 40506, USA \label{Kentucky}
\and
Universitäts-Sternwarte München, Fakultät für Physik, LudwigMaximilians-Universität München, Scheinerstrasse 1, 81679 München, Germany \label{Munich}
\and
National Astronomical Observatory of Japan, 2-21-1 Osawa, Mitaka, Tokyo 181-8588, Japan \label{NAOJapan}
\and
NASA-Goddard Space Flight Center, Code 662, Greenbelt, MD, 20771, USA \label{Goddard}
\and
Purple Mountain Observatory, Chinese Academy of Sciences, 10 Yuanhua Road, Nanjing 210023, China \label{PurpleMObs}
\and
European Space Agency (ESA), European Space Astronomy Centre (ESAC), Camino Bajo del Castillo s/n, 28692 Villanueva de la Cañada, Madrid, Spain \label{ESA}
\and
Instituto Nazionale di Astrofisica (INAF), Osservatorio Astronomico di Padova, Vicolo dell’Osservatorio 5, 35122, Padova, Italy \label{Padova}
\and
Department of Astronomy and Astrophysics, University of California, Santa Cruz, 1156 High Street, Santa Cruz, CA 95064, USA \label{SCruz}
\and
Jet Propulsion Laboratory, California Institute of Technology, 4800 Oak Grove Drive, Pasadena, CA 91001, USA \label{JPL}
\and
Astronomy Department, California Institute of Technology, 1200 E. California Blvd, Pasadena, CA 91125, USA \label{Caltech}
\and
Institute for Astronomy, University of Hawai’i at Manoa, 2680 Woodlawn Drive, Honolulu, HI 96822, USA \label{Honolulu}
\and
School of Physics and Astronomy, Tel Aviv University, Tel Aviv 69978, Israel \label{TelAviv}
\and
Max-Planck-Institut f{\"u}r extraterrestrische Physik, Gie\ss{}enbachstra\ss{}e 1, 85748 Garching, Germany \label{MPIEP}
\and
Excellence Cluster ORIGINS, Boltzmannsstra\ss{}e 2, 85748, Garching, Germany \label{ECO}
\and
Department of Astronomy, University of Michigan, 1085 S. University Ave., Ann Arbor, MI 48109, USA \label{UMichigan}
\and
Department of Astronomy, University of Massachusetts, Amherst, MA 01003, USA \label{Amherst}
}

\date{Released on TBD / Accepted date: TBD}

\abstract{
We present COSMOS2025, the COSMOS-Web catalog of photometry, morphology,  photometric redshifts and physical parameters for more than 700,000 galaxies in the Cosmic Evolution Survey (COSMOS) field. This catalog is based on our \textit{James Webb Space Telescope} 255\,h COSMOS-Web program, which provides deep near-infrared imaging in four NIRCam (F115W, F150W, F277W, F444W) and one MIRI (F770W) filter over the central $\sim 0.54 {\, \rm deg}^2$ ($\sim 0.2 {\, \rm deg}^2$ for MIRI) in COSMOS. These data are combined with ground- and space-based data to derive photometric measurements of NIRCam-detected sources using both fixed-aperture photometry (on the space-based bands) and a profile-fitting technique on all 37 bands spanning \SIrange{0.3}{8}{\micron}. We provide morphology for all sources from complementary techniques including profile fitting and machine-learning classification. We derive photometric redshifts, physical parameters and non-parametric star formation histories from spectral energy distribution (SED) fitting. The catalog has been extensively validated against previous COSMOS catalogs and other surveys. Photometric redshift accuracy measured using spectroscopically confirmed galaxies out to $z\sim9$ reaches $\sigma_{\rm MAD} = 0.012$ at $m_{\rm F444W}<28$ and remains at $\sigma_{\rm MAD} \lesssim 0.03$ as a function of magnitude, color, and galaxy type. This represents a factor of $\sim 2$ improvement at 26 AB mag compared to COSMOS2020. The catalog is approximately 80\% complete at $\log(M_{\star}/{\rm M}_{\odot}) \sim 9$ at $z \sim 10$ and at $\log(M_{\star}/{\rm M}_{\odot}) \sim 7$ at $z \sim 0.2$, representing a gain of  1\,dex compared to COSMOS2020. COSMOS2025 represents the definitive COSMOS-Web catalog. It is provided with complete documentation, together with redshift probability distributions, and it is ready for scientific exploitation today. 
}


\titlerunning{COSMOS-Web galaxy catalog}
\authorrunning{Shuntov et al.}
\maketitle


\section{Introduction} \label{sec:intro}


One of the most fundamental objectives of astronomy is to map and understand the contents of the Universe. Our knowledge of the cosmos has grown in lock-step with technological progress. Extragalactic surveys have followed closely this explosive growth in technological capability, from the first deep surveys with photographic plates in the 1970s to ground-based `deep-field' imaging with electronic detectors which routinely reached $B\sim28$ mag at the end of the 20th century. Each technological advance opened a new window in wavelength, depth, or resolution. Space-based observatories, starting with the \textit{Hubble Space Telescope} (\hst) pioneered a series of surveys starting with the Hubble Deep Field North \citep[HDF-N][]{Williams1996} and continuing with the Great Observatories Origins Deep Survey \citep[GOODS,][]{Giavalisco2004}, the Hubble Ultra-Deep Field \citep[HUDF,][]{Beckwith2006, Ellis2013, Illingworth2013, Teplitz2013} and the Cosmic Assembly Near-infrared Deep Extragalactic Legacy Survey \citep[CANDELS,][]{Grogin2011, Koekemoer2011}. These data revealed a distant Universe which was much more complex than had been seen in low-resolution ground-based observations. These surveys, often combined with ground-based spectroscopic measurements, revealed the cosmic history of star formation, reionization and stellar mass assembly together with the transformation of the physical and morphological properties of galaxies \citep[e.g.,][for reviews]{MadauDickinson2014, Conselice_2014, Robertson2021}.

However, the majority of these surveys, although reaching great depths, covered only small areas, totalling only $0.2\, \si{\deg}^2$ ($800\, {\rm arcmin}^2$). Despite the discovery of some remarkable objects \cite[e.g.,][]{Oesch2016} these surveys were limited in their ability to probe a large range of galaxy populations and environments. Moreover, their volume at lower redshifts is small, meaning that it is impossible to trace the evolution of a well-defined population of objects over a significant range in cosmic time. 

The Cosmic Evolution Survey \citep[COSMOS,][]{scoville_cosmic_2007} was designed to bridge the gap between shallow, wide-area ground-based surveys and deep, pencil-beam space-based surveys. Thanks to one of the largest-ever  allocations of HST time, a contiguous $2\, \si{\deg}^2$ patch of the sky was covered with the Advanced Camera for Surveys (ACS) in the F814W band \citep{Koekemoer2007}. At the same time, a series of complementary observing programs were started to provide the necessary wavelength coverage across the electromagnetic spectrum. COSMOS demonstrated the feasibility of weak lensing measurements in tomographic redshift bins to map the distribution of dark matter over large areas \citep{Massey_2007}, enabled the mapping of large-scale structure at intermediate redshifts \citep{Scoville2013} and demonstrated how mass and environment drive the evolution of galaxies across cosmic time \citep{peng_mass_2010}. 

The \textit{James Webb Space Telescope} (\JWST) represents the next major technological step forward, providing orders-of-magnitude increases in sensitivity in $1-5\, \si{\micron}$ with the Near Infrared Camera \citep[NIRCam,][]{RiekeNIRCAM} and $6-30\, \si{\micron}$ from the Mid Infrared Instrument \citep[MIRI,][]{WrightMIRI} delivering an unprecedented combination of sensitivity, spatial resolution, and field-of-view. One of the largest contiguous \JWST\ programs, in both area and time allocated, is the COSMOS-Web survey (GO\#1727, PIs: Casey \& Kartaltepe, \citealt{Casey2023CW}). COSMOS-Web covers the central part of the COSMOS field, covering $\sim 0.54\, \si{deg}^2$ ($1920\, \si{arcmin}^2$) in four NIRCam filters (F115W, F150W, F277W, F444W) and $\sim0.2\, \si{deg}^2$ ($720\, \si{arcmin}^2$) in one MIRI filter (F770W). This combination of depth and area together with existing multiwavelength data makes the COSMOS-Web a unique survey to study galaxy evolution across a large range of cosmic history, capturing some of the rarest and most extreme objects in the Universe. 

To optimally exploit the rich  multi-wavelength data in COSMOS with \JWST\ imaging from COSMOS-Web, traditional fixed-aperture photometric techniques are impractical. The point-spread function (PSF) in COSMOS ranges from $\ang{;;1}$ in ground-based bands to $\ang{;;0.2}$ in NIRCam. However, techniques where a model profile is fitted to each source after convolution with the PSF are particularly well-suited to COSMOS data. These methods enable consistent measurement of total fluxes for sources at different wavelengths observed with different point-spread functions. In the COSMOS2020 catalog \citep{weaver_cosmos2020_2022} these methods for deep multiwavelength photometry were validated by comparing measurements made using the profile fitting code \texttt{The Farmer} \citep{WeaverFarmer2023} with the \textsc{Classic} aperture photometry. These showed excellent agreement with the advantage of the profile fitting method that the data do not need to be PSF-homogenized before measurement. 

Today, building on the legacy of previous COSMOS catalogs, we present a new ultra-deep  COSMOS catalog derived from our unique \JWST\ NIRCam data. 
Sources are detected on the four NIRCam bands using dual-extraction `hot and cold' detection method \citep{2004ApJS..152..163R,leauthaud_weak_2007} that achieves high completeness and purity. 
Measurements are performed using aperture photometry for the space-based \hst/ACS F814W, \JWST/NIRCam and \JWST/MIRI bands and a novel model-fitting approach based on \texttt{SourceXtractor++} \citep{bertin20, kummel20, Kummel2022} to measure total photometry in 37 bands as well as NIR morphology for over 700,000 galaxies. We also provide additional morphological measurements from bulge-disk decomposition, independent model-fitting code \texttt{GaLight} \citep{Ding2020, Birrer2021}, as well as machine learning-based morphological classification \citep{MHC2024}. We used the rich multi-band coverage at $\sim 0.3 - 8 \, \si{\micron}$ to derive photometric redshifts and physical parameters, including novel non-parametric star formation histories (SFH) from SED fitting using \lephare{} \citep{arnouts_measuring_2002, ilbert_accurate_2006} and \cigale{} \citep{Boquien19}. This, in combination with the large area, enables us to identify numerous galaxy candidates at $z>10$ \citep[e.g.,][]{Casey2024, Franco2024}.

This paper is organized as follows. In Section~\ref{sec:data}, we describe the data. In Section~\ref{sec:photometry-extraction}, we describe the methods used for source detection and photometric and morphological measurements. In Section~\ref{sec:photometry-validation}, and \ref{sec:morphology} we describe, compare, and validate these measurements. In Section~\ref{sec:photo-z}, we present the photometric redshifts derived from SED fitting with \lephare{}. Finally, in Section~\ref{sec:physical-prop}, we present the physical properties measured from both \lephare{} and \cigale{}. Our conclusions are presented in Sect.~\ref{sec:conclusions}. 

We adopt a standard $\Lambda$CDM cosmology with $H_0=70$\,km\,s$^{-1}$\,Mpc$^{-1}$, $\Omega_{\rm m,0}=0.3$ and $\Omega_{\Lambda,0}=0.7$. All magnitudes are expressed in the AB system \citep{oke_absolute_1974}, for which a flux $f_\nu$ in $\mu$Jy
($10^{-29}$~erg~cm$^{-2}$s$^{-1}$Hz$^{-1}$) corresponds to AB$_\nu=23.9-2.5\,\log_{10}(f_\nu/{\rm \mu Jy})$.

\section{Observations and Data Reduction} \label{sec:data}

\begin{figure}[t!]
\includegraphics[width=1\columnwidth, trim=0.2cm 0.8cm 0cm 1cm, clip]{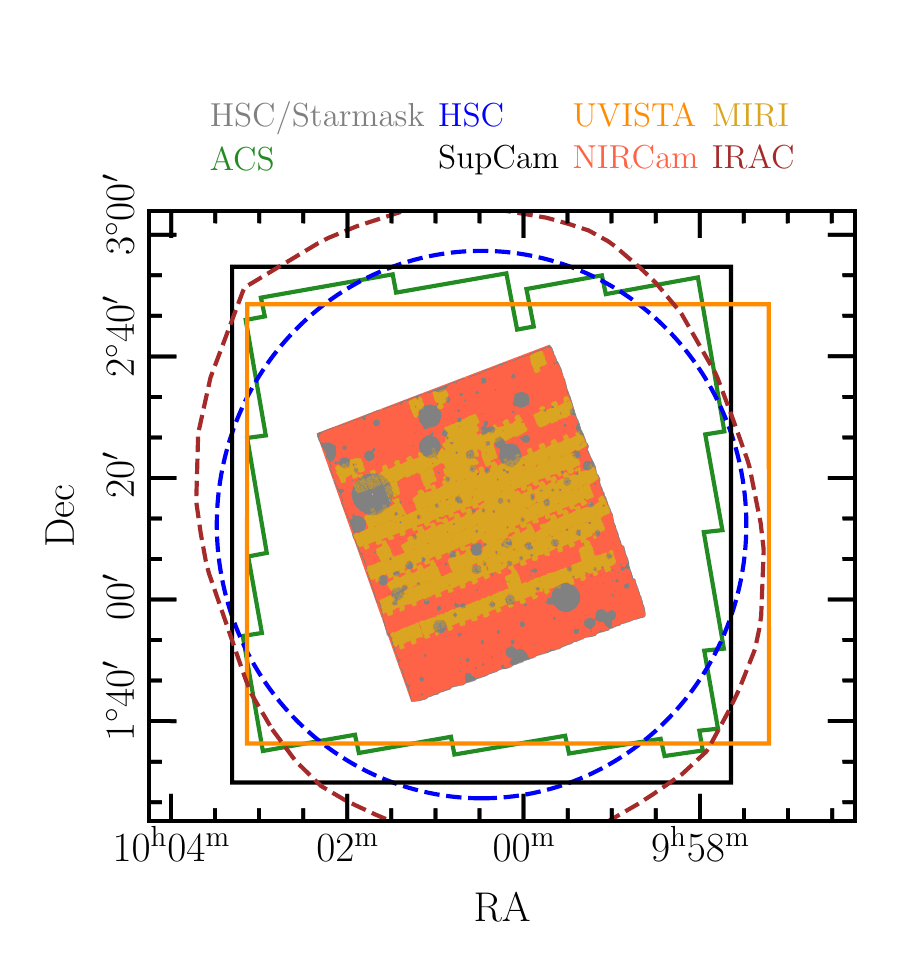}
\caption{Survey footprint in the COSMOS field. The regions show the footprints of different imaging instruments color coded accordingly. The \JWST\ footprint from NIRCam and MIRI is shown in the colored areas. The gray regions mark the area affected by HSC bright star masks.
\label{fig:survey-footprint}}
\end{figure}

In this section we provide a brief overview of the imaging data we used to construct the galaxy catalog. Figure~\ref{fig:survey-footprint} shows the footprint of the different surveys in the COSMOS field. The COSMOS-Web catalog is built from the \JWST\ NIRCam imaging whose footprint is shown in red, while the MIRI imaging is marked in yellow. The footprint of legacy imaging surveys included in the catalog are shown with the regions and fully encompass COSMOS-Web. In Fig.~\ref{fig:depth-of-bands} we show the depths of all bands as a function of wavelength and in Table~\ref{tab:band_infos} we provide information about the bands, including their $5\,\sigma$ empty aperture depths.

\subsection{JWST data} \label{subsec:cweb}
COSMOS-Web is a 255h JWST Cycle 1 Treasury Program (\#1727) covering 0.54\,deg$^2$ of contiguous four-filter NIRCam imaging (F115W, F150W, F277W, F444W) and 0.2\,deg$^2$ of MIRI imaging (F770W) acquired in parallel \citep{Casey2023CW}. The mosaic is roughly square, measuring $\ang{;46}\times \ang{;46}$, and is divided into 152 visits that use the 4TIGHT dither. 

Data reduction of the NIRCam and MIRI data will be described in detail in two forthcoming papers (Franco et al., Harish et al. in prep), which we summarise briefly. The COSMOS-Web NIRCam observations were processed using the \textit{JWST} Calibration Pipeline \citep{jwstcalibrationpipeline2023}, with additional optimizations for image quality and astrometric precision. Raw NIRCam exposures were retrieved from the Mikulski Archive for Space Telescopes (MAST) and processed with pipeline version 1.14.0, supplemented by custom corrections appropriate for deep  \textit{JWST} imaging \citep[e.g.,][]{Bagley2024}. These corrections included mitigation of $1/f$ noise \citep{Schlawin2020}, background subtraction, artifact removal (including wisp correction and claw removal), and identification and masking of defective pixels. Calibration was performed using the Calibration Reference Data System (CRDS) \texttt{pmap-1223}, which corresponds to the NIRCam instrument mapping \texttt{imap-0285}. The final science mosaics were produced at a pixel scale of $\ang{;;0.03}~\textrm{pixel}{^{-1}}$, ensuring optimal spatial resolution for accurate photometric measurements.
Astrometric refinement was conducted using the \textit{JWST}/HST Alignment Tool (\texttt{JHAT}; \citealt{Rest2023}), aligning the NIRCam images to a reference catalog constructed from \textit{HST}/ACS F814W mosaics \citep{Koekemoer2007}, with astrometry calibration tied to Gaia Early Data Release 3 (EDR3; \citealt{GaiaDR3}). This procedure resulted in a median absolute positional offset of $<5$\, mas, with a median absolute deviation (MAD) of $<12$\,mas across all filters. Similar to the NIRCam data reduction, the MIRI observations were processed using version 1.12.5 of the JWST Calibration Pipeline \citep{jwstcalibrationpipeline2023}, with an additional custom background subtraction step to mitigate the strong sky and thermal background present in our data (see also \citealt{Yang2023b,Perez-Gonzalez2024b}). Calibration was performed using CRDS context version \texttt{pmap-1130}, and the astrometric alignment was based on the reference catalog from the \textit{HST}/ACS F814W mosaics \citep{Koekemoer2007}. The overall positional accuracy in RA and DEC was 0.35 and 6 mas, respectively, with a MAD of 28 mas.

\subsection{Ultraviolet data}
We use the same $U-$ band imaging from the Canada-France-Hawaii Telescope's (CFHT) MegaCam instrument as in COSMOS2020.  These include data  for the CFHT Large Area $U-$ band Deep Survey \citep[CLAUDS,][]{sawicki_cfht_2019} and for COSMOS. Processing of these images are described in \cite{weaver_cosmos2020_2022} and \cite{sawicki_cfht_2019}.

\subsection{Ground-based optical data}
 \label{sec:opticaldata}
In common with previous COSMOS catalogs, most optical data comes from the Subaru telescope, using either the  Hyper Suprime-Cam (HSC) or Suprime-Cam instruments \citep{2002PASJ...54..833M, 2018PASJ...70S...1M}. We use the third public data release (PDR3) of the HSC Subaru Strategic Program (HSC-SSP) comprising the $g,r,i,z,y$ broad, as well as \textit{NB}0816, \textit{NB}0921, \textit{NB}1010, narrow bands \citep{Aihara2022}. Compared to the PDR2 used in COSMOS2020, PRD3 is more uniform and slightly deeper with improved sky subtraction and better photometric and astrometric calibrations \citep{Aihara2022}. We used the public PDR3 data access tools\footnote{\url{https://hsc-release.mtk.nao.ac.jp/doc/index.php/data-access__pdr3/}} to retrieve the processed images and weight maps from the `UltraDeep' layer. Since we use only the $\sim0.54\,\rm{deg}^2$ central area, the depth is uniform over the COSMOS-Web footprint. 

We also include medium and narrow bands from Suprime-Cam used in COSMOS2015 \citep{laigle_cosmos2015_2016}
and COSMOS2020 \citep{taniguchi_cosmic_2007,taniguchi_subaru_2015}. These include 11 medium-bands (\textit{IB}$427$, \textit{IA}$484$, \textit{IB}$505$, \textit{IA}$527$, \textit{IB}$574$, \textit{IA}$624$, \textit{IA}$679$, \textit{IB}$709$, \textit{IA}$738$, \textit{IA}$767$, \textit{IB}$827$), and two narrow-bands (\textit{NB}$711$, \textit{NB}$816$). These are the same images as in COSMOS2020, which are also PSF homogenized on a tile-level on the individual images (see Section~3.1.2 in \citealt{weaver_cosmos2020_2022}). 

\subsection{HST optical data}
\label{sec:HST-data}
We use the full COSMOS \hst\ ACS F814W dataset \citep{Koekemoer2007}. Specifically we use \hst\ ACS F814W mosaics produced from data that was recalibrated using updated reference files and calibration pipelines, including improved treatments for charge-transfer efficiency, updated flat fields, biases, and dark current reference files (appropriate for the observation dates corresponding to the data). These mosaics have also had their astrometric alignment updated, with their absolute astrometry directly aligned to the Gaia-DR3 reference frame\footnote{\url{https://www.cosmos.esa.int/web/gaia/dr3}} where these improvements follow the approaches developed and described in detail by \citet{Koekemoer2011}, updated as needed to accommodate current calibrations.

\subsection{Ground-based near-infrared data}
\label{sec:ground-NIR}
UltraVISTA \citep{mccracken_ultravista_2012} was a large public survey carried out on the VISTA \citep{2015A&A...575A..25S} telescope with the VIRCAM \citep{2006SPIE.6269E..0XD} instrument.  The survey took place between 2010--2023. We use the final UltraVISTA DR6 which comprises the entire UltraVISTA public survey dataset as well as guaranteed time observations which were taken in 2010. DR6 comprises almost 100,000 images taken over 2290 hours of observing time. Full details are given in the documentation\footnote{\url{https://archive.eso.org/cms/eso-archive-news/sixth-and-last-release-of-ultravista-public-survey-data.html}}. Compared to previous releases, depths are now completely uniform over the $\ang{1.5}\times\ang{1.2}$ UltraVISTA survey area. 

\subsection{Other data}
\label{sec:IRAC}
We include data from the \textit{Spitzer} telescope's mid-infrared \citep{Werner2004} IRAC camera in channels 1, 2, 3 and  4. These data were reprocessed as part of the Cosmic Dawn Survey \citep[DAWN][]{McPartland2025, moneti_euclid_2021}. 
Although photometry is extracted in these bands, this information is not used for photometric redshift and physical parameter estimation from SED fitting.

\begin{figure}[t!]
\includegraphics[width=1\columnwidth]{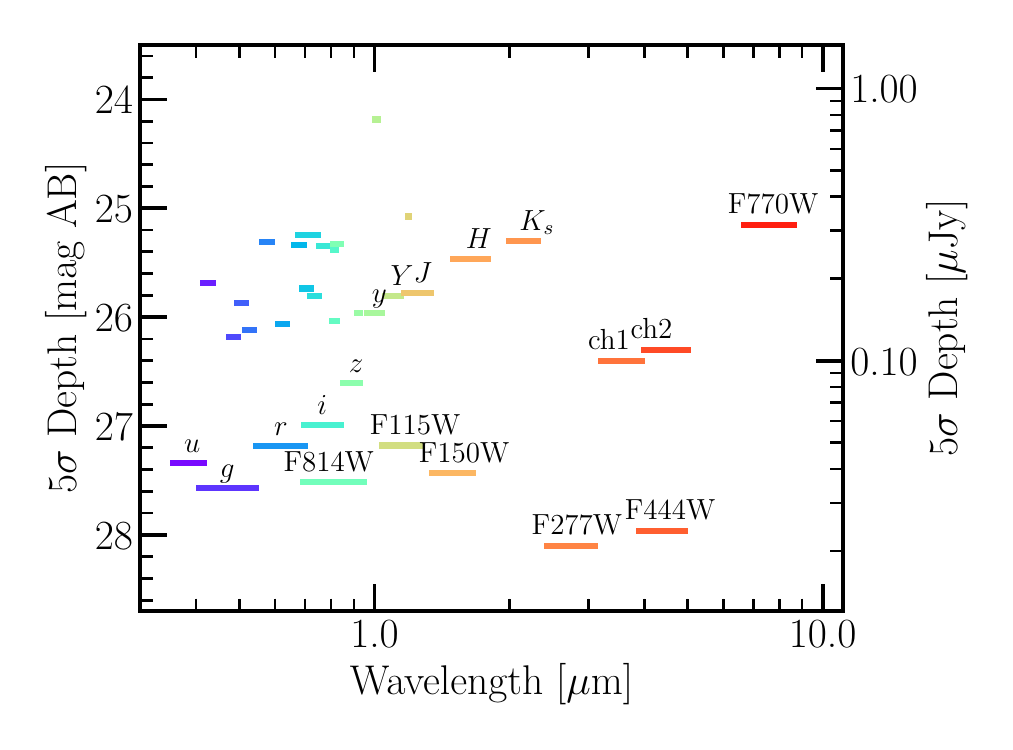}
\caption{$5\,\sigma$ depths in all bands. These are measured from the variance of the background measured in empty apertures averaged over the field of view. These apertures are $\ang{;;0.15}$ in diameter in the NIRCam and HST bands,  $\ang{;;0.5}$ in MIRI and $\ang{;;1}$ for the ground-based data. The segment length corresponds to the filter width.
\label{fig:depth-of-bands}}
\end{figure}

\begin{table}[t!]
\centering
\caption{UV-optical-IR data in COSMOS2025}
\setlength{\tabcolsep}{8pt}
\begin{threeparttable}
\begin{tabular}{lcccc}
 \hline \hline
Instrument & Band & Central\tnote{a} & Width\tnote{b} & Depth\tnote{c} \\
/Telescope &  & $\lambda$ [\AA{}] & [\AA{}] &  \\
(Survey) &  &  &  &  \\
 \hline
 NIRCam & F115W & 11622 & 2646 & 27.2  \\
 & F150W & 15106 & 3348 & 27.4  \\
 & F277W & 28001 & 6999 & 28.1  \\
 & F444W & 44366 & 11109 & 28.0 \\
 \hline
 MIRI & F770W & 77108 & 20735 & 25.2  \\
 \hline
MegaCam & $u^{*}$ & 3858 & 598 & 27.3  \\
/CFHT  \\
 \hline
ACS/$HST$ & F814W & 8333 & 2511 & 27.5 \\
 \hline
HSC & $g$ & 4847 & 1383 & 27.6 \\
/Subaru & $r$ & 6219 & 1547 & 27.2  \\
HSC-SSP & $i$ & 7699 & 1471 & 27.0  \\
PDR3 & $z$ & 8894 & 766 & 26.6 \\
 & $y$ & 9761 & 786 & 26.0  \\
 & \textit{NB}$0816$ & 8168 & 110 & 26.0  \\
 & \textit{NB}$0921$ & 8168 & 133 & 26.0  \\
 & \textit{NB}$1010$ & 10100 & 94 & 24.2  \\
 \hline
Suprime-Cam & \textit{IB}$427$ & 4266 & 207 & 25.7  \\
/Subaru & \textit{IA}$484$ & 4851 & 229 & 26.2  \\
 & \textit{IB}$505$ & 5064 & 231 & 25.9  \\
 & \textit{IA}$527$ & 5261 & 243 & 26.1  \\
 & \textit{IB}$574$ & 5766 & 273 & 25.3  \\
 & \textit{IA}$624$ & 6232 & 300 & 26.1 \\
 & \textit{IA}$679$ & 6780 & 336 & 25.3  \\
 & \textit{IB}$709$ & 7073 & 316 & 25.7  \\
 & \textit{IA}$738$ & 7361 & 324 & 25.8  \\
 & \textit{IA}$767$ & 7694 & 365 & 25.3  \\
 & \textit{IB}$827$ & 8243 & 343 & 25.3  \\
 & \textit{NB}$711$ & 7121 & 72 & 25.2  \\
 & \textit{NB}$816$ & 8150 & 120 & 25.3  \\
 \hline
VIRCAM & $Y$ & 10216 & 923 & 25.8  \\
/VISTA & $J$ & 12525 & 1718 & 25.8  \\
UltraVISTA & $H$ & 16466 & 2905 & 25.5  \\
DR6  & $K_s$ & 21557 & 3074 & 25.3  \\
 & \textit{NB}$118$\tnote{d} & 11909 & 112 & 25.1  \\
 \hline
 IRAC & ch1\tnote{d} & 35686 & 7443 & 26.4 \\
/\textit{Spitzer} & ch2\tnote{d} & 45067 & 10119 & 26.3 \\
 & ch3\tnote{d} & 57788 & 14082 & 23.2 \\
 & ch4\tnote{d} & 79958 & 28796 & 23.1 \\
 \hline
\end{tabular}
\begin{tablenotes}
\item[a] Median of the transmission curve.
\item[b] Full width of the transmission curve at half maximum.
\item[c] $5\sigma$ depth computed in empty apertures with diameters of $1.0\arcsecond$ for the ground-based, $0.15\arcsecond$ for the space-based \textit{JWST}/NIRCam and \textit{HST}/ACS and $0.5\arcsecond$ for \textit{JWST}/MIRI images, averaged over the NIRCam area.
\item[d] Not used in SED fitting. 
\end{tablenotes}
\end{threeparttable}

\label{tab:band_infos}
\end{table}

%

\section{Object detection and photometry} \label{sec:photometry-extraction}

\begin{figure*}[t!] 
\begin{center}
\includegraphics[width=0.65\textwidth]{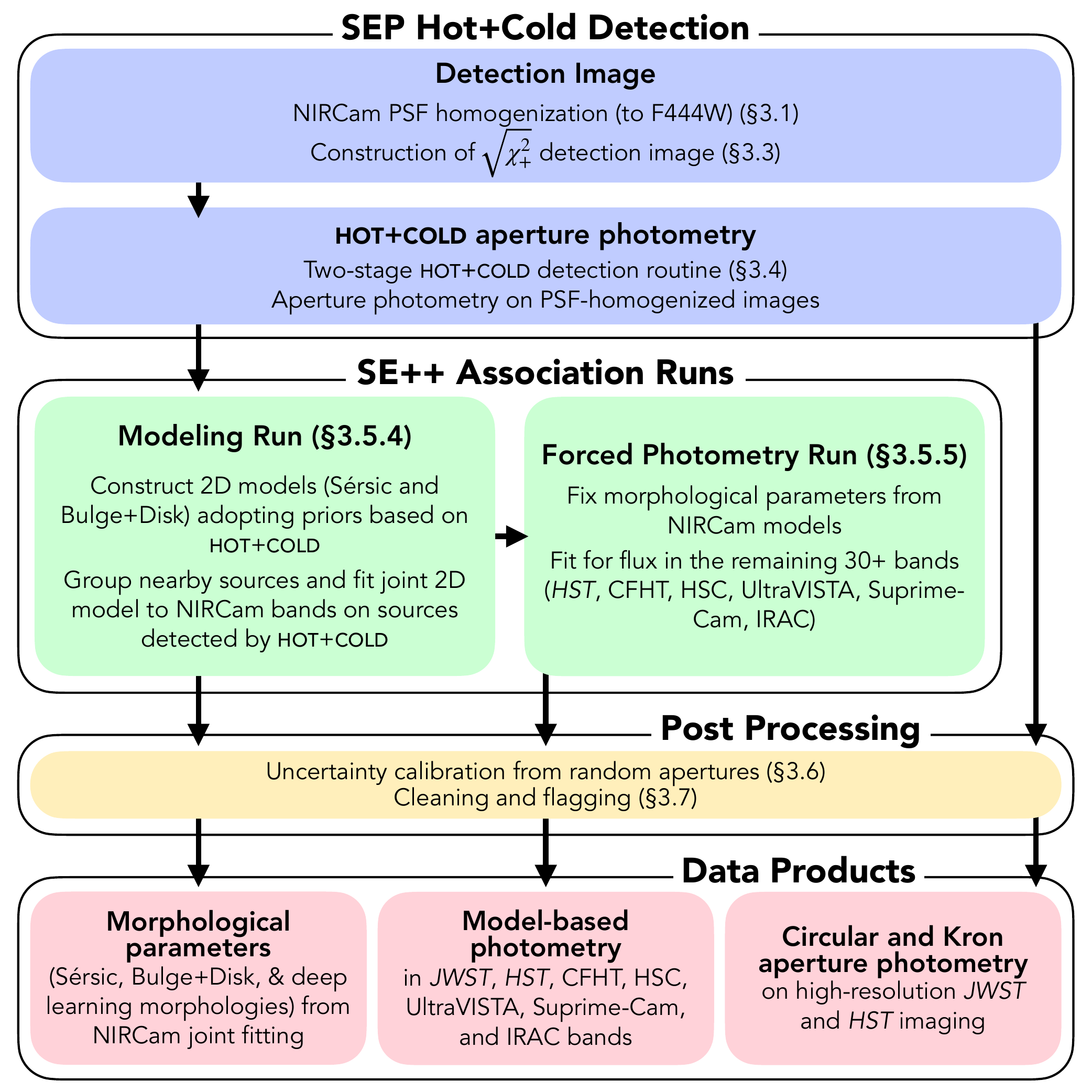}
\caption{A summary of the key steps in preparing the photometric catalog. First, source lists are generated using the `hot and cold' detection technique and used to perform aperture photometry on PSF homogenized space-based images. Next, sources are grouped and a joint two-dimensional model light profile is extracted for all sources. Using the morphological parameters extracted from these fits, forced photometry is then performed on all bands. Finally, in the post-processing step, the photometric errors for each source are adjusted based on variance measurements in empty apertures and the source lists are cleaned and flagged. The resulting data products include morphological measurements, model-based photometry in 37 bands, and simple aperture photometry in 6 space-based bands, for more than 700,000 galaxies. 
}
\label{fig:flowchart}
\end{center}
\end{figure*}

To construct our catalog, we develop a set of techniques that make optimal use the rich multiwavelength data in COSMOS. Two sets of photometric measurements are made, the first using apertures on PSF-homogenized \JWST\ and \hst{} images. The second uses a two-dimensional light profile model fitted on all available data, both ground and space. In both cases, the photometric extraction is carried out on sources detected using a `hot and cold' detection technique described in Sect.~\ref{sec:hotcold-catalog}. Our approach is summarized in Figure~\ref{fig:flowchart}.

\subsection{Point-spread function reconstruction and homogenization} \label{sec:PSF-reconstruct-homogenize}

Accurately measuring the point-spread function (PSF) is crucial for flux measurements, profile fitting and morphological measurements. In the following, we describe the our methodology in PSF reconstruction and image homogenization.

\subsubsection{Point-spread function reconstruction} \label{sec:PSFEx-reconstruction}

To reconstruct the PSF we use \texttt{PSFEx}\footnote{\url{https://github.com/astromatic/psfex}} \citep{Bertin2011}. \psfex can  empirically reconstruct the PSF at different positions on the detector focal plane and at various pixel scales, determining how much sky area is encoded in each pixel. \cite{berman2024efficient} show that \psfex provides the best performance in reconstructing the NIRCam PSF in combined images. In contrast to empirical PSF fitters like \psfex, forward modeling approaches such as WebbPSF \citep{2014SPIE} produce PSF models for each single exposure. However, applying these forward modeling approaches to mosaics is non-trivial \citep{jwstcalibrationpipeline2023,harvey2024weak}. 
 
We first run \texttt{SExtractor} on all bands to construct catalogs containing vignettes of all stars. We use the \texttt{MU\_MAX} and \texttt{MAG\_AUTO} diagram \citep{leauthaud_weak_2007} to select point sources. We account for the spatial variation of the PSF in two ways. First, we construct a separate model for each of the 20 tiles of the full mosaic (cf. Franco et al. in prep) and all bands. Second, for each tile, we use order$-1$ polynomial interpolation to fit spatial variations of the PSF across the field, as implemented in \psfex. Finally, we used a sampling step of 1/4th of the PSF FWHM for each band.
 
For validation and quality control, we use the mean relative error diagnostics developed in \cite{berman2024efficient} to quantify pixel-level mismatches between sources and PSF models. Finally, we examine residuals in the second adaptive moments \citep{hirata2003shear,mandelbaum2005systematic} of input sources and PSF models. This is done using \texttt{Galsim}'s HSM module \citep{rowe2015galsim}. Specifically, the \texttt{FindAdaptiveMom}'s function returns the size $(\sigma_{\textsc{HSM}})$ and shape $(g_1, g_2)$ moments of an object. In this case, the objects are sources from a training catalog or a PSF model. While the NIRCam and MIRI PSFs are not well approximated by elliptical Gaussians, these second moment statistics are still useful for detecting significant size and shape mismatches when used alongside our other quality metrics. 

Finally, we found no signs of systematic modeling errors using the mean relative error diagnostics. Examples of these figures for the NIRCam PSFs can be found in Appendix \ref{sec:appendix-psf}, \ref{fig:psf-MRE}. These figures show that our PSF models adequately capture the key features in the sources we are trying to model. This is supported by the strong agreement between the second moments in sources and PSF models. The average size error was less than $10\%$ across the NIRCam filters, which we use to model each source (c.f. \S\ref{sec:modeling-run}). The errors in shape were similarly low.

\subsubsection{Point-spread function homogenization} \label{sec:PSF-homogenization}
The PSF FWHMs vary significantly between different bands. To ensure consistency in our aperture photometry measurements in different bands, we construct a set of PSF-homogenized mosaics for the \JWST/NIRCam and \hst{} data. We adopt the lowest resolution NIRCam F444W PSF as our target. We followed a procedure similar to \cite{weaver_cosmos2020_2022}, using the Python tool \texttt{pypher} \citep{pypher} to generate convolution kernels with a band-dependent regularization factor. This parameter was optimized to minimize residuals between the convolved PSF model and the target while preventing harmonic artifacts from Fourier transformations. All \JWST/NIRCam and \hst{} tile mosaics were convolved with their respective kernels, ensuring a uniform PSF across all images, and accounting for spatial variations between tiles.
We also generate a convolution kernel to match the NIRCam F444W image to the MIRI F770W PSF, which we use to examine the fraction of flux lost when performing aperture photometry on the MIRI data, as described in more detail Section~\ref{sec:hotcold-aperphoto}.
While convolution can introduce correlated noise in the  backgrounds, potentially affecting flux measurements, this effect is accounted for in the photometric error estimates (see \S\ref{sec:uncertainty-calibration}).

\subsection{Masking} \label{sec:masking}

\begin{figure}[t!] 
\includegraphics[width=8.8cm]{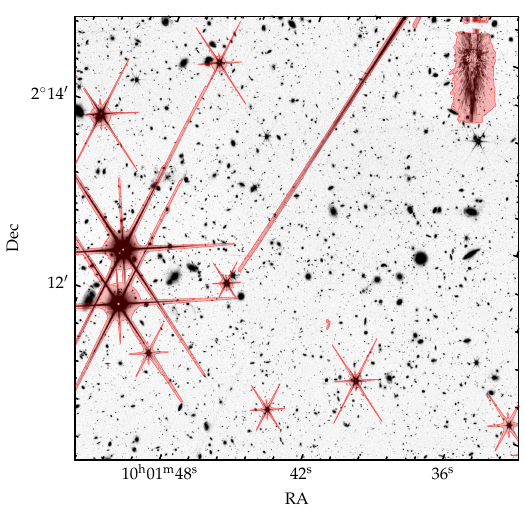}
\caption{Cutout of the NIRCam $\chi^2_+$ detection image in tile B10, with the custom star mask overlaid. The mask covers bright stars and their large diffraction spikes. At the top right is a large `Dragon's Breath' artifact (type II), which arises from the bright ($K\sim 6.5$ Vega mag) star to the north. This feature, and other artifacts, are manually masked.
\label{fig:starmask}}
\end{figure}

Parts of our mosaics contain bright low-$z$ galaxies, diffuse scattered light from bright stars, and imaging artifacts. Here, photometric measurements cannot be made accurately, and these regions must be identified and masked. For bright stars, we developed a semi-automatic procedure to identify and mask regions of the NIRCam mosaics affected by diffraction spikes. This requires a sufficiently large NIRCam PSF model. Since \texttt{WebbPSF} is not optimized beyond $\ang{;;30}$ and does not account for scattered light, we use a larger PSF model provided by STScI.\footnote{PSF models from Bryan Holler, \url{ https://stsci.app.box.com/s/2na9re7blhxq0sk8lcqxluxj45owhndm}} 
This model is convolved with a Gaussian kernel and normalized to one. It is then converted into a series of region files, capturing the extent of the PSF model down to a range of limiting thresholds. This allows us to adopt the most appropriate mask for each star, accounting for their brightness. For each star brighter than $\text{mag}_{F150W} \leq 17.0$, we automatically select the most appropriate region based on the segmentation map of the star, translate the region to the star position, and rotate to match the position angle of the observations. From there, we manually adjust the regions around each star, scaling or rotating if necessary to ensure that the entirety of the emission is masked.  Imaging artifacts and a few nearby diffuse galaxies are manually added to the mask. 
Figure~\ref{fig:starmask} shows a cutout of the detection image in the B10 tile, with the overlaid star mask.

\subsection{Detection image} \label{sec:detection-image}

\begin{figure}[t!]
\includegraphics[width=0.99\columnwidth]{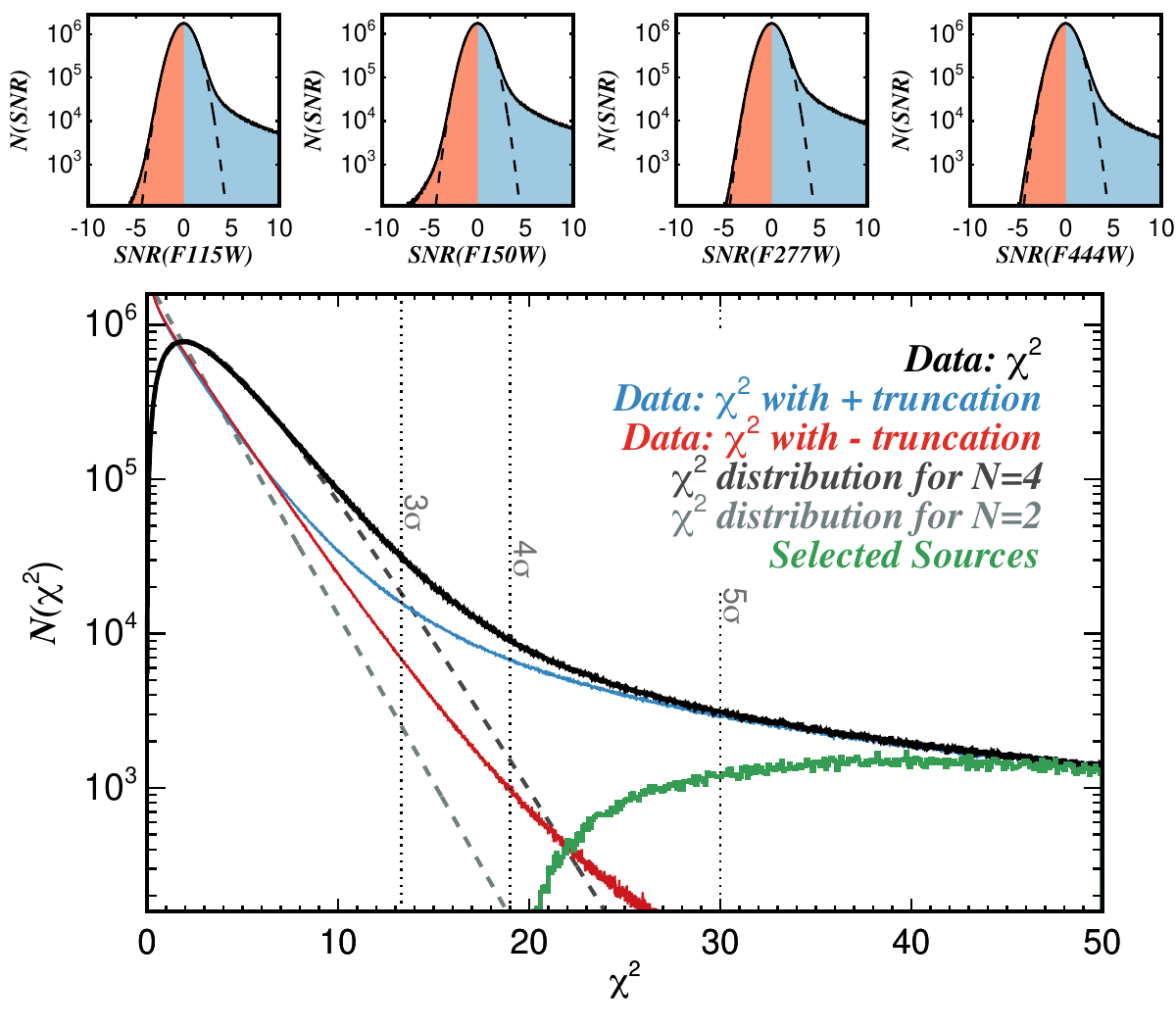}
\caption{Distribution of pixels in the PSF-homogenized, noise-equalized NIRCam maps (top panels) and $\chi^2$ distribution for the quadrature sum of the four NIRCam SNR maps (bottom panel). We show the distributions without truncation (black), with positive truncation $\chi^2_+$ (blue), and with negative truncation $\chi^2_-$ (red).  The $\chi^2$ distribution for degrees of freedom $N=4$ and $N=2$ are shown in dark and light dashed gray lines.  The distribution of sources' maximum $\chi^2_+$ values that are recovered by the \hotcold\ catalog are shown in the green histogram. Vertical lines mark the equivalent of $3$, $4$, and $5\sigma$ thresholds for $\chi^2$ distributions with $N=2$.}
\label{fig:chi2}
\end{figure}

We construct a detection image as the coaddition of all four COSMOS-Web NIRCam bands following the `$\chi^2$' technique \citep{szalay_simultaneous_1999,Drlica-Wagner2018}. This method optimally combines multiple images to maximize the detection signal-to-noise. In the first step, we generate PSF-homogenized science mosaics by matching the PSF to that of the lowest-resolution F444W band (see \S\ref{sec:PSF-homogenization}). Then, for each band and tile, we construct noise-equalized images (\texttt{NSCI}) by multiplying the PSF-homogenized science mosaics (\texttt{SCI}) by the square root of the weight map (\texttt{WHT})\footnote{The weight maps are derived as the inverse variance of the read noise.}, i.e., 
\begin{equation} 
    \texttt{NSCI} = \texttt{SCI} \sqrt{\texttt{WHT}}. 
\end{equation}
Next, we measure the root-mean-square (RMS) of the noise-equalized images by iteratively fitting the negative tail of the pixel distribution (i.e., below $1\,\sigma$) with a Gaussian function. By dividing the PSF-homogenized, noise-equalized images by their measured RMS values, we produce signal-to-noise ratio (SNR) maps for each filter. 
A nominal $\chi^2$ detection image would be created by adding these maps in quadrature \citep[e.g.,][]{szalay_simultaneous_1999}. 
However, marginally negative SNR pixels across multiple individual bands may lead to false detections in a $\chi^2$ image. 
To mitigate this, we modify slightly the construction of the $\chi^2$ image by truncating the individual SNR maps (setting negative pixels to zero) before combination in the $\chi^2$ image. 
Finally, we use the square root of the $\chi^2$ image as our detection image. This ensures that our first-pass measurement of galaxy shapes are calculated on a detection image with pixel values proportional to flux density and not the square of the flux density.

Figure~\ref{fig:chi2} illustrates the variations in the  $\chi^2$ distribution for this data set.
Without truncation, the pixel values in the detection image follow a $\chi^2$ distribution with four degrees of freedom for the four NIRCam bands. With truncation, the distribution instead resembles a $\chi^2$ distribution with two deegrees of freedom because the SNR maps halve the number of contributing bands. The effective $3\,\sigma$, $4\,\sigma$, and $5\,\sigma$ detection thresholds for the $\chi^2$ distribution with $N=2$ are indicated. We adopt an initial threshold of $\chi^2 = 13.3$, corresponding to $3\,\sigma$; the equivalent $\sqrt{\chi^2_{+}}$ value is 3.69. Sources whose peak $\sqrt{\chi^2_{+}}$ equals this threshold are excluded from the catalog unless they also satisfy the additional \hotcold\ aperture criteria described in Sect.~\ref{sec:hotcold-catalog}. Under this criterion, the limiting SNR  ratio in F277W (the deepest NIRCam filter) is approximately $4\,\sigma$, barring detection in any other NIRCam filter.

\subsection{The \hotcold\ aperture catalog}  \label{sec:hotcold-catalog}

\begin{figure*}[t!]
\includegraphics[width=0.92\columnwidth]{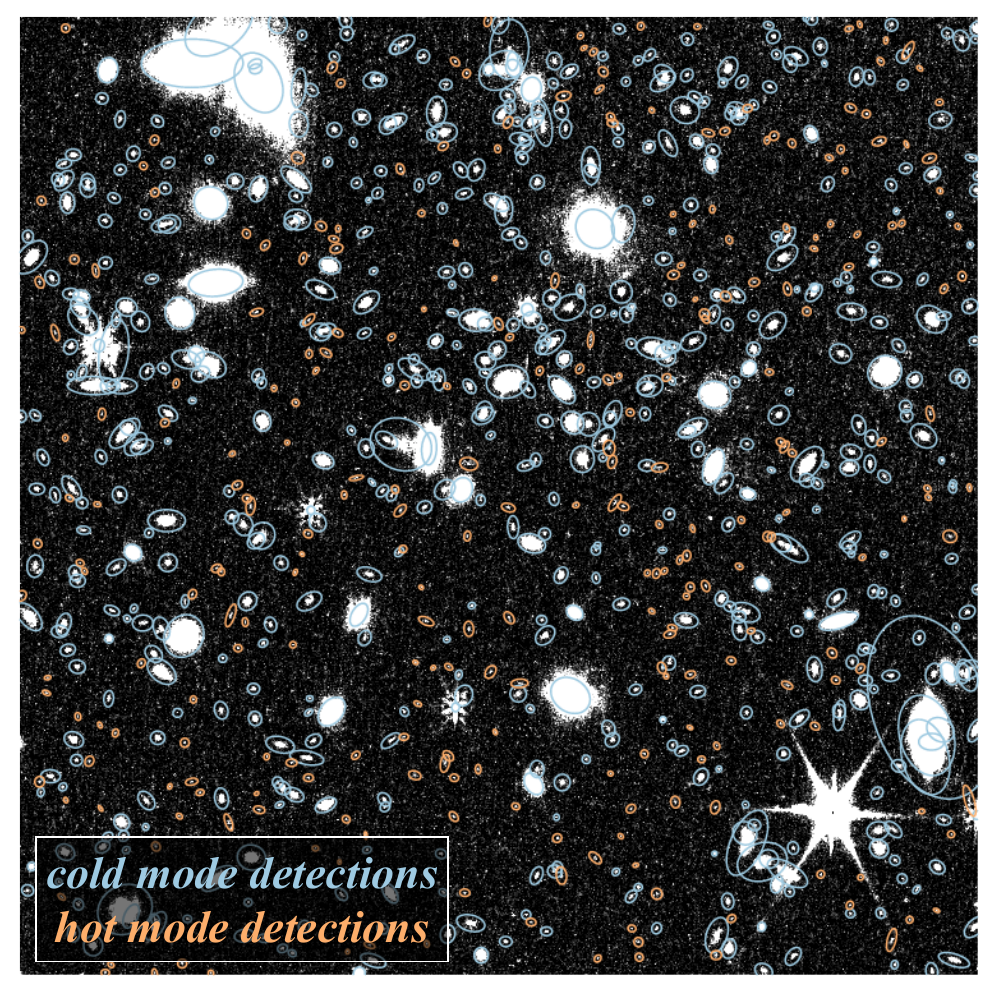}
\includegraphics[width=1.08\columnwidth]{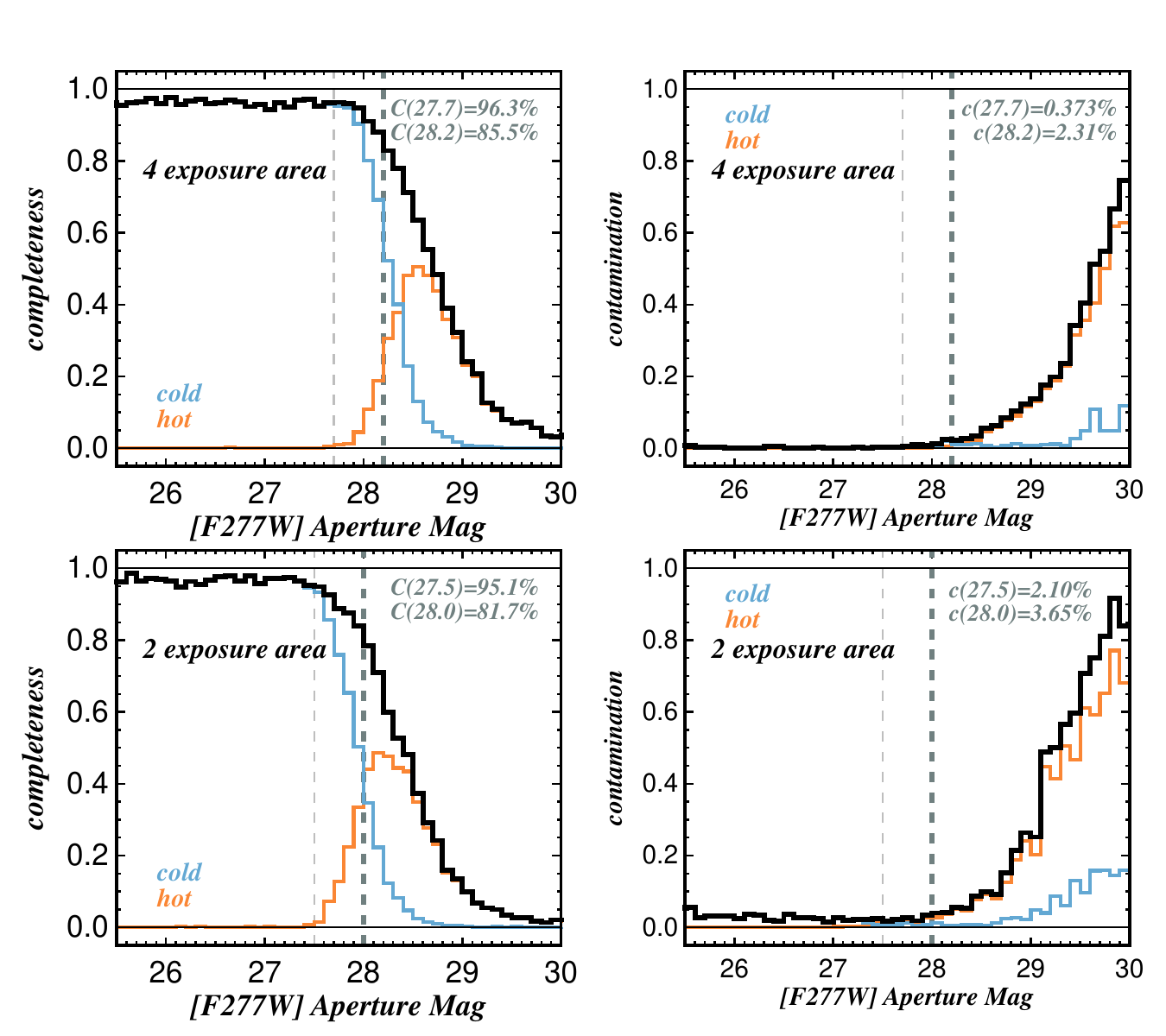}
\caption{An illustration of our `hot and cold' detection technique.  \textit{Left panel:} A $90''\times 90''$ cutout of the COSMOS-Web $\chi^2_{\rm +}$ detection image, showing Kron ellipses of all detected sources found from cold mode (blue) and hot mode (orange) SEP runs.  Note the star in the lower-right corner that has been manually masked. \textit{Right panel:}  we measure catalog completeness and contamination using  an independent catalog from PRIMER (which reaches one magnitude deeper).  Completeness is measured as a fraction of PRIMER sources recovered as a function of F277W magnitude (the deepest COSMOS-Web filter) for cold-mode (blue) and hot-mode (orange) sources. Contaminants are measured as sources found in COSMOS-Web but not detected in PRIMER imaging.  We split these measurements into `2 exposure' (bottom row) and `4 exposure' (top row) areas where the imaging depth varies significantly. Completeness at the $8\,\sigma$ ($5\,\sigma$) survey depth exceeds 95\%\ (80\%) and contamination at these depths does not exceed 2\% (4\%).}
\label{fig:hotcold}
\end{figure*}

Source detection for the COSMOS-Web catalog is performed using a \hotcold\ scheme, which we outline in the following section. 
We additionally produce an aperture photometry catalog on PSF-homogenized space-based images. 

\subsubsection{Source detection} \label{sec:hotcold-detection}

Source detection and aperture photometry is performed using \texttt{SEP} \citep{K.Barbary2016}, a python implementation of \texttt{SExtractor} \citep{bertin_sextractor_1996}. 
We employ a hot and cold detection scheme, similar to catalogs produced for the CANDELS survey \citep[e.g.,][]{galametz2013, guo2013, nayyeri_candels_2017, stefanon2017}. 
The premise of the hot and cold detection strategy is that the optimal detection parameters are fundamentally different for ultra-deep, pencil-beam surveys and shallower, wide-field surveys. While deep and narrow surveys can use aggressive detection parameters, pushing the catalog to the detection limit, shallow and wide surveys generally must be less aggressive to properly deblend bright sources. The hot and cold strategy involves first extracting objects with a high threshold optimized for bright, extended sources (the `cold mode') and then running a separate extraction optimized for faint, isolated sources (the `hot mode'). 

For the cold mode, we adopt a detection threshold of $\sqrt{\chi^2_+} = 4.66$, which corresponds to $4\sigma$ ($N=2$). 
The detection image is convolved with a top-hat  filter kernel with a diameter of 9 pixels, to optimize detection for extended sources.  We use \verb|minarea = 15|,  \verb|deblend_nthresh = 64|, \verb|deblend_cont = 0.001|, and \verb|clean_param = 2.0|. For the  hot-mode, we use a detection threshold of $3.698$, which corresponds to $3\sigma$ ($N=2$). We use a Gaussian filter kernel with a FWHM of 3 pixels, and adopt \verb|minarea = 8|,  \verb|deblend_nthresh = 32|, \verb|deblend_cont = 0.01|, and \verb|clean_param = 0.5|. 

The two catalogs are then merged as follows. First, bright stars are removed from the cold-mode catalog via the star masks described in \S\ref{sec:masking}. In particular, any source with $>80\%$ of its segmentation map pixels overlapping the star mask is removed from the catalog; this mostly applies to the stars themselves or galaxies falling entirely on the diffraction spikes. Sources with between 0 and 80\% of their pixels overlapping the mask are not removed from the catalog, but are flagged with the keyword \texttt{flag\_star}. Next, an elliptical mask is defined based on the cold-mode sources from \verb|a_image| and \verb|b_image|, using a scale factor of 6 and a minimum radius of 10 pixels. Flagged sources are not used to construct the elliptical mask, since their shape parameters can be unreliable. Any hot mode source that overlaps the elliptical mask, or intersects with the cold-mode segmentation map, is removed; otherwise, it is added to the final catalog. 
The result is a single catalog in which the majority of sources are taken from the cold mode, except isolated/faint sources detected in the hot mode (Fig.~\ref{fig:hotcold}). 

The left panel of Figure~\ref{fig:hotcold} provides an illustration of our  source detection method, where cold mode sources are generally brighter and allowed to crowd with each other more significantly.  Hot mode sources can only be detected in regions outside the cold mode elliptical mask. In general, hot mode sources are on the margins of the detection limits, fainter than 28.0 (27.6) mag in 4-exposure (2-exposure) depth areas of the mosaics.

\subsubsection{Detection completeness and contamination} \label{sec:hotcold-complete-contam}

The right panel of Fig.~\ref{fig:hotcold} shows the detection completeness and contamination as a function of magnitude in the deepest F277W band, split between the hot- and cold-mode detections. This is measured by cross-matching with an independent reduction of the PRIMER-COSMOS Survey \citep[GO\#1837]{PrimerDunlop2021} within our larger COSMOS-Web mosaic. Processed using the same imaging pipeline, none of the COSMOS-Web imaging was used in the construction of the PRIMER mosaics or vice versa.
PRIMER-COSMOS has eight bands of NIRCam coverage over 140\,arcmin$^2$; we generate a PRIMER-COSMOS hot and cold catalog with the same technique used for COSMOS-Web, by producing a PSF-homogenized positive-truncated $\sqrt{\chi^2_{+}}$ image (with 8 degrees of freedom for the 8 filters). The PRIMER-COSMOS catalog is generally 1-1.2 magnitudes deeper than COSMOS-Web in matched filters, so our completeness/contamination measurements here are not particularly sensitive to variations in depth within the PRIMER-COSMOS mosaics themselves.

Completeness is measured as a fraction of PRIMER sources recovered by F277W magnitude bin in overlapping regions not masked in either PRIMER or COSMOS-Web.
F277W magnitudes are measured in a \ang{;;0.3} diameter circular aperture without aperture correction.
The blue and orange histograms indicate the relative fraction of those sources detected in the cold vs. hot mode.
Completeness is measured independently in regions of the COSMOS-Web mosaics at 4-exposure depth vs. 2-exposure depth and are remarkably consistent with expectation as a function of source SNR.
Contamination is measured as fraction of COSMOS-Web sources that are not recovered in the PRIMER-COSMOS catalog above a nominal 4$\sigma$ detection threshold; such contaminants are deemed to be pure instrumental noise but are exceedingly rare above the nominal COSMOS-Web detection limits.

Of 784,016 sources in the COSMOS-Web catalog, 566,521 are recovered by cold mode detection (72.3\%) and 217,495 are added by hot mode detection (27.7\%).
Of the 580,496 sources brighter than F277W$<$28.2 (in 0$\farcs$3 diameter circular apertures), 533,861 are cold mode (92.0\%) and 46,635 are hot mode (8.7\%).  
While hot mode sources dominate the tail of very faint sources, it is clear that a significant fraction of real sources can only be recovered via this extra detection mode, even well above the nominal 5$\sigma$ depth of our imaging.

\subsubsection{Aperture photometry} \label{sec:hotcold-aperphoto}

Aperture photometry is measured for \hst/ACS F814W, \JWST/NIRCam F115W, F150W, F277W, F444W, and \JWST/MIRI F770W.
Only these bands are processed for the initial \hotcold\ aperture photometry catalog due to their high spatial resolution (FWHM$\le$\ang{;;0.3}).
All bands except F770W are PSF-homogenized to F444W, as described in \ref{sec:PSF-homogenization}.
We provide photometry in circular apertures (\texttt{aper}) and Kron elliptical apertures (\texttt{auto}). 
The former is provided for convenience and simple measurements; in most cases, the \texttt{auto} photometry is the more reliable measure of the total flux, except for highly blended sources. 

Aperture photometry is measured in circular apertures with diameters of \ang{;;0.2}, \ang{;;0.3}, \ang{;;0.5}, \ang{;;0.75}, and \ang{;;1}. No aperture or PSF corrections are applied. The auto photometry is measured in elliptical apertures computed using a Kron \citep{Kron1980} factor of $k=1.6$ and a minimum circular radius of 1.1 pixels (\texttt{kron1}). 
These small Kron apertures are intended to capture realistic colors while maximizing signal-to-noise. 
We correct measurements in these small apertures to the total flux by additionally performing photometry on the F444W image using the default Kron factor $k=2.5$ (\texttt{kron2}). 
This aperture correction is applied multiplicatively to the fluxes and uncertainties for all filters.
We then apply an additional correction for the fraction of PSF flux that falls outside this larger Kron aperture. 
The MIRI/F770W fluxes are handled following \citet{Finkelstein2024}: we perform an additional Kron aperture measurement on a F444W image which has been PSF-matched to F770W, and compare the result to our nominal F444W photometry. We correct our F770W measurements using this ratio, which accounts for the fraction of flux lost due to the larger PSF, assuming the F444W image traces the profile of the object well.

\subsection{The SE++ catalog} \label{sec:TheSE++catalog}

Photometric measurements are made in 37 bands, summarized in Table \ref{tab:band_infos} and Fig.~\ref{fig:depth-of-bands}, using the nominal images (non PSF-homogenized). There are significant band-to-band PSF FWHM variations between ground- and space-based images in the COSMOS-Web dataset, ranging from $\sim 0.04 \arcsec$ to $\sim 0.6 \arcsec$. For this reason, we adopt a multi-band model-fitting approach using \texttt{SourceXtractor++} where parametric models convolved with the corresponding-band PSF (described in \S\ref{sec:PSFEx-reconstruction}) are fitted to all detected sources in all available bands.

\subsubsection{\texttt{SourceXtractor++}} \label{sec:SE++desc}

\texttt{SourceXtractor++} \citep[hereafter \SEpp]{bertin20,kummel20, Kummel2022} is a multi-band, multi-object model-fitting engine developed for the \textit{Euclid} mission \citep{Laureijs2011, EuclidMellier2024} and a successor to \texttt{SExtractor2.0} \citep{bertin_sextractor_1996}. The photometric and morphological parameter recovery was tested extensively in the \textit{Euclid Morphology Challenge} where it achieved the highest scores \citep{HubertEMC2022, MerlinEMC2022}. \SEpp\ is an optimized photometric measurement program, including flexible profile fitting (combining S\'ersic, constant, and point-source models) with simultaneous coupled-parameter multi-band fitting. Additionally, it directly uses the native WCS information, avoiding the need for image resampling. This makes \SEpp\ highly efficient for accurate multi-band photometric and morphological measurements. Importantly, since \SEpp\ fits a source model convolved with the corresponding PSF in each band, we use the native images that are not PSF homogenized in contrast with the \hotcold\ catalog. Our catalog preparation technique is described in the following sections. 

\begin{figure*}[t!]
\includegraphics[width=1\textwidth]{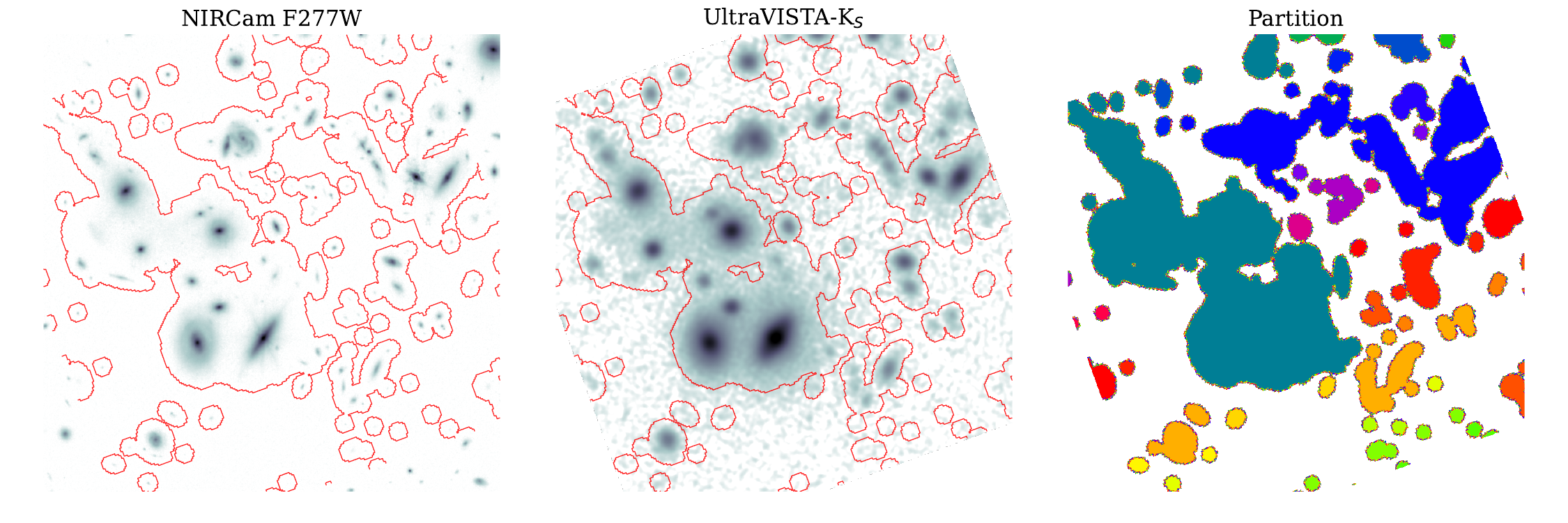}
\caption{Cutout of a relatively crowded region in the high-resolution NIRCam (left), the low-resolution ground-based \UVISTA $K_{S}$-band image (middle), and the partition image (right). Groups are defined using the partition image such that sources falling in the blobs with connecting pixels (here showed in the same color) are assigned the same \texttt{group\_id}.}
\label{fig:DetectionSegmentation_example}
\end{figure*}

\subsubsection{Grouping and iterative fitting} 
In \SEpp\, neighboring sources are grouped together and fitted simultaneously. This is essential to accurately retrieve the flux and shape of the sources without contamination from neighboring sources. Additionally, incorporating lower-resolution ground-based images -- where light profiles from neighboring sources overlap significantly -- requires more conservative, larger groups than would be required than if only NIRCam images were included. We build the groups in two steps. First, we draw ellipses around all sources by taking the positions, Kron radius, axis ratio and position angle from the input \hotcold\ catalog. We scale the Kron radius (measured on the NIRCam images) with a factor that corresponds to the ratio of the NIRCam and \UVISTA\ PSF FWHM, and clip it between values of 0.7 and 3.0 $\SI{}{\arcsec}$. Next, we construct a $0.3\,\sigma$ threshold segmentation map (masking the stars) on the \UVISTA\ $K_{S}$ band. This is necessary to account for the overlapping wings of relatively bright nearby sources (see Fig.~\ref{fig:DetectionSegmentation_example}) and ensure that these are grouped and fitted together. Finally, we combine the images from the two steps to create the final partition map, where groups with connecting pixels are given a unique \texttt{group\_id}. Sources from the input \hotcold\ catalog whose centroids fall in the same group are assigned the same \texttt{group\_id}. 

Figure~\ref{fig:DetectionSegmentation_example} illustrates this technique, showing a relatively crowded region in NIRCam F277W, \UVISTA\ $K_{S}$ and the resulting partition image. The red contours encircle the sources that are grouped together, while the partition image shows the groups color-coded by their \texttt{group\_id}.

Another feature of \SEpp\ is the iterative fitting of grouped sources. Because these are processed simultaneously, many parameters must be fitted. Iterative fitting deals with this by fitting the brightest source\footnote{This is found by \SEpp\ from an initial internal run that measures iso-photometry.} first, while masking the others. It then subtracts 100\%\footnote{set by the parameter \texttt{set\_deblend\_factor(1.0)}} of the estimated flux of the brightest object, and repeats the procedure for the following brightest source. One such iteration over all sources in the group is called a `meta iteration'. We allow three `meta iterations'\footnote{\texttt{set\_meta\_iterations(3)}} with a threshold of 10\% relative change in the $\chi^2$ to consider the meta iterations converged\footnote{\texttt{set\_meta\_iteration\_stop(0.1)}}. This procedure deals effectively with groups of blended sources, increasing the reliability of the fitted models and the derived photometry and morphology.

\begin{figure*}[t!]
\setlength{\abovecaptionskip}{-1mm}
\includegraphics[width=1\textwidth]{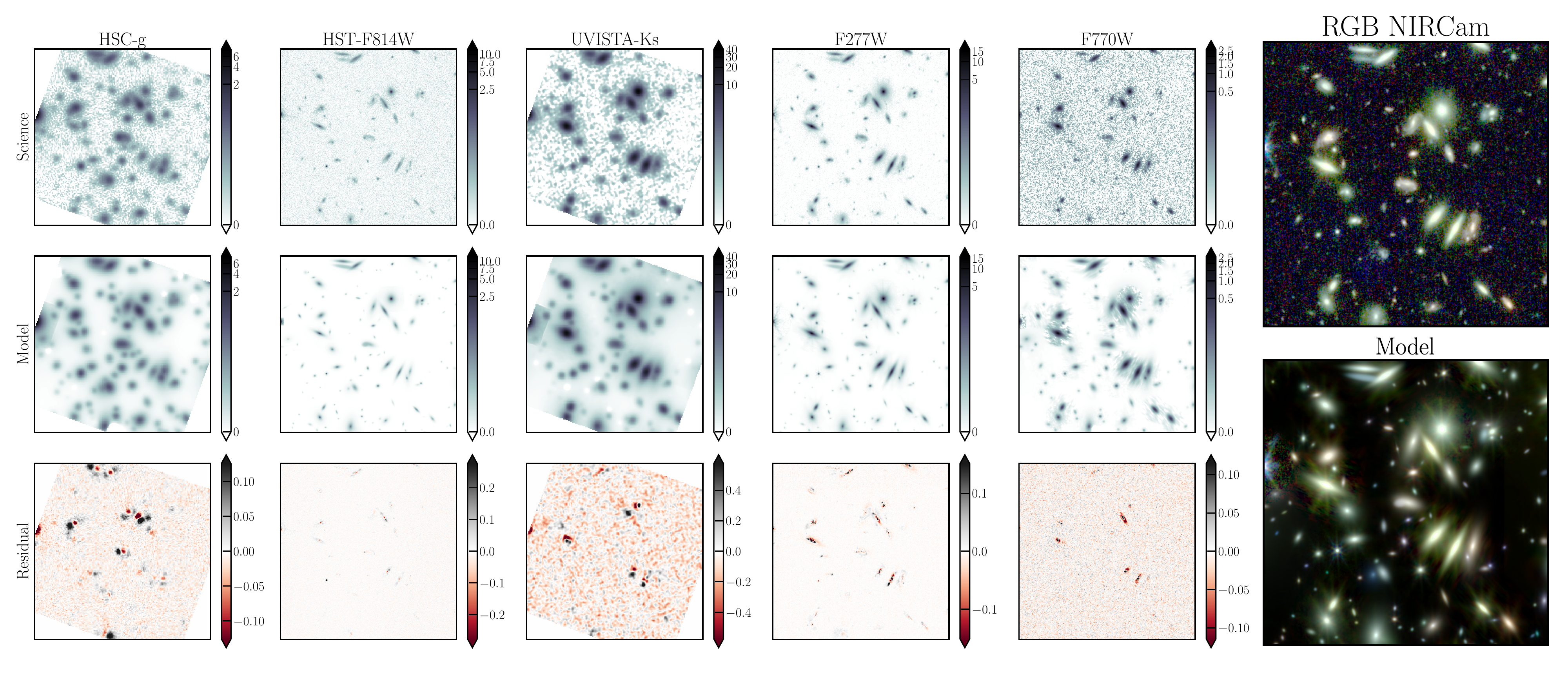}
\caption{Demonstration of the performance of our cataloging methodology with \texttt{SourceXtractor++} over a relatively crowded and randomly selected $\ang{;;20} \times \ang{;;20}$ area. The top/middle/bottom rows show the science/model/residual images in five wide filters: HSC $g$, \textit{HST}/ACS F814W, \UVISTA\ $K_S$, NIRCam F277W and MIRI F770W. The science and model images are log scaled, while the residual is linearly scaled between $\pm 6$ standard deviations of the residual cutout image. The RGB science and model images are made using F444W, F277W and F150W.}
\label{fig:SE++demo}
\end{figure*}

\subsubsection{Two-stage association mode run}
We used the `association' or `\texttt{assoc}' mode in \SEpp\ \texttt{v0.22} to extract the photometric catalogs using a measurement catalog derived from our hot and cold detection scheme (\S \ref{sec:hotcold-detection}). Specifically, we used the \texttt{no-detection} functionality of the \texttt{assoc} mode available in \SEpp\ \texttt{v$\geq$0.21}, to bypass the need to provide a detection image. The input \texttt{assoc} file provides initial values for the centroids and the model parameters (flux, half-light radius and axis ratio) that are subsequently fitted for each source on the measurement images.

We adopt a two-stage \texttt{assoc} mode run where in the first, modeling stage, we fit the structural parameters (including the total flux) of the parametric models on the four NIRCam bands simultaneously. In the second, forced photometry stage, we fit only for the photometry in the remaining bands. In the following, we describe the two stages in detail.

\subsubsection{Modelling run} \label{sec:modeling-run}

In the modelling run, we fit user-defined models on every detected source in the input \texttt{assoc} catalog. For each source or a group of sources, the fit is done inside a square frame whose size is defined using the source Kron size with a  $\sim25\%$ margin. We inspected that these model frames are large enough to account for the total light of the source by imposing a minimum frame size and appropriate scaling factors to the Kron size\footnote{In the assoc mode this is regulated by the \texttt{assoc-source-sizes} configuration parameter.}. Models are convolved with the PSF, rasterized following the pixel grid and WCS, and fitted to the science image of each corresponding band by using a modified least-squares function that aims at minimizing the residual; we used the \texttt{WHT} maps for this. The $1\,\sigma$ uncertainty estimates of the model parameters are obtained from the covariance matrix of the fit, which is computed by inverting the approximate Hessian matrix of the loss function at the best-fit values.

To enhance the scientific utility of the catalogs, we perform two independent runs of \texttt{SE++} fitting all sources with the following models:

\begin{enumerate}
\item \textbf{S\'ersic} models \citep{Sersic1963}, defined by the S\'ersic index $n_{\rm S}$, effective radius $R_{\rm S, eff}$, axis ratio $(a/b)_{\rm S}$, position angle $\theta_{\rm S}$, and total flux $f_{\rm S, tot}$. We parametrize and fit the ellipticities $e_1$ and $e_2$, which are directly related to the axis ratio. We place priors on $R_{\rm S, eff}$, $n_{\rm S}$, and $e_1$, $e_2$, as shown in Fig.~\ref{fig:priors};

\item \textbf{Bulge + Disk} models, built as composites of an exponential disk ($n_{\rm S}=1$) and a de Vaucouleurs bulge ($n_{\rm S}=4$) light profiles. We parametrize them by the effective radii of the bulge, $R_{\rm B, eff}$, and disk, $R_{\rm D, eff}$; the axis ratios, $(a/b)_{\rm B}$ and $(a/b)_{\rm D}$; a common position angle $\theta_{\rm BD}$, shared by both components, which we assume to be coaxial; the total flux of both components, $f_{\rm BD, tot}$; and the bulge-to-total ratio, $B/T = f_{\rm B, tot}/f_{\rm BD, tot}$. From this, we also derive the total fluxes of each component, $f_{\rm B, tot}$ and $f_{\rm D, tot}$. We apply priors to $R_{\rm B, eff}$, $R_{\rm D, eff}$, $(a/b)_{\rm B}$, and $(a/b)_{\rm D}$ as shown in Fig.~\ref{fig:priors}. The $B/T$ prior follows a bell curve ranging from $5\times 10^{-5}$ to $1$, with a mean and spread that increase with wavelength.

\end{enumerate}

For each source, the same set of structural parameters ($R_{\rm eff}, \, n, \, a/b, \, \theta$) is fitted on the four NIRCam modeling bands (F115W, F150W, F277W, F444W), and we do not let them vary with wavelength. This makes the resulting parameters an effective average of all fitted bands, weighted by the corresponding weight map. As such, the structural parameters correspond to the averaged morphology over the $1-5$ \SI{}{\micro \meter} wavelength range.

\subsubsection{Forced-photometry run} \label{sec:forced-photomerty}

In the second, forced-photometry stage, we run on \MIRI\ F770W, \textit{HST}/ACS F814W and the remaining ground-based and IRAC data by fixing the structural parameters from the modeling run and fitting only for the flux and a small coordinate offset. These are initialized with the best-fit parameters from the modeling run; in the case of the flux, the initial value for all bands is a mean of the flux in the four modeling bands from the first run. The total flux is fitted for all bands independently, without imposing any prior, while the coordinate offset ($\pm0.05 \, \SI{}{\arcsec}$) is fitted for all bands jointly. However, we note that this coordinate offset is not enough to capture the proper motion of stars due to the relatively large difference between the observation epochs of different bands; this is out of the scientific scope of this work.

Figure~\ref{fig:SE++demo} demonstrates the source extraction with \SEpp\ on a randomly selected $\ang{;;20} \times \ang{;;20}$ area, by showing the science, model and residual images of five wide filters from both modelling and forced photometry run. The models accurately reproduce the flux distribution in the science images, yielding clean residual images. This validates our methodology in measuring photometry for every NIR-detected source across many ground- and space-based bands. However, some residuals remain in the HSC $g$ because low-redshift galaxy sizes typically decrease with wavelength \citep[e.g.,][]{Vulcani2014}, whereas we fix sizes to measurements at \SIrange{1}{5}{\micron}.

\subsection{Calibration of uncertainties} \label{sec:uncertainty-calibration}

Up to this point in our analysis, the photometric errors accounted only for photon noise from detected sources and not for Poisson noise from the background. Consequently, the uncertainties in the catalog are underestimated for faint objects. We compute correction factors for both the \hotcold\ catalog (based on PSF-homogenized mosaics) and for \SEpp\ catalog (based on native-resolution images). For aperture photometry in each band, we place 200,000 random circular apertures in a range of diameters from \ang{;;0.03} to \ang{;;1} on the noise-equalized version of each mosaic, avoiding regions containing detected sources (as defined by the segmentation maps). For a fixed aperture size and filter, we empirically determine the background Poisson noise by fitting the negative tail of the distribution of measured fluxes with a Gaussian function. Following \citet{Labbe2003, Gawiser2006, Whitaker2011, Skelton2014, Rieke2023, Finkelstein2024}, we model the Poisson background noise as a function of aperture area (number of pixels, $N$) by fitting the relation:
\begin{equation}
\sigma_N = \alpha \, N^{\beta/2},
\end{equation}
where $\sigma_N$ is the noise in an aperture containing $N$ pixels, and $\alpha$ and $\beta$ are fit parameters.

Calibrating model-based photometric uncertainties is  more complex. \SEpp\ determines the best-fit photometric and structural parameters of a model by minimizing a loss function, which is a $\chi^2$ difference between the model and observed data. The uncertainties on these parameters are derived using the approximate Hessian matrix of the loss function at the best-fit values, which approximates the covariance matrix of the fitted parameters. However, these uncertainties are similar to standard aperture photometry in that they only directly account for photon noise from the source and are typically underestimated by about a factor of two \citep{HubertEMC2022, MerlinEMC2022}. 

Model-based photometry is not measured in a fixed area, unlike circular (or elliptical) apertures, complicating the uncertainty calibration. However, that uncertainty should scale with the effective size over which photometry is measured, and so we first must generate an estimate of the effective aperture area for each source given its model characteristics.
To do this, we generate models on a three-dimensional grid of $R_{\rm eff}$, $n_{\rm S}$ and $a/b$ with 20 points along each dimension. For each point on the grid and for each band, we generate a Sérsic model on the image pixel grid and convolve it with the corresponding PSF. 

We then generate elliptical apertures which contain 90\%\ of the model flux and count the effective pixel area in that elliptical aperture, $N_{\rm eff}$. The adoption of 90\%\ contained flux was calibrated empirically against circular sources in the field whose aperture photometry is quite similar to their model-based photometry.  Then for each source in our catalog, we measure $N_{\rm eff}$ by interpolating numerically from the $R_{\rm eff}$, $n_{\rm S}$, $a/b$ grid using \texttt{scipy}'s \texttt{RegularGridInterpolator}. We then adopt the same power law scaling measured using circular apertures ($\sigma_{N_{\rm eff}} = \alpha \, N_{\rm eff}^{(\beta/2)}$) to generate an estimate of the Poisson background noise for model-based photometric measurements. 

All photometric uncertainties already present in the hot and cold \ and \SEpp\ catalogs are then added in quadrature with this calculation of Poisson background noise (or random aperture noise per source and per band), $\sigma_N$. We note that generally, the Poisson background noise calculation dominates the error budget of sources close to the detection threshold $\lesssim$3$\sigma$).


\subsection{Cleaning and flagging} \label{sec:cleaning-flagging}

The catalog may contain spurious sources or objects with incorrect photometry. In this Section, we describe the procedures we  adopted to identify these problematic sources and identify them with the \warnfl keyword. 

\subsubsection{Hot pixels} \label{sec:hot-pix}

Despite extensive efforts to mitigate imaging artifacts, small clusters of 1–4 pixels with unusually high S/N remain in our NIRCam mosaics. Although masking these `hot pixels' on the images would be the best approach, this task is impractical for COSMOS-Web, therefore hot pixels remain in the final mosaics. We therefore flag hot pixels at the catalog level.

PSF homogenization (for the construction of the detection image) renders hot pixels less obviously identifiable; thus, many are included as sources in the initial \hotcold\ catalog.  Nominally, one should make a concerted effort to remove them from \hotcold\ before proceeding to the next stage of catalog construction.  This is because, in theory, known false sources present in the catalog could impact the model photometry of neighbor sources.  However, in making this catalog, we have discovered that removing them at a later stage (after running \SEpp) produces nearly identical results. This is because the S\'ersic model fits built on NIRCam imaging for hot pixels are intrinsically small ($<$\,1 pixel).  Therefore, we proceed by flagging them only after the full catalog is made.

Hot pixels are identified using a curve of growth analysis, where concentric circular apertures measure the enclosed flux on native resolution NIRCam images.  Sources for which concentric apertures indicate a morphology more compact than the intrinsic PSF in any of the four NIRCam bands are flagged as hot pixels.  Specifically, we calculate the ratio of flux densities, $F_\nu$, in circular apertures that are 0$\farcs$1 and 0$\farcs$25 in radius, dubbed apertures 0 and 1 respectively, i.e. $R_{\rm [filt]}\equiv F_{\rm [filt],0}/F_{\rm [filt],1}$.  Note this is applied to non-PSF homogenized aperture photometry, which is computed as described in \S~\ref{sec:modeling-run}. We then derive an error on $R_{\rm [filt]}$, $\sigma_{\rm R[filt]}$ using error propagation (neglecting covariance terms). We then empirically calibrate the threshold ratios for each filter using their native resolution, above which we only expect hot pixels to reside.  These thresholds $R_{\rm thresh}$ are 0.75, 0.70, 0.39, and 0.36 for F115W, F150W, F277W, and F444W respectively.  Sources are flagged as hot pixels if they are more than 1$\sigma$ above this threshold in any of the four filters, i.e. $R_{\rm source,[filt]} - \sigma_{\rm R[filt]} \ge R_{\rm thresh[filt]}$. This curve-of-growth criterion results in 38,916 sources flagged as hot pixels (of which 12,563 are brighter than [F277]$<$28.2).

After iteratively visually inspecting sources flagged as hot pixels, it was determined that a subset were real galaxies and could be recovered by requiring additional criteria for flagging.  For example, if best-fit Sérsic model radii are found to be smaller than a single NIRCam pixel (0$\farcs$03) then the hot pixel flag is retained; if the fit is larger, it is likely a true astrophysical source whose emission may manifest like a hot pixel in one of the four NIRCam filters, but modeled across all filters (\S~\ref{sec:modeling-run} ), is substantially larger than a pixel.  Modifying the hot pixel criteria with this size criterion reduces the number of sources flagged as hot pixels to 13,241 (of which 6,940 are brighter than [F277]$<$28.2).

To summarize, hot pixels are identified as sources fulfilling the curve of growth criteria for an  NIRCam filter and have \SEpp+ S\'ersic model fits with radii $<$0$\farcs$03. Hot pixels are assigned \warnfl$=1$.

\subsubsection{Assuring consistency between space- and ground-based bands} 
Inconsistent photometry between space- and ground-based bands is identified in the \SEpp\ catalog in cases where the flux ground-based bands disagree with those from NIRCam bands at the similar wavelengths. This mainly happens for relatively faint sources near the detection limits which are blended with brighter neighboring sources in the low-resolution ground-based bands. Due to association confusion, \SEpp\ can fit a source that has a biased flux several magnitudes brighter than what is measured at similar wavelength in the high-resolution NIRCam where there is no association confusion. The improved and conservatively large grouping described in Sect.~\ref{sec:TheSE++catalog} helps alleviate many such association confusions, but a fraction remains. We flag sources with inconsistent ground‑ and space‑based photometry by identifying those where the lower flux uncertainty in UltraVISTA $Y$, $J$, or $H$ bands exceeds twice that of the nearest NIRCam band, as follows:

\begin{equation}\label{condition_warn_fl_1}
\begin{split} 
& (2\times F_{\rm F115W} < (F_{Y} - \delta F_{Y} ) ) \, \& \, \delta {\rm mag}_{Y} < 0.5 \\
& (2\times F_{\rm F115W} < (F_{J} - \delta F_{J} ) ) \, \& \, \delta {\rm mag}_{J} < 0.5  \\
& (2\times F_{\rm F150W} < (F_{J} - \delta F_{J} ) ) \, \& \, \delta {\rm mag}_{J} < 0.5  \\
& (2\times F_{\rm F150W} < (F_{H} - \delta F_{H} ) ) \, \& \, \delta {\rm mag}_{H} < 0.5
\end{split}
\end{equation}
If a source satisfies these four conditions then it is assigned \warnfl$=2$. Unsurprisingly, many \warnfl$=2$ are nearby bright stars that in the ground-based bands contaminate the flux of many neighboring sources. Outside HSC star masks, about $1.6 \%$ of the sources have \warnfl$=2$ (Table~\ref{table:selection-numbers}).

Additionally, we identify a second case where the ACS/NIRCam flux is bright enough so that the sources should be detected in the HSC and UltraVISTA bands but are not. In this case, the model magnitude is zero or negative but the $\ang{;;}$ diameter aperture flux in the corresponding band is positive $F_{[\rm filt]} (1'')>0$.  

\begin{equation}\label{condition_warn_fl_2}
\begin{split} 
& F_{i} \leq 0 \, \& \, \delta F_{i} > 0  \, \& \, {\rm mag}_{\rm F814W} < 28 \, \& \, \delta F_{\rm F814W} > 0 \\
& F_{H} \leq 0 \, \& \, \delta F_{H} > 0  \, \& \, {\rm mag}_{\rm F150W} < 26 \, \& \, \delta F_{\rm F150W} > 0 \\
& F_{K_S} \leq 0 \, \& \, \delta F_{K_S} > 0  \, \& \, {\rm mag}_{\rm F277W} < 26 \, \& \, \delta F_{\rm F277W} > 0 
\end{split}
\end{equation}
If a source satisfies these three conditions, then it is assigned \warnfl$=3$. Table~\ref{table:selection-numbers} quantifies the number of sources affected by these flags.

\subsubsection{Other artifacts} 
We identify and flag other artifacts as follows: 
Sources with detection in only one NIRCam or MIRI band are predominantly `snowballs' and hot pixels and are flagged with \warnfl$=4$. These have a  magnitude error lower than 0.2 mag in one of the NIRCam or MIRI bands, but a magnitude error larger than 0.5 mag in all other bands (or below the limiting magnitude of the band). Sources with unrealistically small radii ($R_{\rm eff}<0.00047$ \ang{;;}) are typically noise detections and are flagged with \warnfl$=5$.
Finally, sources with an unrealistic flux ratio between $0{\farcs}25$ and $0{\farcs}1$ that is smaller than the ratio identified for stars are also identified to be snowballs and hot pixels and are flagged with \warnfl$=6$. More precisely, the flux ratio condition is $F_{\rm F277W}(0{\farcs}25)/F_{\rm F277W}(0{\farcs}1) < 1.6$ and $F_{\rm F444W}(0{\farcs}25)/F_{\rm F444W}(0{\farcs}1) < 1.9$.

In summary, we provide the \warnfl\ to flag artifacts and sources with potentially problematic photometry and derived photo-$z$. For the most secure sources that do not satisfy any of the criteria described above, we assign \warnfl$=0$, and we advise users to use this sample for most scientific applications. Sources with \warnfl$=2,3$ are to be handled with care; they have NIRCam and ACS photometry that is not problematic, but the issues with the ground-based photometry prevents robust SED fitting. We advise users to discard sources with all the other flags \warnfl$=1,4,5,6$ from scientific analysis, or to carefully inspect them.

\subsubsection{HSC star mask}
In our ground-based imaging, and in particular the HSC data, the larger PSF of bright stars compared to \JWST\ data means more sources are  by photometric contamination. We flag these sources using the HSC star masks \citep{Coupon_HSC_Masks_2018} from the COMOS2020 catalog \citep{weaver_cosmos2020_2022}. These are conservative and flag all sources with flux contamination from stars in all ground-based bands. Table~\ref{table:selection-numbers} shows the number of sources remaining in the catalog after applying the HSC star mask flag. We flag sources whose ground-based photometry is affected by bright stars with \texttt{flag\_star\_hsc}$\ =1$. Compared to the total number, $17\%$ of the sources are affected by the HSC star mask, therefore having unreliable ground-based photometry.

\begin{table}[t!]
\begin{threeparttable}
\begin{center}
\setlength{\tabcolsep}{4pt}
\caption{Description of the quality and star mask flags}
\begin{tabular}{c|c|c}
\hline
\hline
Case & $N$\tnote{a} & Flag  \\
\hline

Total number of sources & 784,016 &   \\

Most secure sources & 694,341 & \warnfl$=0$   \\

Hot pixels & 13,241 & \warnfl$=1$   \\

Ground vs. space incons. 1 & 51,793\tnote{c} & \warnfl$=2$   \\

Ground vs. space incons. 2  & 157\tnote{d} & \warnfl$=3$   \\

Single NIRCam detec. & 17,520 & \warnfl$=4$   \\

Radius too small &  133 & \warnfl$=5$   \\

Small flux ratio  & 6,831 & \warnfl$=6$   \\

\hline
\hline
Star mask & $N$\tnote{b} &  Flag \\
\hline
HSC  & 653,932 & \texttt{flag\_star\_hsc}$=0$   \\
\hline
\end{tabular}
\label{table:selection-numbers}
\begin{tablenotes}
\item[a] Number of objects affected by the flag.
\item[b] Number of objects remaining.
\item[c] Of which $39,137$ are inside the HSC star mask.
\item[d] Of which $71$ are inside the HSC star mask.
\end{tablenotes}
\end{center}
\end{threeparttable}
\end{table}

\section{Photometric validation and comparisons} \label{sec:photometry-validation}

In this Section, we compare our two sets of photometric catalogs. For \SEpp\ we provide total photometric quantities in all 37 bands, while for the hot and cold catalogs we provide PSF-homogenized aperture photometry along with aperture-to-total correction in the four NIRCam, one MIRI, and the HST/F814W band. For the \SEpp\ catalog, we also compare the ground-based photometry to the COSMOS2020 catalog. We use the total photometry derived from the S\'ersic model-fitting of \SEpp\ as the primary photometric reference.

\subsection{Magnitude number counts} \label{sec:mag-number-counts}

\begin{figure*}[t!]
  \centering
\begin{subfigure}[b]{0.42\textwidth}
            \includegraphics[width=1\hsize]
            {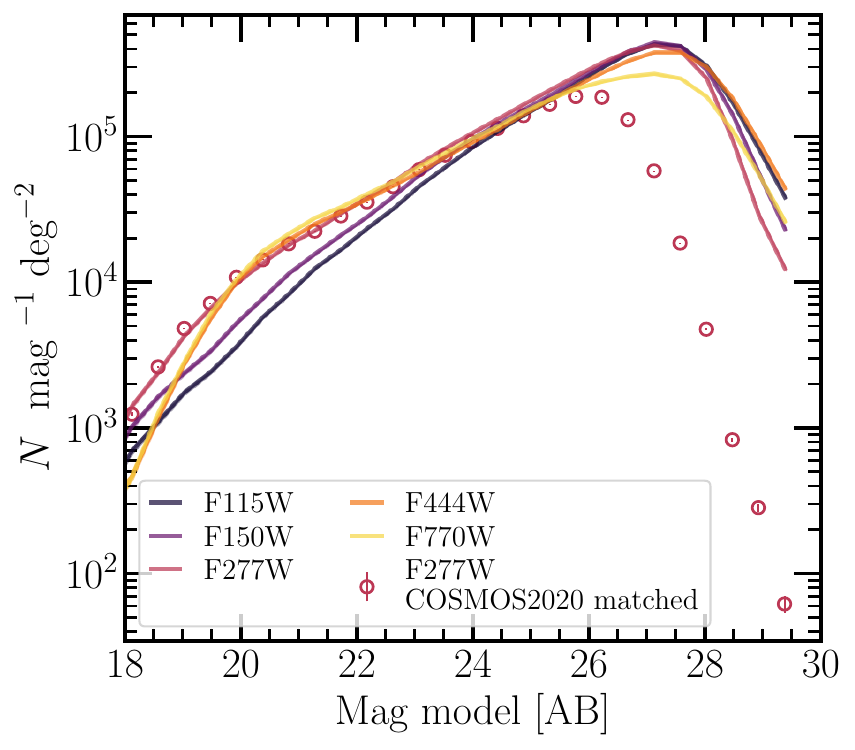}     
  \end{subfigure}
  \begin{subfigure}[b]{0.42\textwidth}
            \includegraphics[width=1\hsize]{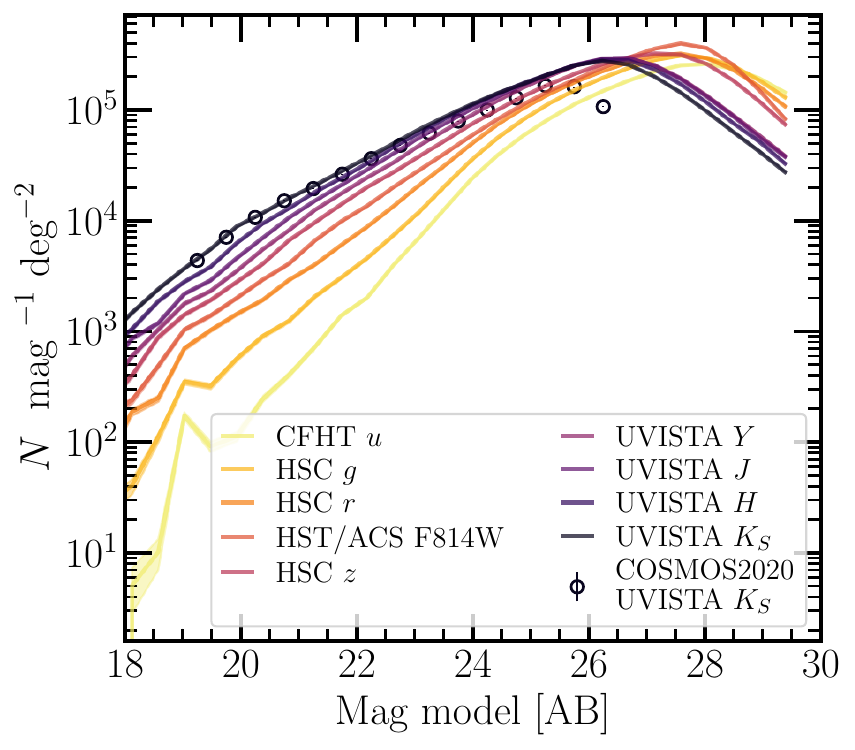}     
  \end{subfigure}
 \caption{Number counts for the \SEpp catalog. The left panel shows counts in NIRCam and MIRI bands. The right panel shows counts in HST/ACS and for ground-based broad-band filters. In both cases, counts are computed in 0.35 magnitude bins and normalized by the effective area after masking, namely $0.43\,{\rm deg}^2$ in all bands except for MIRI, which covers $0.15 \, {\rm deg}^2$.)}
  \label{fig:mag-ncounts}
\end{figure*}

Figure~\ref{fig:mag-ncounts} shows the galaxy number counts computed using \SEpp\ photometry in the \JWST  and \textit{HST}/ACS and the ground-based broad bands. Galaxies are selected using the star-galaxy classification described in Section~\ref{sec:star-galxy-separation}. 

Overall, the slope of the magnitude number counts at intermediate magnitudes shows the expected trend \citep{metcalfe_galaxy_2001} with respect to wavelength, with bluer bands having steeper slopes. As expected  \citep{1993ApJ...415L...9G}, at longer wavelengths, the slope of the counts flattens at intermediate magnitudes. The amplitude of the break becomes more important at redder wavelengths. In a recent paper, \cite{Manzoni2025} provide an excellent summary of the physical origins of these changes.

We also compared with number counts from COSMOS2020. In the left panel of Fig.~\ref{fig:mag-ncounts} we show the F277W number counts for sources in our catalog that match and are detected in COSMOS2020 that turn over at $\sim 25.5$ mag, comparable to the $K_S$ depth of COSMOS2020. At $m_{\rm F277W}\gtrsim25$ the COSMOS2020-matched counts become more shallow which is likely due to the deeper and redder selection function of COSMOS-Web. In the right panel of Fig.~\ref{fig:mag-ncounts} we directly compare the $K_S$ number counts from COSMOS-Web and COSMOS2020 \citep[taken from][]{weaver_cosmos2020_2022}. This includes all sources detected in the respective catalogs. There is a relatively good agreement with two noticeable differences. First, the COSMOS-Web $K_{S}$ counts turn over at a fainter magnitude $\sim 26.5$ compared to $\sim25.5$ for COSMOS2020 due to the deeper UltraVISTA DR6 used in COSMOS-Web versus DR4 used in COSMOS2020. Second, the COSMOS-Web $K_{S}$ counts are slightly steeper, likely due to the deeper and redder selection function.

\subsection{Comparison between \SEpp\ and hot and cold catalogs}

\begin{figure*}[th!]
\centering
\includegraphics[width=\textwidth]{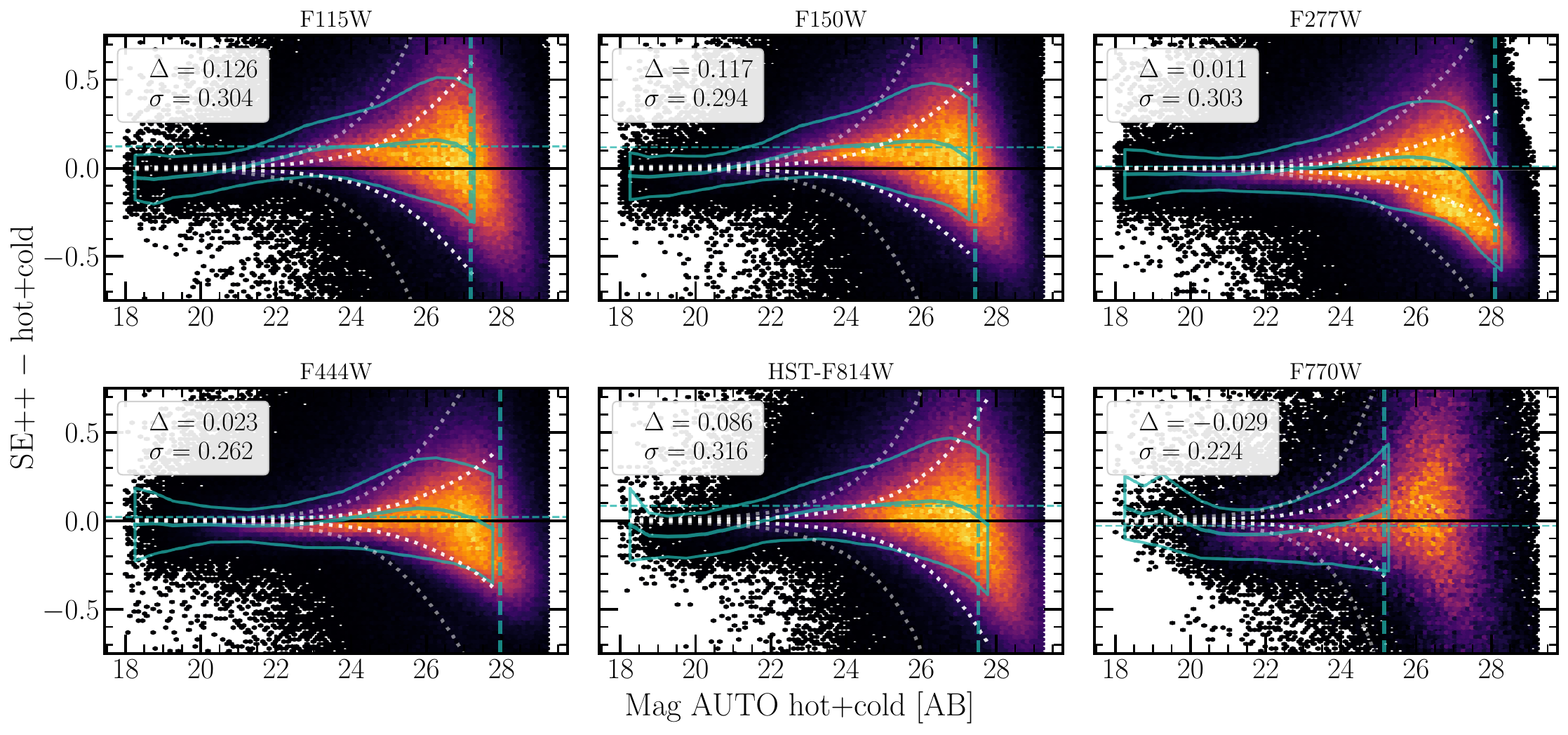}
\caption{
Photometric comparison between \SEpp\ model magnitudes and \texttt{auto} magnitudes from the \hotcold\  catalog, as a function of \texttt{auto} magnitude. The \texttt{auto} photometry is corrected to total as described in \S\ref{sec:hotcold-aperphoto}. The panels show the following bands: F115W (top left), F150W (top centre), F277W (top right), F444W (bottom left), HST F814W (bottom center), and F770W (bottom right). In each panel, the solid teal line indicates the running median offset, while the shaded envelope denotes the $3\,\sigma$-clipped standard deviation. The dotted white curves show the $\pm1\,\sigma$ and $\pm3\,\sigma$ envelopes of the combined photometric uncertainties. The vertical dashed teal line marks the $5\,\sigma$ depth in each band. Summary statistics report the median offset $\Delta$ and standard deviation $\sigma$ for sources brighter than this limit.}
\label{fig:Mag-Compar-SE++hotcold}
\end{figure*}

We compare the \SEpp\ model magnitudes and \texttt{auto} magnitude fluxes in the \hotcold\ catalog in Fig.~\ref{fig:Mag-Compar-SE++hotcold}. The \texttt{auto} photometry is corrected to total as described in \S\ref{sec:hotcold-aperphoto}. We also show the regions  corresponding to the $\pm1\,\sigma$ and $\pm3\,\sigma$ photometric uncertainties, obtained by adding in quadrature the errors from both photometric data sets.

Overall, there is excellent agreement between the two sets of photometric measurements, with a running median offset ($\Delta = {\rm mag}_{\rm SE++} - {\rm mag}_{\hotcold}$) $<0.13$ mag and a median scatter of $\sim0.3$ mag. This offset and scatter is larger for the NIRCam SW F115W and F150W bands and smallest for the NIRCam F277W and F444W, remaining within the $\pm 1\, \sigma$ photometric uncertainty envelope. The offset shows that \SEpp\ magnitudes are overall slightly fainter for faint and brighter for bright sources. These differences can come both from inadequate modeling by \SEpp\ or imperfect PSF homogenization and uncertain aperture-to-total factors applied in \hotcold. Despite these secondary differences, the comparison shows that the two photometry sets are consistent within the uncertainties.

\subsection{Comparison with COSMOS2020} \label{sec:C2020-compar}
To validate model-fitting ground-based photometry from \SEpp, we compare with the COSMOS2020 catalog \citep{weaver_cosmos2020_2022}. COSMOS2020 is a galaxy catalog covering $\sim 2 \, {\rm deg}^2$ in the COSMOS field and provides both aperture photometric measurements (\textsc{Classic} catalog) and from model-fitting (\texttt{The Farmer} catalog). We compare sources matched in coordinates within $0\farcs6$. To ensure a comparison of consistent quantities, i.e.,  model-derived total fluxes, we compare with \texttt{The Farmer} photometry. 

There are two noticeable differences in the methodologies of COSMOS2020/\texttt{The Farmer} and \SEpp\ that need to be kept in consideration in the comparison. The first one is the different models that are fitted to sources. The \SEpp\ catalog fits S\'ersic models to all sources, where the structural parameters are fitted on the NIRCam bands simultaneously. \texttt{The Farmer}, on the other hand, uses a decision tree to decide which one out of five discrete models to fit. This is described in detail in \cite{WeaverFarmer2023}, but briefly, these models are a point-source, a circularly symmetric and exponential light profile at a fixed effective radius, an exponential profile (S\'ersic model with $n_{\rm S}=1$), a de Vaucouleurs profile (S\'ersic model with $n_{\rm S}=4$) and a composite of an exponential and a de Vaucouleurs profile. The structural parameters are fitted on a \texttt{chi-squared} combination of $izYJHK_{s}$ bands and the photometry is measured on all bands individually in a forced photometry approach, similar to the \SEpp\ catalog. The second difference is the input dataset. These are presented in detail in Section~\ref{sec:data} of this paper and in Section 2 of \cite{weaver_cosmos2020_2022}, but briefly, the main difference is that we use HSC PDR3 data processed with the HSC pipeline, as opposed to HSC PDR2 and a custom pipeline in \cite{weaver_cosmos2020_2022}. Furthermore, for the UltraVISTA data we use DR6, as opposed to DR4 in COSMOS2020. Keeping these differences in data and methodology in mind, we compare the magnitudes and colors of both independent catalogs.

\begin{figure*}[ht!]
\includegraphics[width=1\textwidth]{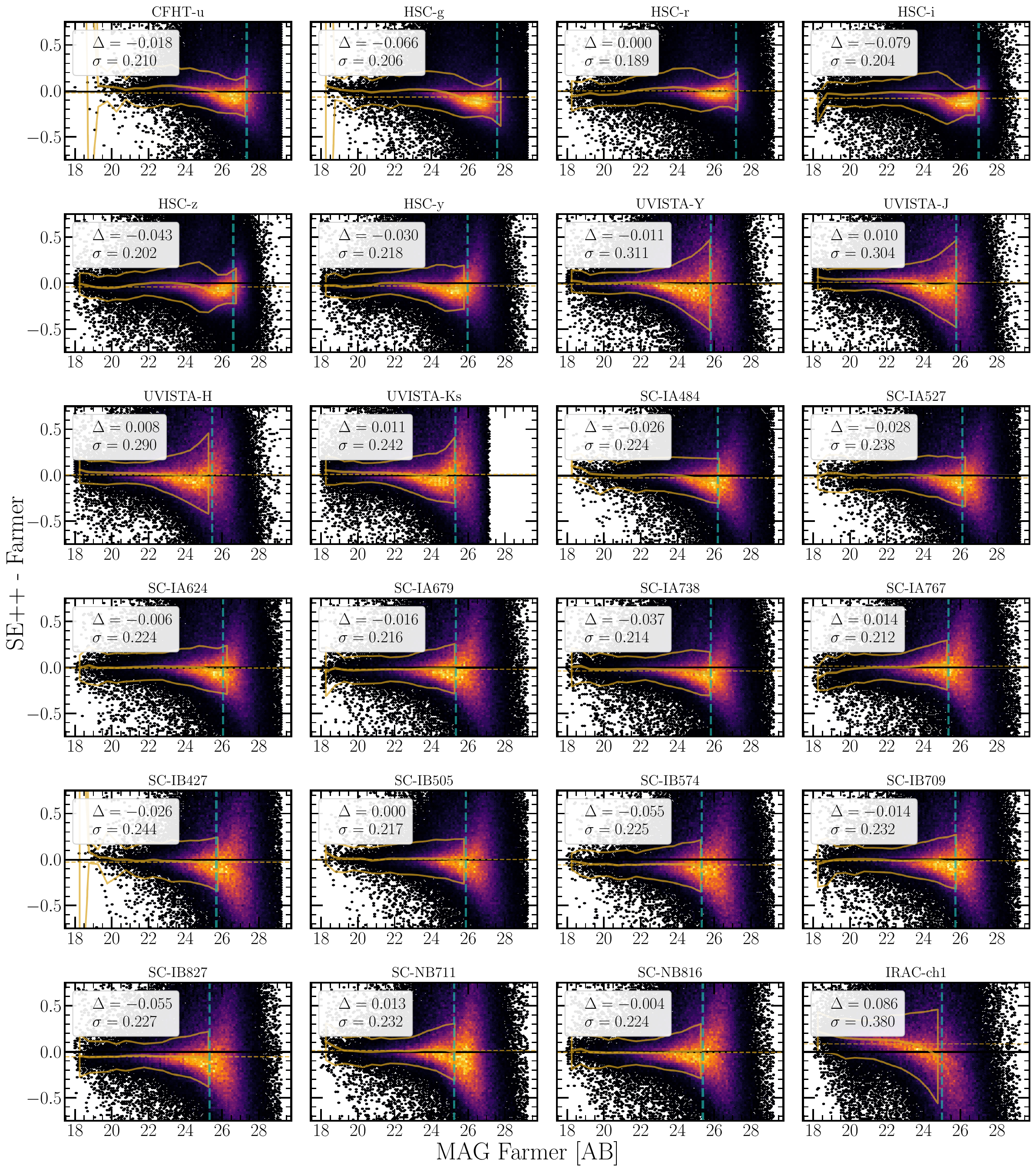}
\caption{Photometric comparison with the COSMOS2020 \textsc{The Farmer} catalog. The density histogram shows the difference between \SEpp\ and \textsc{The Farmer} as a function of \textsc{The Farmer} magnitude. The solid yellow line shows the median offset, while the envelope marks the $3\,\sigma$-clipped standard deviation. The dashed line shows the median offset computed down to the $5\, \sigma$ depth magnitude, shown by the teal vertical dashed line. The legend shows the average offset and standard deviation for magnitudes brighter than the $5\, \sigma$ depth.}
\label{fig:Mag-Compar-C20F}
\end{figure*}

\subsubsection{Photometric comparisons with COSMOS2020}
In Figure~\ref{fig:Mag-Compar-C20F} we compare the total magnitudes from \SEpp\ and \texttt{The Farmer} in all the broad, medium and narrow ground-based bands and one IRAC (ch1) band. The density histogram shows the magnitude difference between \SEpp\ and \texttt{The Farmer} as a function of magnitude. We mark the median offset with a solid yellow line, and the $3\,\sigma$-clipped standard deviation with an envelope; in the legend we show the averaged offset ($\Delta$) and standard deviation ($\sigma$) for magnitudes brighter than the $5\, \sigma$ depth.

Overall, there is excellent agreement between the two sets of photometric measurements with averaged offset $\Delta \lesssim0.08$ mag, and with $\sigma \lesssim 0.3$ for all bands. Additionally, there is no significant magnitude trend of the rolling offset higher than $1\, \sigma$ significance. There are, however, some noticeable second-order trends that are more prominent for the HSC bands. This refers to the kink around $25-26$ mag towards brighter \SEpp\ magnitudes (but offset less than $\sigma$). This could be due to several reasons, from the different image background modeling to the difference in PSF and source modeling in \texttt{The Farmer}, as mentioned above. 

Finally, the offset and scatter is largest for the IRAC bands at fainter \SEpp\ magnitudes. This is likely due to our simplistic treatment of the IRAC PSF where we model the high spatial variability with a second-order polynomial (as implemented by \psfex). Further complication is the confusion by forcing models that are significantly below ($\sim 1.5$ mag) the depth of the IRAC bands (compared to the NIRCam detection) coupled with the severe source blending. We do not investigate or correct this in further detail, since the IRAC bands are largely redundant, given the \JWST\ coverage. We only include them in the photometry for legacy value and do not use them in the photo-$z$ and physical parameter inference.

\begin{figure*}[ht!]
\centering
\includegraphics[width=1\textwidth]{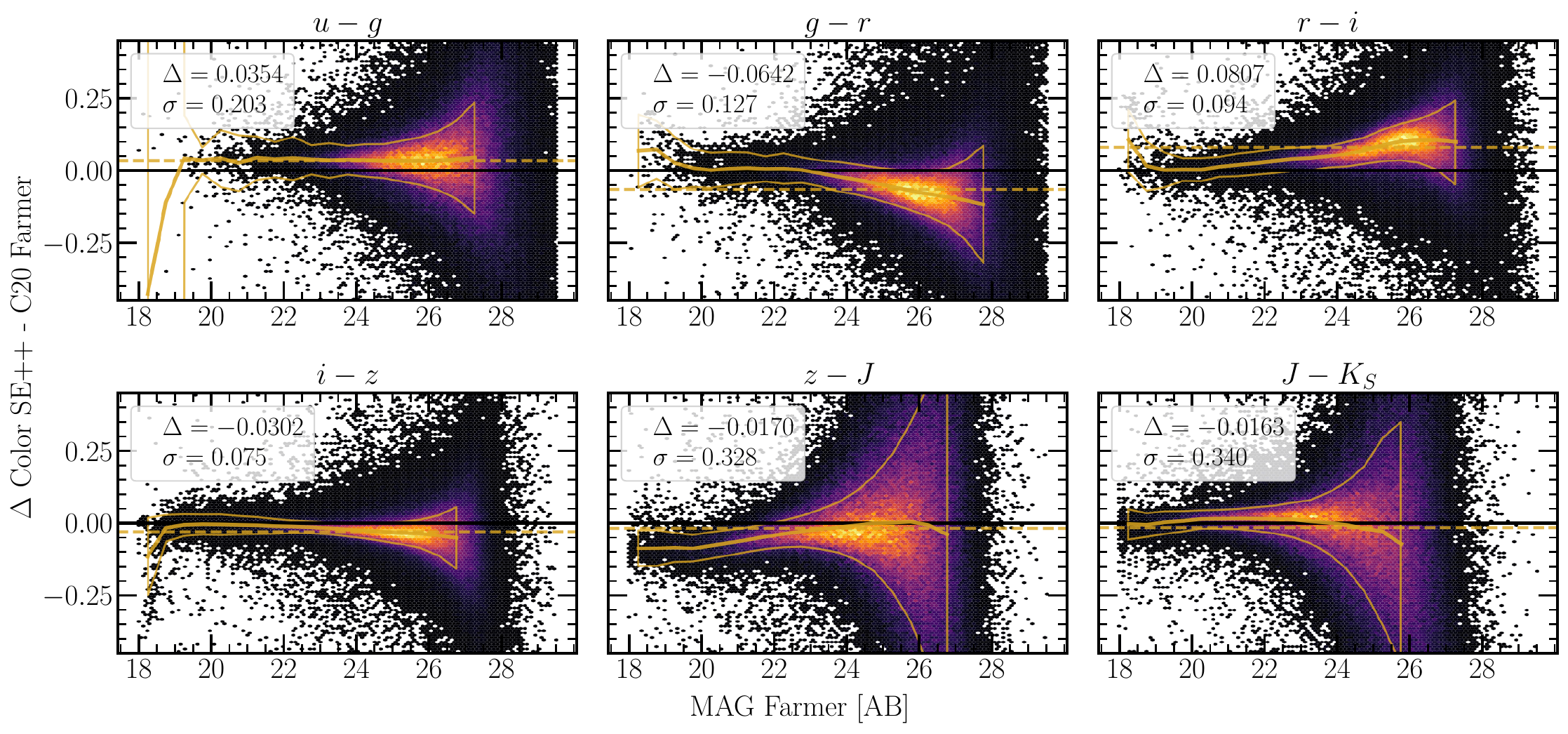}
\caption{Color comparison in the ground-based bands with the COSMOS2020 \textsc{Farmer} catalog. Similarly to Fig.~\ref{fig:Mag-Compar-C20F}, the density histogram shows the color difference between \SEpp\ and the \textsc{Farmer} as a function of magnitude. The solid yellow lines and envelopes show the median offset and the $3\,\sigma$-clipped standard deviation. The legend shows the averaged offset (also marked in dashed yellow line) and the standard deviation for magnitudes brighter than the $5\,\sigma$ depth.}
\label{fig:Color-Compar-C20F}
\end{figure*}

\subsubsection{Color comparisons with COSMOS2020} 
Because colors provide the most stringent constraints in SED fitting, validating them is crucial. We compare colors from \SEpp\ with \texttt{The Farmer} photometry in COSMOS2020 for $u-g$, $g-r$, $r-i$, $i-z$, $z-J$, and $J-K_{S}$ bands in Fig.~\ref{fig:Color-Compar-C20F}. The density histogram shows the color difference as a function of magnitude, and the solid yellow line and envelope show the rolling median offset and $3\,\sigma$-clipped standard deviation. 

In overall, there is excellent agreement in the colors with an averaged color offset less than $0.08$ mag, and a standard deviation of $<0.4$. The difference is highest for the $r-i$ color, where \SEpp\ shows redder colors with about $1\, \sigma$ significance for the faint end. In the case of $z-J$, there is also a $1\, \sigma$ significant offset towards bluer \SEpp\ colors for bright sources. Additionally, there are small second-order curvatures for the faintest magnitudes, more notably for the HSC bands, similar to the magnitude comparison, and likely the same origins. This can be due to slight differences between the HSC DR2 versus DR3 data and photometric calibrations. Nonetheless, the overall excellent agreement in both photometry and colors means that the \SEpp\ photometry passes this test.

\section{Morphological measurements} \label{sec:morphology}

\begin{figure*}[t!]
\centering
\includegraphics[width=0.98\textwidth]{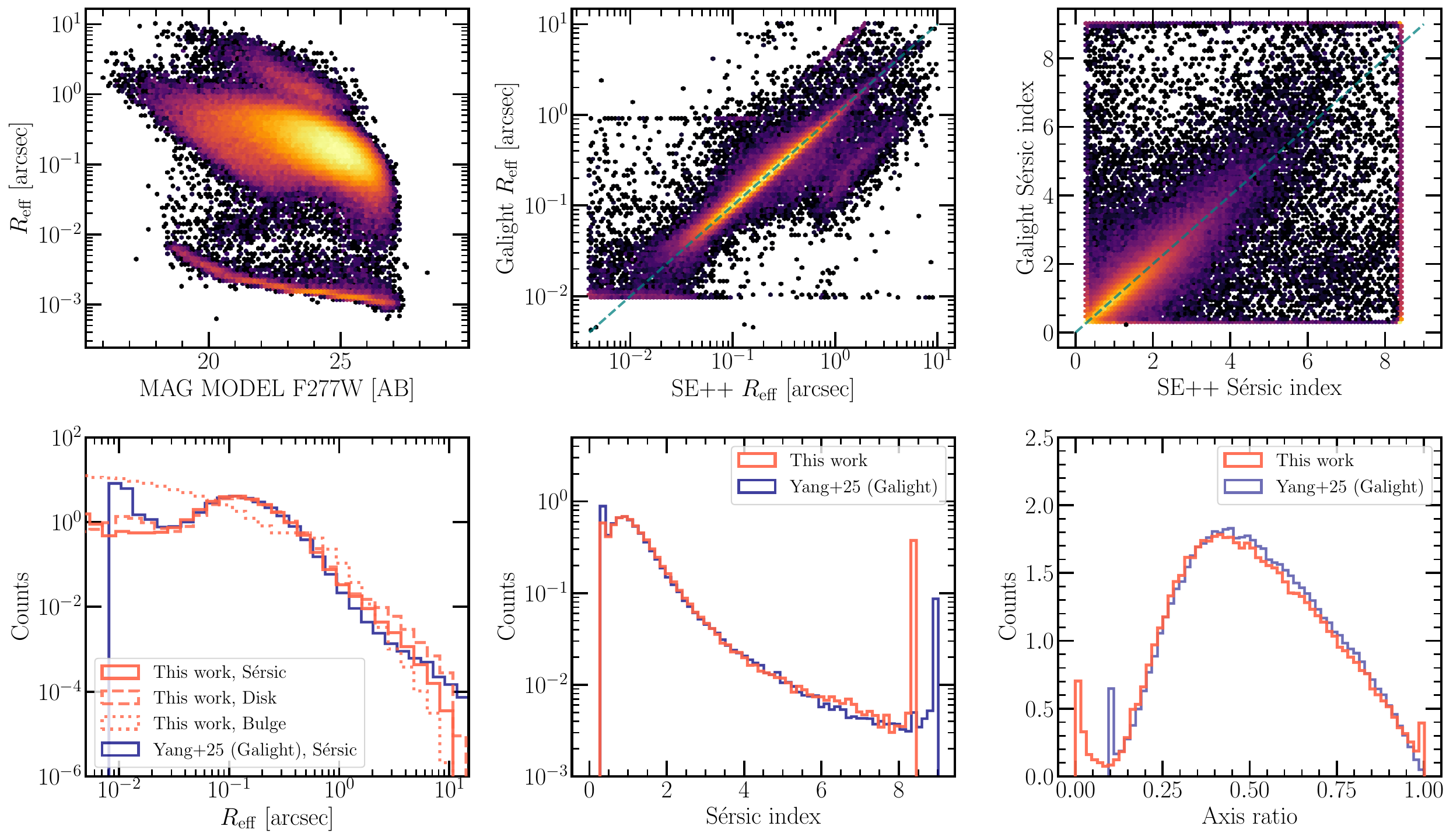}
\caption{Distributions and comparisons of several morphological quantities measured by \SEpp\ and \texttt{Galight} for sources with with S/N $> 10$. \textit{Top row}: Density histogram of the effective radius from the S\'ersic fit as a function of F277W magnitude, comparison of $R_{\rm eff}$ (middle) and $n_{\rm S}$ (right) between \SEpp\ and \texttt{Galight}. \textit{Bottom row}: Histograms of the S\'ersic total, bulge and disk effective radii (left),  S\'ersic index (middle) and axis ratio (right) from \SEpp\ and \texttt{Galight}.}
\label{fig:Morpho-distributions}
\end{figure*}

\begin{figure*}[t!]
  \centering
\begin{subfigure}[b]{0.9\textwidth}
            \includegraphics[width=1\hsize]{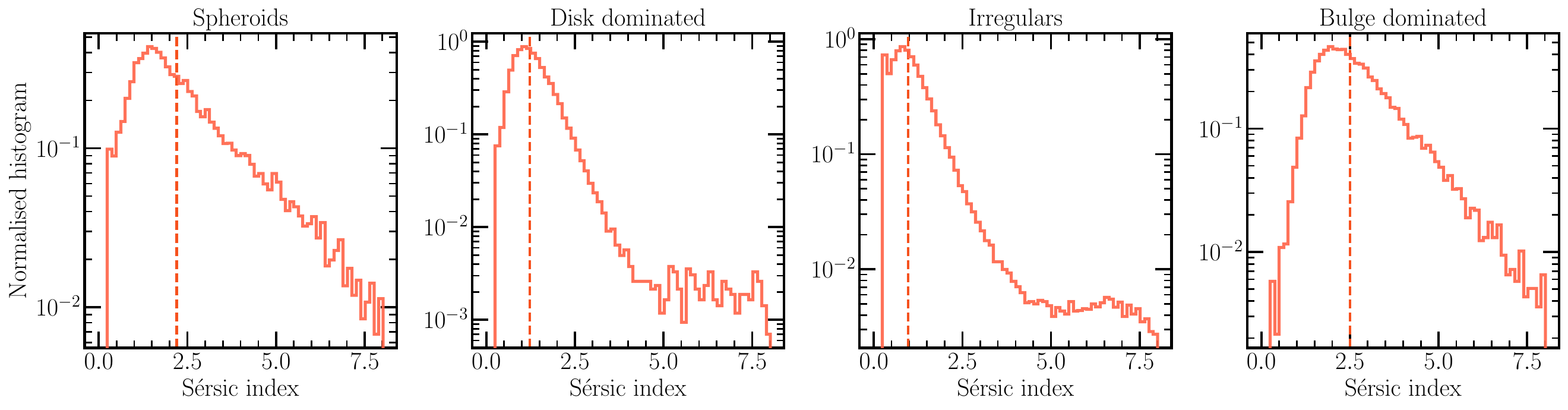}     
  \end{subfigure}
  \begin{subfigure}[b]{0.9\textwidth}
            \includegraphics[width=1\hsize]{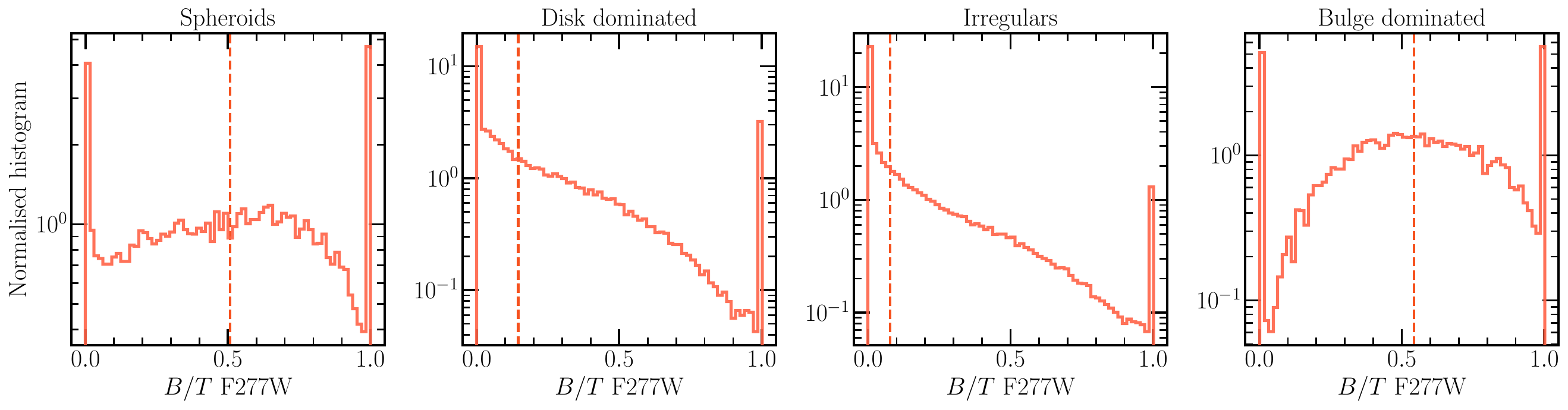}     
  \end{subfigure}
 \caption{Distributions of the S\'ersic index (top row) and $B/T$ in F277W (bottom row) for spheroid, disk-dominated, irregular and bulge-dominated galaxy types classified with machine learning. The dashed vertical line marks the median. These include sources with ${\rm S/N}>10$ classified as galaxies.}
  \label{fig:Morpho-compar-ML}
\end{figure*}

A major advantage of model fitting is that we obtain morphological measurements for all detected sources. Furthermore, to maximize the scientific applications of the catalog, we provide independent measurements from \texttt{Galight} and morphological classification from machine learning. We present and compare these for sources with ${\rm S/N}>10$ and \warnfl$={0,2,3}$ in the following.

\subsection{\SEpp}
As described in Section~\ref{sec:TheSE++catalog}, we measure morphologies from the two independent \SEpp\ fits: 1) S\'ersic models, resulting in the S\'ersic index $n_{\rm S}$, total effective radius $R_{\rm S,eff}$ and axis ratio $(a/b)_{\rm S}$; and 2) Bulge + Disk models, measuring bulge and disk effective radii $R_{\rm B,eff}$, $R_{\rm D,eff}$, axis ratios $(a/b)_{\rm B,eff}$, $(a/b)_{\rm D,eff}$ and bulge-to-total flux ratio $B/T$. These structural parameters are effectively averaged over the $1-5$ \SI{}{\micro \meter} wavelength range, with $B/T$ being measured for all bands. Fig.~\ref{fig:Morpho-distributions} shows the distributions of several structural parameters: $R_{\rm S,eff}$ vs F277W magnitude in the top left, $R_{\rm S,eff}$ histogram for the S\'ersic, bulge and disk components in the bottom left, $n_{\rm S}$ histogram in the bottom middle and $(a/b)_{\rm S}$ in the bottom right panel. 

The $R_{\rm S,eff}$ distribution peaks around \ang{;;0.1}, and shows a clearly isolated locus of unresolved, point-like sources below \ang{;;1}. Unresolved sources are expected to have their measured size independent of magnitude; however, the slight slope shown in Fig.~\ref{fig:Morpho-distributions} is due to the uniform prior on $R_{\rm S,eff}$ with a lower limit of $-0.01\times$ the Kron radius from \hotcold. There is also a cloud of $\sim 2\%$ of sources with sizes larger than the main locus, a lot of which are also scattered towards large $R_{\rm S,eff}$ compared to the independent measurements from \texttt{Galight} (\S \ref{sec:galight}). These are predominantly faint (mag $\gtrsim 27$) in \hotcold\, in many cases fainter than the depth of one or more of the four modelling bands, and may have brighter companions nearby. This can drive a large solution for the model $R_{\rm S,eff}$. 

The S\'ersic index distribution peaks at $n_{\rm S}\sim 1$ and exponentially decreases at higher values. There is an accumulation of solutions at the minimum and maximum allowed values (0.3 and 8.5), mainly because of degeneracy of the solutions for unresolved sources. The axis ratio shows a distribution peaking at $(a/b)_{\rm S}\sim 0.4$ and skewed towards lower values. These distributions have the expected shape, agreeing with those obtained independtly with \texttt{Galight} and serve as a validation of the morphological measurements.

\subsection{\texttt{Galight}} \label{sec:galight}
An independent structural measurement using \texttt{Galight}, which is built on \texttt{Lenstronomy} \citep{Ding2020, Birrer2021}, complements \SEpp\ measurements by providing structural parameters measured in four NIRCam bands individually, rather than averaged over a broad wavelength range as in \SEpp.
The S\'ersic parameter constraints include the half-light radius between \ang{;;0.01} arcsec and the cutout frame size, the S\'ersic index between 0.3 and 9, and the axis ratio between 0.1 and 1, with initial guesses derived from the SE++\ catalog. \texttt{Galight} adopts the same source detection maps and PSFs as SE++, ensuring consistency. These modeling details are described in \cite{Yang2025}. Furthermore, \texttt{Galight} offers two additional independent morphology measurements,  one using two S\'ersic models and another employing a single S\'ersic  with a central point source model. This catalog and its release are described in Yang et al. in prep.

\subsection{Machine learning}
Galaxy morphological classification can be efficiently carried out by machine learning (ML) models such as supervised convolutional neural networks (CNN) by extracting image features that are correlated with galaxy morphology \citep[][for a review]{MHC-Lanusse2023}. We apply the supervised ML models from \cite{MHC2024} to classify galaxies into four broad morphological classes: spheroid, disk-dominated, bulge-dominated and irregular. These are described in detail in \cite{MHC2025}. Briefly, a supervised CNN model is trained with labels from the CANDELS survey and domain-adapted to JWST using adversarial domain adaptation. We used the same architecture, input image size ($32\times32$ pix) and normalization as \cite{MHC2024} and train the network using COSMOS-Web images as target domain. Uncertainties in the classification are estimated with an ensemble of 10 separate trainings for each of the three F150W, F277W and F444W filters, with different initial conditions and training sets. The final probability for each morphological class in each of the three filters is computed as the average of the outputs from the 10 networks. For validation, \cite{MHC2024} shows that this ML model achieves about $80-90 \%$ agreement with visual classification (tested on CEERS, but with similar performance in COSMOS-Web, \citealt{MHC2025})

\subsection{Comparison between the methods}

Comparison of the morphological measurements between the three methods is shown in Fig.~\ref{fig:Morpho-distributions} and Fig.~\ref{fig:Morpho-compar-ML}, showing generally excellent agreement. The effective radii show a relatively tight correlation, centered at the one-to-one ratio and with no significant offsets. The striping is a result of the parameter bounds and, for \texttt{Galight} the fact that the stamp size scales with the $R_{\rm eff}$ solution by \SEpp. The large scatter in the S\'ersic index highlights the difficulty of measuring this parameter, but we find no significant offset between \SEpp\ and \texttt{Galight}. The parameter distributions also agree well.

In Fig.~\ref{fig:Morpho-compar-ML} we show the histograms of the S\'ersic index and $B/T$ in F277W for four morphological classes from the ML technique for  spheroid, disk-dominated, irregular and bulge-dominated. Galaxies are assigned to a morphological class based on the highest probability from the four categories, measured in F277W. In general, there is good agreement with the distributions of $n_{\rm S}$ and $B/T$ and the corresponding morphological classes. For example, spheroid and bulge-dominated systems show distributions skewed towards larger $n_{\rm S}$ and $B/T$ with medians of $n_{\rm S}\sim2.5$ and $B/T\sim0.55$. On the other hand, irregular, and disk-dominated systems tend to have lower values with medians of $n_{\rm S}\sim1.2$ and $B/T\sim0.15$. Additionally, as expected, irregulars tend to be more extended than disk-dominated, while spheroids less compact than bulge-dominated. The dispersion is, perhaps, relatively high, which can be due to both uncertain ML classifications and \SEpp\ measurements.

\section{Photometric redshifts} \label{sec:photo-z}

\subsection{Method} \label{sec:photoz-methods}

We use the template-fitting code \lephare{} \citep{arnouts_measuring_2002,ilbert_accurate_2006} to measure photometric redshifts. While this code was used for previous COSMOS catalogs \citep{laigle_cosmos2015_2016,weaver_cosmos2020_2022}, we modified significantly the configuration for the COSMOS-Web catalog. The main changes are outlined below. 

The photometric redshifts of galaxies in the previous versions of the COSMOS catalog were estimated with a limited set of templates from \citet{ilbert_mass_2013} which reduces the risk of degeneracies in the color-redshift space. Given the improvement in the multi-color coverage and sensitivity from NIRCam imaging, we decided to increase the size of the galaxy template library with a more representative library generated using the \citet{bruzual_stellar_2003} Stellar Synthetic Population models (hereafter BC03). We select the same set of templates as the ones used to derive physical parameters in \citet{ilbert_ssfr_2015}. The advantage of such an approach is that physical parameters are derived simultaneously, with a library more diverse and representative than previous COSMOS catalogs. These templates are built assuming six different Star Formation Histories (SFH), including four exponentially declining and two delayed analytic forms, as well as two metallicities (solar and half-solar). To minimse computational time, we restricted the number of ages at 43, ranging from 0.05 to \SI{13.5}{\giga\year}. We add emission lines to the BC03 templates by adopting the recipe in \citet{saito_EL-COSMOS_2020}. The emission line fluxes are allowed to vary together by a factor of two around the fiducial value during the fit.

Dust attenuation is a crucial ingredient in the modeling of galaxy SEDs. Given the variety of galaxy populations present in the COSMOS-Web catalog, we decided to explore a wide range of dust attenuation applied to the templates, with a maximum $\rm{E(B-V)}\leq1$. Dusty galaxies are known to be present in the field \citep[e.g.,][]{McKinney_scubaDive_2024} and exploring such range of attenuation is necessary to recover their photo-$z$. We included three different attenuation curves \citep{calzetti_dust_2000, arnouts_NRK_2013, salim_dust_2018}. The high-$z$ analog dust attenuation curve from \citet{salim_dust_2018} includes a bump at 2175{\AA}. We find that this bump reduces systematic differences between modeled and observed magnitudes at fixed redshift, compared to the \citet{prevot_typical_1984} curve previously used. The BC03 templates do not include dust emission in the modelised spectra. Since the COSMOS-Web wavelength coverage probes 7.7$\mum$ with  MIRI imaging, the Polycyclic aromatic hydrocarbon (PAH) emission could already contribute to the SED of $z<0.2$ galaxies. Therefore, we modeled this dust emission. We assumed that the energy absorbed in ultraviolet-to-optical range is fully remitted in IR by the dust. We rescaled the dust emission template library from \citet{bethermin_2012} based on \citet{magdis_2012} to the expected IR luminosity using energy balance. The intergalactic medium (IGM) absorption is accounted for by using the analytical correction of \citet{madau_opa_1995}.

\begin{figure}[t!]
\includegraphics[width=1\columnwidth]{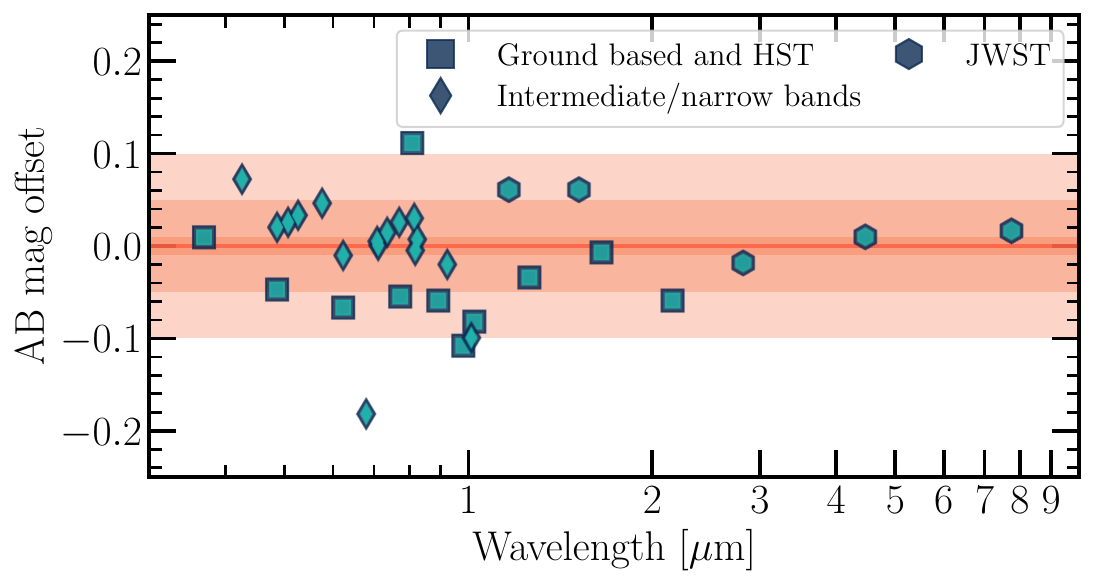}
\caption{Magnitude offsets as a function of band wavelength derived as the median difference between the model and observed magnitudes with \lephare. The shaded regions mark $\pm 0.1$, $\pm 0.05$ and $\pm 0.01$ mag. \JWST\, ground-based/\hst\ and intermediate/narrow bands are marked with different symbols for clarity. These are used as absolute calibration of the photometry when fitting with \lephare.}
\label{fig:mag-offsets}
\end{figure}

Before performing a fit on the full catalog, we adapted the absolute calibration of the photometry following \citet{ilbert_accurate_2006}. We restricted the fit to the galaxies having a spectroscopic redshift. The median of the difference between the predicted and the observed magnitudes is used to modify the absolute calibration in each band. Then, the fit is redone iteratively after having applied the correction. The procedure converged in two iterations. The values of the offsets are shown in Fig.~\ref{fig:mag-offsets}. These offsets usually remain below 0.05 mag. We note that the low values obtained in the NIRcam bands are driven by the better sensitivity in these bands relative to ground-based data, which anchors the fit at these wavelengths. 

The fitting procedure outputs several quantities. For each element $i$ in the library, we derived a probability P$_i$ associated to the $\chi^2_i$.  We produce a likelihood distribution which is the sum of all individual P$_i$ at a given redshift. This distribution is considered as our posterior redshift Probability Distribution function (PDF$_z$), 
assuming a flat prior for all parameters. We adopt the median of the PDF$_z$ as a fiducial redshift point-estimate. The associated uncertainties encompass 68\% of the PDF$_z$. We also provide the redshift corresponding to the template which produces the minimal $\chi^2$ over the full library. This redshift does not necessarily correspond to the median of the PDF$_z$ in case of a multi-peaked PDF.

\begin{figure*}[t!]
\includegraphics[width=1\textwidth, trim=0cm 0.8cm 0cm 2cm, clip]{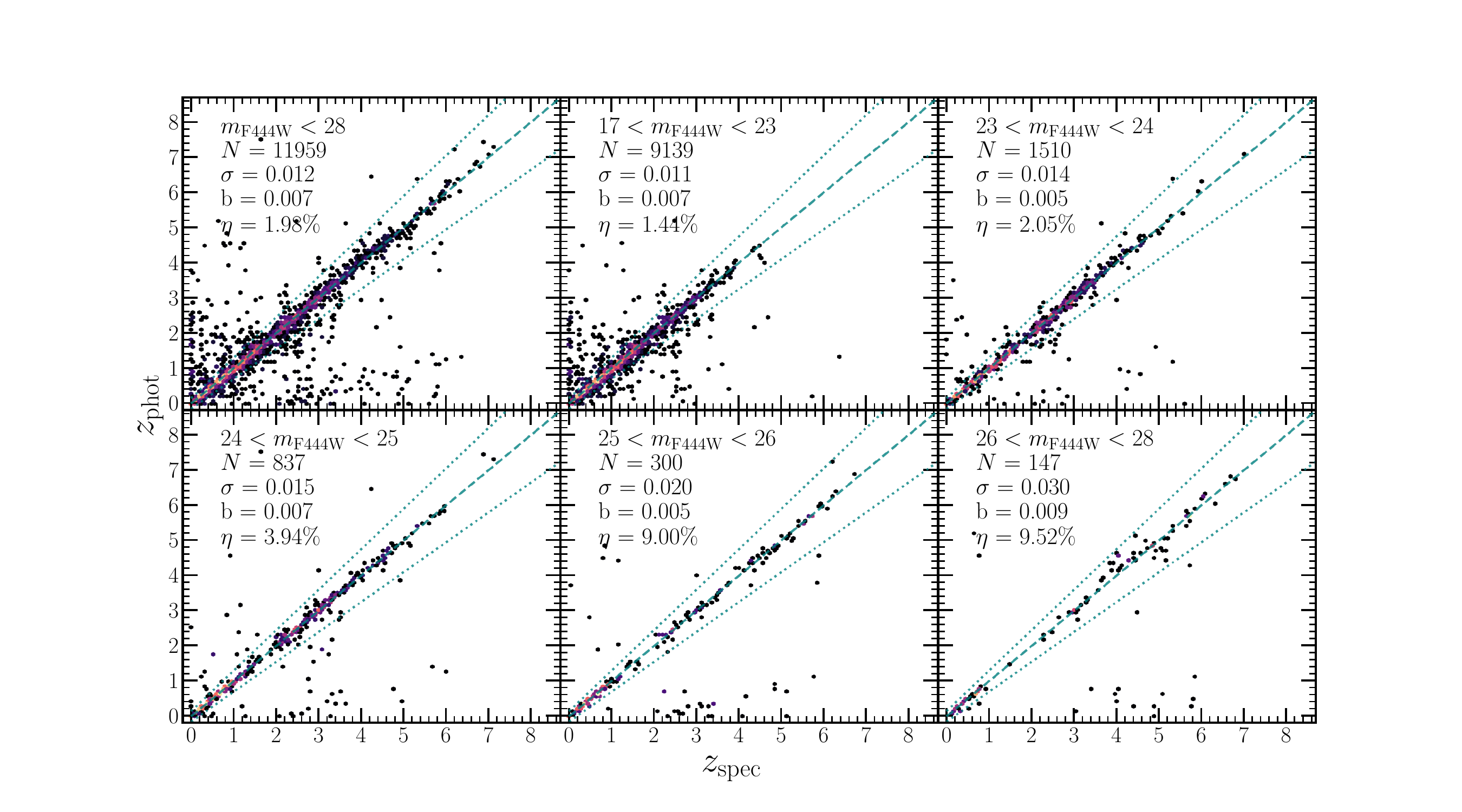}
\caption{Photometric vs. spectroscopic redshift comparison for m$_{\rm F444W} < 28$ (top left) and for different F444W magnitude-selected samples. The dashed teal line marks the one-to-one relation, while the dotted lines mark correspond to \zphot $> \pm 0.15\,(1+z_{\rm spec})$. The number of sources in the magnitude selection ($N$), median absolute deviation ($\sigma$), overall bias ($b$) and the outlier fraction $\eta$ are noted in each panel.
}
\label{fig:photoz-specz}
\end{figure*}

When running \lephare{}, we fit the observations with stellar and AGN templates in parallel to the galaxy templates. For the stellar library, we adopt the same library as \citet{Kauffmann2022} which includes brown dwarf templates. This is necessary to identify degeneracies with high redshift galaxies. For the AGN library, we adopt the templates from \citet{salvato_photoz_2011} which includes hybrid between galaxies and AGN emission. However, we emphasize that a specific work needs to be carried out to measure the photometric redshifts of sources dominated by an AGN which required additional steps not performed in this paper (e.g., correction for source variability, applying prior based on the X-ray emission).

\subsection{Photometric redshift performance} \label{sec:photoz-preformances}

For the absolute photometry calibration and assessing the photo-$z$ performance, we use a sample of about 12,000 spectroscopic redshifts with a high ($>97\%)$ confidence level out to $z=8$. These are compiled from most spectroscopic programs in COSMOS (both public and private) and are presented in detail in \cite{Khostovan2025}.

For the performance assessment, we use the standard metrics, such as the median absolute deviation (MAD), defined as
\begin{equation} \label{eq:sigma-mad}
\sigma_{\text{MAD}} = 1.48 \times \text{ median}\left(\dfrac{|\Delta z - \text{median}(\Delta z)|}{(1+z_{\rm spec})}\right),
\end{equation}
where $\Delta z = z_{\rm phot} - z_{\rm spec}$. Sources whose photo-$z$ solutions deviate by $|\Delta z| > 0.15 (1+z_{\rm spec})$ are classified as outliers \citep{Hildebrandt2012}. Finally, the overall redshift bias is defined as the median of $\Delta z$.

\begin{figure*}[t!]
\includegraphics[width=1\textwidth, trim=0cm 3cm 0cm 5cm, clip]{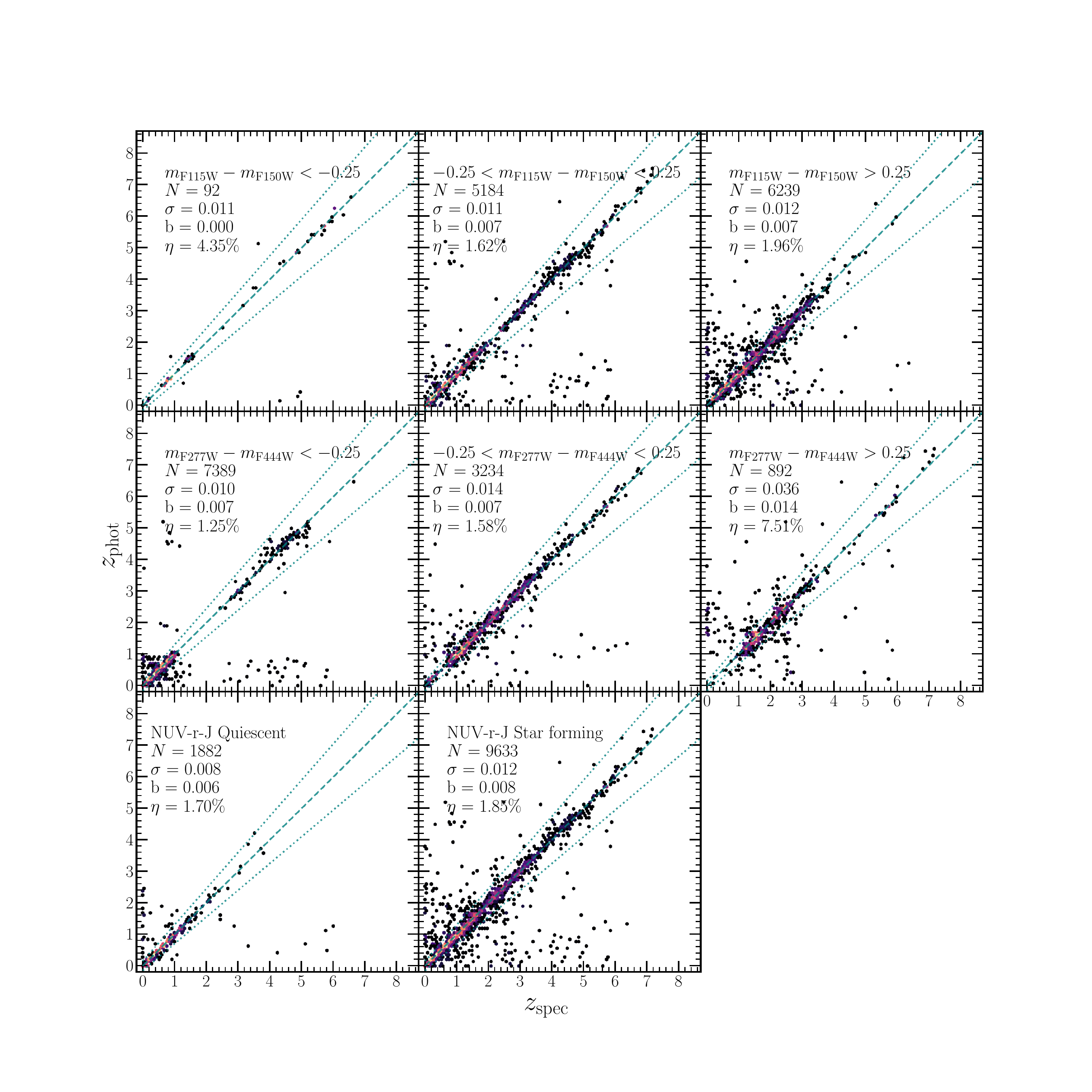}
\caption{Photometric vs. spectroscopic redshift comparison for different color-selected samples. The top row shows F115W$ - $F150W, while the middle row shows F277W$-$F444W color-selected samples at $<-0.25$ (left), $-0.25 < {\rm color} <0.25$ (middle), and $>0.25$ (right column). The bottom row shows the performance for quiescent and star-forming galaxies selected using the $NUVrJ$ diagram, i.e., $(NUV - r) > 3\times(r - J) + 1 \, {\rm and} \, (NUV - r) > 3.1$. The dashed teal line marks the one-to-one relation, while the dotted lines mark correspond to \zphot $> \pm 0.15\,(1+z_{\rm spec})$. The number of sources in the magnitude selection ($N$), median absolute deviation ($\sigma$), overall bias ($b$) and the outlier fraction $\eta$ are noted in each panel.
}
\label{fig:photoz-specz-vs-color}
\end{figure*}

We evaluate the photo-$z$ performance for different magnitude, color, and galaxy type selected samples in Figs.~\ref{fig:photoz-specz} and \ref{fig:photoz-specz-vs-color} and Table~\ref{table:photoz-performance}. In general, the photo-$z$ show excellent performance, with $\sigma_{\rm MAD} = 0.012$, $<2 \%$ outliers and $b=0.007$ for galaxies brighter than $m_{\rm F444W} = 28$, the $5\,\sigma$ detection limit in F444W (top left panel of Fig.~\ref{fig:photoz-specz}). The performance decreases as a function of magnitude, going from $\sigma_{\rm MAD} = 0.011$ and $\eta=1.44\%$ for $m_{\rm F444W}<23$, to $\sigma_{\rm MAD} = 0.030$  and $\eta=9.52\%$ for the faintest sample at $26<m_{\rm F444W}<28$. For the fainter samples, there is an increasing number of sources with \zphot$\lesssim1.5$ and \zspec$\gtrsim2$, which can be attributed to a misidentification between the Lyman and Balmer breaks for sources with lower signal-to-noise ratio. The bias does not show a particular magnitude dependence.

\begin{table}[t!]
\begin{threeparttable}
\centering
\caption{Photo-$z$ performance estimated using high-quality spectroscopic redshifts with $>97\%$ confidence.}
\begin{tabular}{c|c|c|c}
\hline
\hline
Mag range & $\sigma_{\text{MAD}}$\tnote{a}  & Outliers\tnote{b} & Bias\tnote{c}  \\
\hline
$17 < m_{\rm F444W} < 23$ & $0.011$ & $1.44\%$ & $0.007$ \\
$23 < m_{\rm F444W} < 24$ & $0.014$ & $2.05\%$ & $0.005$ \\
$24 < m_{\rm F444W} < 25$ & $0.015$ & $3.94\%$ & $0.007$ \\
$25 < m_{\rm F444W} < 26$ & $0.020$ & $9.00\%$ & $0.005$ \\
$26 < m_{\rm F444W} < 28$ & $0.030$ & $9.52\%$ & $0.009$ \\
\hline
\hline
Color range & $\sigma_{\text{MAD}}$\tnote{a}  & Outliers\tnote{b} & 
Bias\tnote{c}  \\
\hline
$m_{\rm F115W} - m_{\rm F150W} < - 0.25 $ & $0.010$ & $4.08\%$ & $0.001$ \\
\hspace{-3mm} $-0.25 <m_{\rm F115W} - m_{\rm F150W} < 0.25 $ \hspace{-2mm}  & $0.011$ & $1.91\%$ & $0.007$ \\
$m_{\rm F115W} - m_{\rm F150W} > 0.25 $ & $0.012$ & $2.05\%$ & $0.007$ \\

$m_{\rm F277W} - m_{\rm F444W} < - 0.25 $ & $0.010$ & $1.47\%$ & $0.001$ \\
\hspace{-3mm} $-0.25 <m_{\rm F277W} - m_{\rm F444W} < 0.25 $ \hspace{-2mm}  & $0.015$ & $1.74\%$ & $0.006$ \\
$m_{\rm F277W} - m_{\rm F444W} > 0.25 $ & $0.035$ & $7.28\%$ & $0.014$ \\
$NUVrJ$ Quiescent\tnote{d} & $0.008$ & $2.26\%$ & $0.006$ \\
$NUVrJ$ Star forming\tnote{d} & $0.013$ & $1.95\%$ & $0.007$ \\
\hline
\end{tabular}
\begin{tablenotes}
\item[a] Defined by Eq.~\ref{eq:sigma-mad}. 
\item[b] Outliers = $|\Delta z| > 0.15 (1+z_{\rm spec})$
\item[c] Bias = median$(\Delta z)$.
\item[d] Quiescent are selected using the standard $NUVrJ$ diagram $(NUV - r) > 3\times(r - J) + 1 \, {\rm and} \, (NUV - r) > 3.1$.
\end{tablenotes}
\label{table:photoz-performance}
\end{threeparttable}
\end{table}

To investigate the photo-$z$ performance for different populations, we analyze samples selected by color and type, shown in Fig.~\ref{fig:photoz-specz-vs-color}. We chose the F115W-F150W and F277W-F444W colors in three selections $<-0.25$ (red), $-0.25 < {\rm color} <0.25$ (middle), and $>0.25$ (blue) using the \SEpp\ model photometry. As a function of the F115W-F150W color, there is no significant change, with all three samples showing excellent performance with $\sigma_{\rm MAD} \approx 0.011$ and $\eta<4\%$. However, as a function of F277W-F444W the performance degrades going from blue to red samples with $\sigma_{\rm MAD} = 0.010$ for the blue, $\sigma_{\rm MAD} = 0.015$ for the middle, and $\sigma_{\rm MAD} = 0.035$ for the red samples. The fraction of outliers increases from $\sim1.5\%$ for the blue and middle to $7.28\%$ for the red sample. There is also a slight bias of $b=0.014$ for the $m_{\rm F277W} - m_{\rm F444W} > 0.25$ sample that does not exceed the outlier definition. This red population is dominated by highly dust-obscured galaxies that can have sub-millimeter counterparts \citep[e.g.,][]{Barrufet2023, Gottumukkala2023}, which are known to be difficult cases for SED fitting \citep[e.g.,][]{Casey2012, Hayward2015}. 

We also assess the performance for star-forming and quiescent galaxies selected using the $NUV-r-J$ diagram, the latter satisfying $(NUV - r) > 3\times(r - J) + 1 \, {\rm and} \, (NUV - r) > 3.1$ \citep{ilbert_mass_2013}. Both show excellent performance with $\sigma_{\rm MAD} = 0.008$ and $\eta = 2.26\%$ for star-forming and $\sigma_{\rm MAD} = 0.013$ and $\eta=1.95 \%$ for quiescent galaxies. These metrics showcase the excellent quality of our photometric redshifts with stable performance for different magnitude, color and type selected populations.

\begin{figure}[t!]
\includegraphics[width=1\columnwidth]{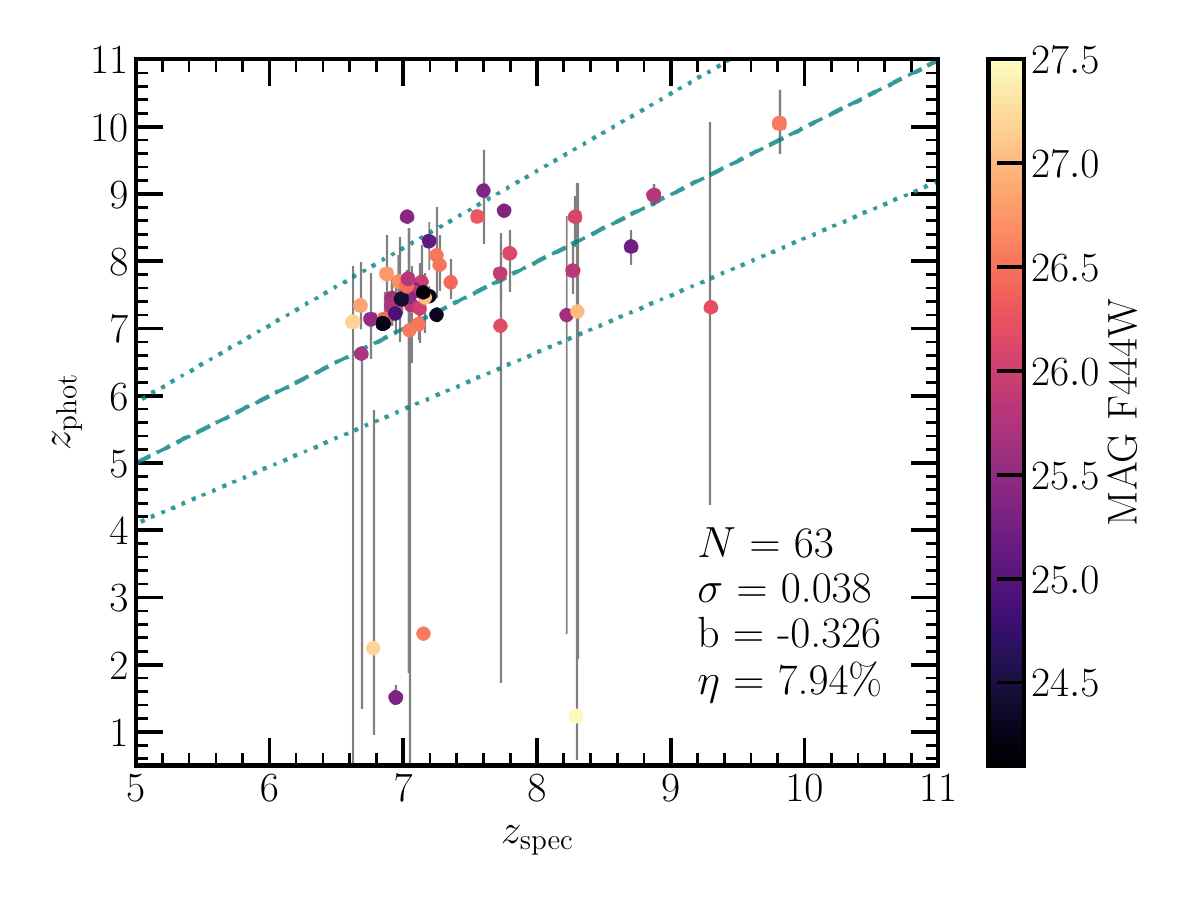}
\caption{Photometric vs. spectroscopic redshift comparison using a sample of \oiii\ emitters from the COSMOS-3D NIRCam/WFSS survey, and a compilation of PRISM spectra available on DJA from CAPERS and transient programs. The dotted lines mark correspond to \zphot $> \pm 0.15\,(1+z_{\rm spec})$. Points are color coded by their F444W magnitude.
}
\label{fig:photoz-specz-c3d}
\end{figure}

To assess the performance at the highest redshifts, we compare with the spec-$z$ of a sample of \oiii\ emitters identified in the first $10\%$ of the data (taken in December 2024) from the COSMOS-3D NIRCam/WFSS survey (ID \#5893 PI: K. Kakiichi). This includes 34 high-confidence line emitters at $6.8 \lesssim z \lesssim9$ (Wang et al. in prep). 
Additionally, we include a compilation of spec-$z$ available on the DAWN \JWST\ Archive \citep[DJA\footnote{\url{https://dawn-cph.github.io/dja/index.html}, \url{10.5281/zenodo.7299500}.},][]{Heintz2024}. These include high confidence (\texttt{grade=3}) PRISM spectra from CAPERS (\#6368, PI: Dickinson) and transient (\#6585, PI: Coulter) programs.
The comparison is shown in Fig.~\ref{fig:photoz-specz-c3d}. There is a relatively good agreement between the photo-$z$ and spec-$z$ with $\sigma_{\rm MAD} = 0.038$ and $\eta = 7.9\%$. However, there is a median bias of $b=-0.3$ towards higher photo-$z$ solutions. This spec-$z$ sample was not used in calibrating the SED fitting with \lephare{} (described in \S\ref{sec:photoz-methods}) which can be one of the reasons for the bias.  Importantly, this high-$z$ spectroscopic sample, once the COSMOS-3D survey is complete, will be highly valuable in recalibrating and improving the photo-$z$ performance for future versions of the catalog, which we plan to build.

\begin{figure}[t!]
\includegraphics[width=1\columnwidth]{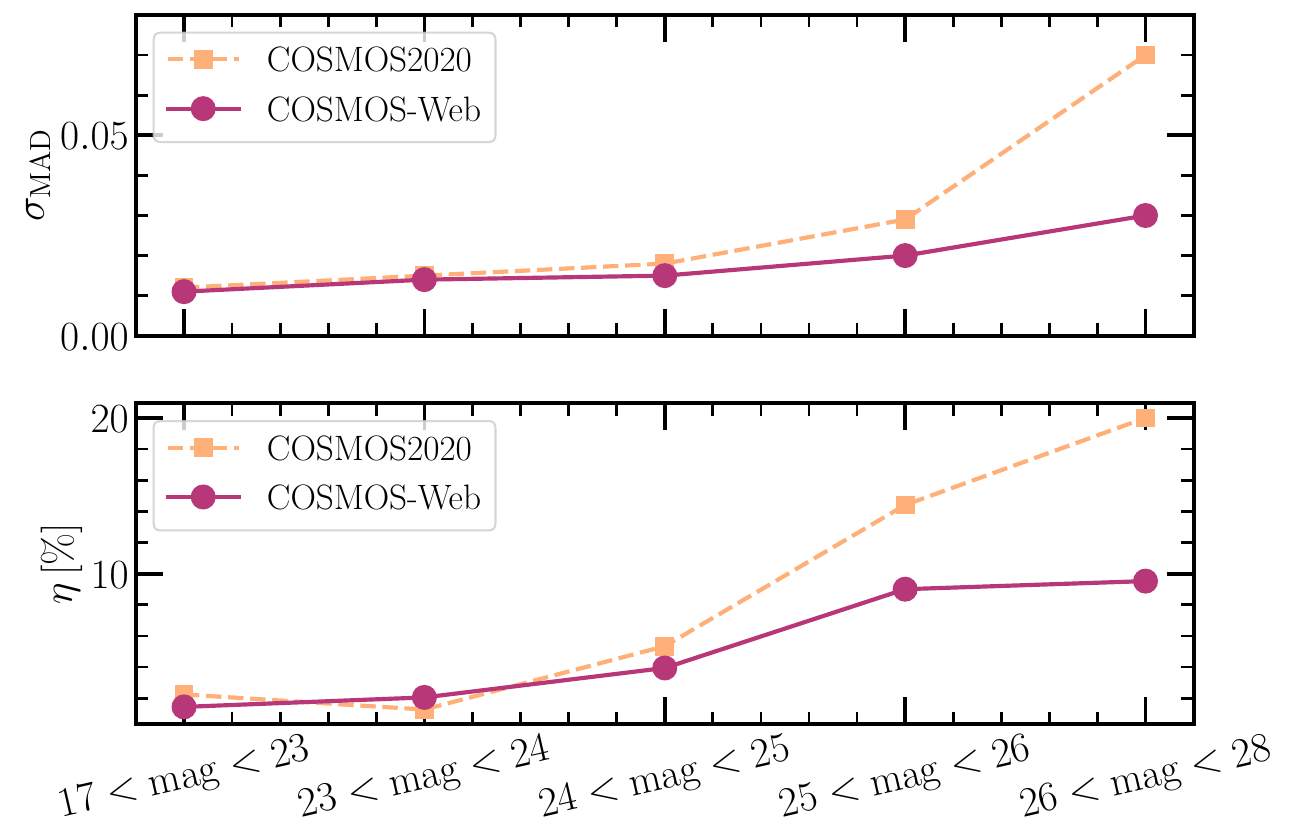}
\caption{Photo-$z$ performance comparison with COSMOS2020. The top and bottom panels show the $\sigma_{\rm MAD}$ and outlier fraction $\eta$ as a function of magnitude bin for COSMOS2020 (dashed lines and square symbols) and COSMOS-Web (solid lines and circle symbols). COSMOS-Web photo-$z$ show better performance in both statistics, especially for fainter magnitudes, where the performance is better by about a factor of 2.
}
\label{fig:zphot-cweb-vs-c20}
\end{figure}

To demonstrate the improvement with respect to the previous COSMOS catalogs, COSMOS2020, Fig.~\ref{fig:zphot-cweb-vs-c20} shows the $\sigma_{\rm MAD}$ and outlier fraction $\eta$ as a function of magnitude for COSMOS-Web and COSMOS2020 catalogs, computed using the same spec-$z$ sample. Since we are matching the spec-$z$ sample to both catalogs, we use the F444W magnitude to select in magnitude ranges for both catalogs. COSMOS-Web photo-$z$ show better performance in both statistics, especially for fainter magnitudes, where the improvement is about a factor of two over COSMOS2020. This is unsurprising, given the much deeper \JWST\ NIR imagaing. 

\begin{figure*}[t!]
\includegraphics[width=1\textwidth]{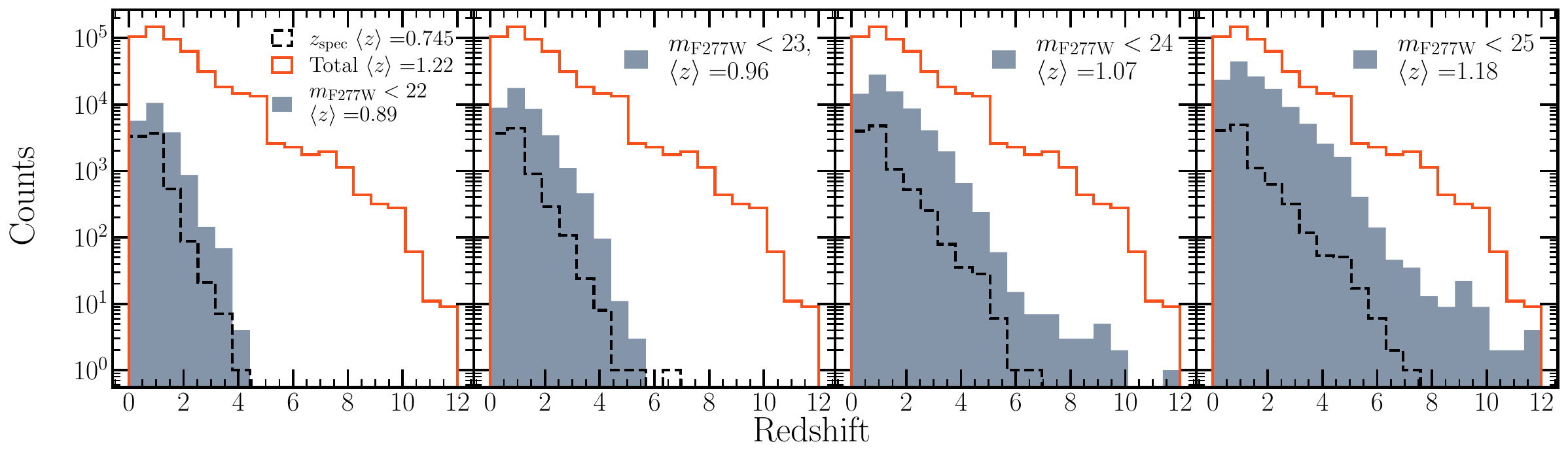}
\caption{Redshift distributions for the total (orange), spec-$z$ (black dashed) and different magnitude-selected (blue filled) samples. The legend indicates the median redshift $\langle z \rangle$ for each sample.}
\label{fig:z-distributions}
\end{figure*}

In Fig.~\ref{fig:z-distributions} we show the redshift distributions for the total, F277W magnitude-selected and spec-$z$ samples. As expected, the median redshift increases for fainter samples. Thanks to the unique combination of survey area and depth at $1-5 \, \si{\micron}$ we detect numerous candidate $z>8$ galaxies. These are studied in detail in the corresponding papers \citep[][also Franco et al. in prep]{Casey2024, Franco2024, Shuntov2024, Paquereau2025}.

\begin{figure}[t!]
\includegraphics[width=0.88\columnwidth]{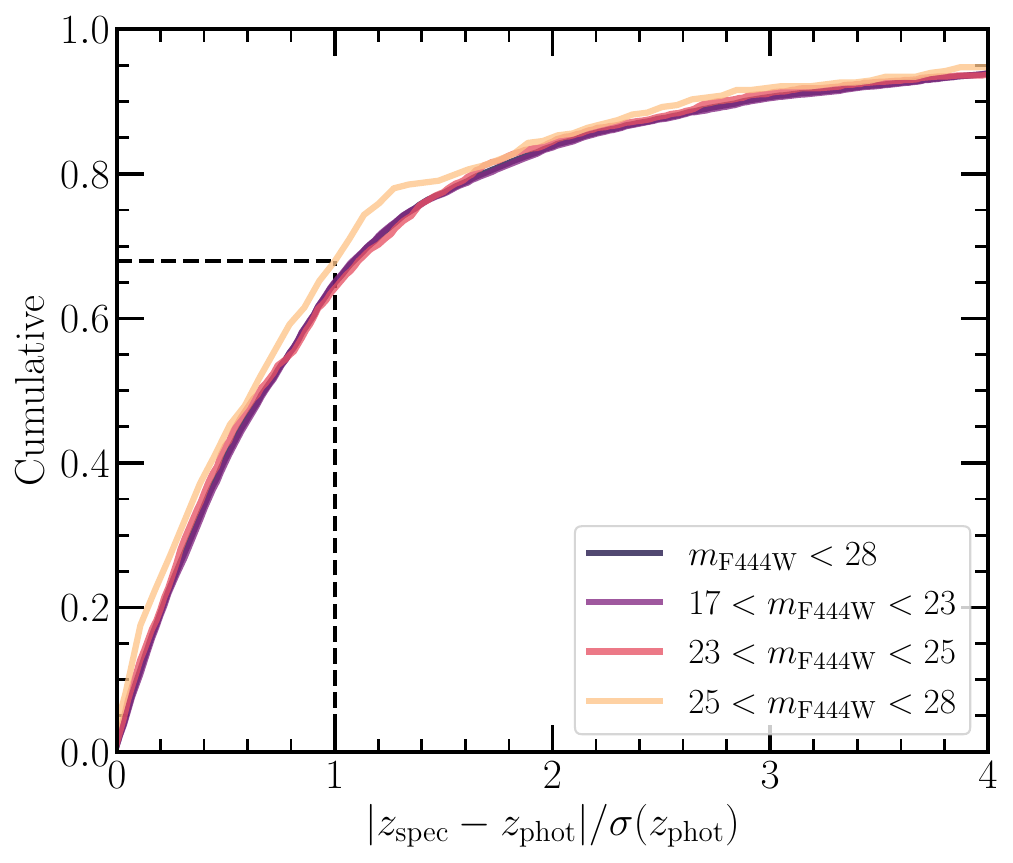}
\caption{Cumulative distribution of the ratio between $|z_{\rm spec} - z_{\rm phot}|$ and the photo-$z$ $1\,\sigma$ uncertainty as a function of F444W magnitude. The $1\,\sigma$ uncertainty is taken as the maximum between ($z_{\rm phot} - z_{\rm phot}^{\rm low 68}$) and ($z_{\rm phot}^{\rm upp 68} - z_{\rm phot}$). For unbiased photo-$z$ uncertainties, the cumulative distributions should reach values $\sim 0.68$ for a ratio of unity (marked by dashed lines).
}
\label{fig:zphot-err-cumulative}
\end{figure}

To evaluate the quality of the photo-$z$ uncertainties, we analyze the cumulative distribution of the ratio between $|z_{\rm spec} - z_{\rm phot}|$ and the photo-$z$ $1\,\sigma$ uncertainty. For well calibrated and unbiased uncertainties, the fraction of spec-$z$ that are within the $1\,\sigma$ uncertainty interval $[z_{\rm phot}^{\rm low 68}\, , z_{\rm phot}^{\rm upp 68}]$ should be $\approx 0.68$. Fig.~\ref{fig:zphot-err-cumulative} shows the cumulative distribution of this ratio for several magnitude selected samples, all of which reach values of $\approx 0.68$ at a ratio of unity.
This means that the flux uncertainties and consequently photo-$z$ uncertainties are well calibrated and unbiased for all magnitudes.

\begin{figure}[t!]
\includegraphics[width=0.88\columnwidth]{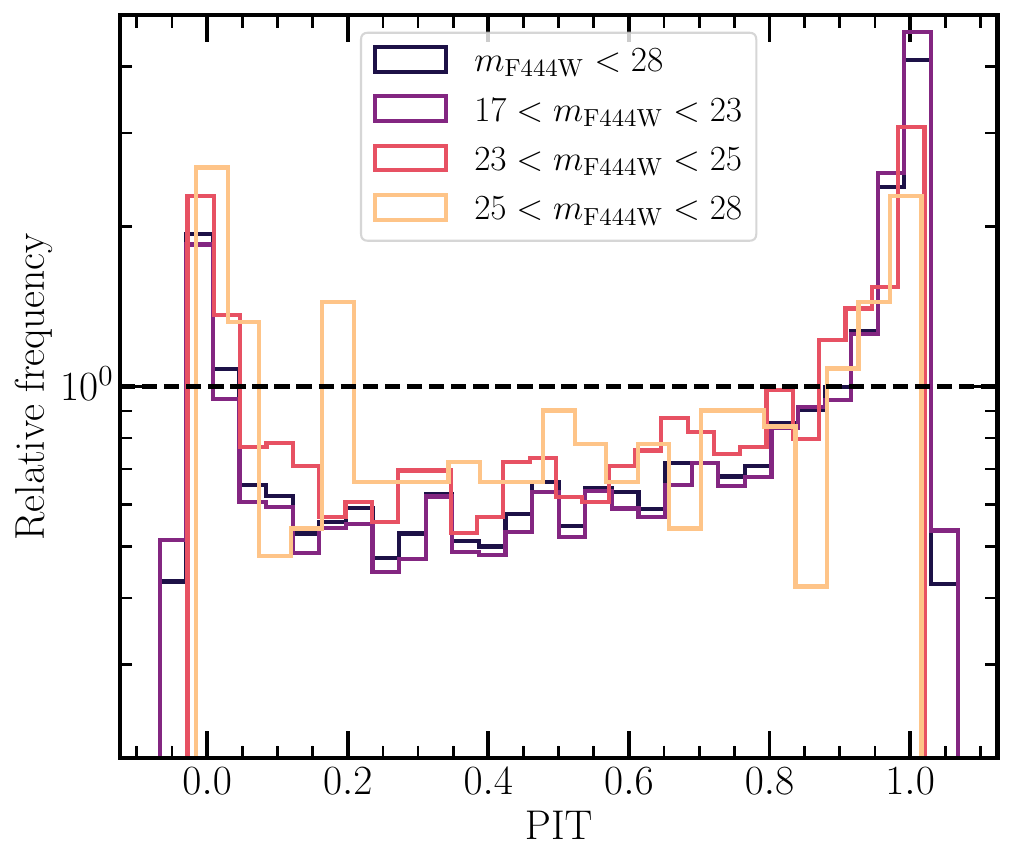}
\caption{Probability integral transform (PIT) of the PDF$_z$ for magnitude-selected samples. The PIT represents the CDF of the PDF$_z$ evaluated at $z_{\rm spec}$.
}
\label{fig:PIT-statistic}
\end{figure}

Finally, we evaluate the quality of the PDF$_z$ using the probability integral transform (PIT) statistic \citep{Bordoloi2010}. The PIT is measured as the Cumulative Distribution Function (CDF), evaluated at the $z_{\rm spec}$ of the source. The histogram of the PIT for all sources is informative on how accurate and well calibrated the PDF$_z$ are. A flat PIT distribution indicates accurate and unbiased PDF$_z$; U-shaped (concave) indicates under-dispersed PDF$_z$; convex corresponds to overdispersed PDF$_z$; distributions with a slope indicate biased PDF$_z$. Fig.~\ref{fig:PIT-statistic} shows the PIT distribution for several magnitude selected samples. All magnitude-selected samples, except the one at $25 < m_{F444W} < 28$ show positive-slope and U-shaped distributions indicating PDF$_z$ that are slightly underdispersed and skewed towards the low-$z$ end.  This means that the PDF$_z$ does not adequately capture the \zspec\ solution, and the peaks at zero and one indicate outliers that can be due to wrong SED templates or problems with the photometry. Additionally, for bright galaxies part of this can be due to the relatively coarse sampling of the redshift range with a step of 0.01, which in some cases can be larger than the photo-$z$ uncertainty. However, since $\sim 75 \%$ of the galaxies have $25 < m_{F444W} < 28$, this means that most of our sources have reasonably well-calibrated PDF$_z$ (although we note that the PIT histogram is noisier because of the smaller spec-$z$ sample of faint galaxies)

\subsection{Separation of stars, QSO/AGN and galaxies} \label{sec:star-galxy-separation}

We combine several criteria to classify galaxies, stars and AGN in the catalog. These criteria are established to maintain a good balance between purity and completeness in the galaxy sample.

First, we describe the quantities necessary to establish these criteria:
\begin{itemize}
    \item minimum $\chi^2$ is derived by \lephare{} for the galaxy, stellar and AGN template libraries. These three values are noted $\chi^2_{\rm{g}}$, $\chi^2_{\rm{s}}$, $\chi^2_{\rm{q}}$, respectively.
    \item $BzK$ color criteria defined by \citet{Daddi_BzK_2004} is very efficient to isolate a stellar sequence in a color-color plane. We adopt colors close to \citet{Daddi_BzK_2004} given our set of filters. We note $Bz$ the magnitude difference between the HSC $g$ and $z$ bands, and $zK$ the magnitude difference between the HSC $z$ and NIRCam F277W bands. This criterion is only applied when the magnitude errors are lower than 0.3 mag in the three bands. 
    \item The effective radius $R_{\rm{eff}}$ derived from the Sérsic model fitted by \SEpp\ is used to select point-like sources. 
    \item We define as r150 the ratio between the flux F$_{\nu}$ measured in circular apertures of radius $0{\farcs}25$ and  $0{\farcs}1$. This criteria is also used to select point-like sources based on the light profile in the F150W filter.
\end{itemize}

Sources which satisfy any of the following criteria are classified as stars in our catalog:
\begin{itemize}
    \item stellar best-fit template $\chi^2_{\rm{s}}<\chi^2_{\rm{g}}$ and $\chi^2_{\rm{s}}<\chi^2_{\rm{q}}$, coupled with compactness criterion $R_{\rm{eff}}<0{\farcs}036$.
    \item $zK<0$ and $gz>2$, which selects a region of the $BzK$ color-color plane in which the stellar locus is well separated from the other sources. We also impose $R_{\rm{eff}}<0{\farcs}036$. 
    \item $\rm{r150}<2$ and $R_{\rm{eff}}<0{\farcs}036$ for magnitude brighter than 27.5 in F277W. We eliminate potential AGN from this selection by imposing $\chi^2_{\rm{s}}<\chi^2_{\rm{q}}$.
\end{itemize}

We flagged sources having an AGN component that could dominate the source emission, and significantly bias the photometric redshift or the physical parameters. Sources which satisfy any of these criteria are classified as AGN in our catalog:
\begin{itemize}
    \item $\chi^2_{\rm{q}}<\chi^2_{\rm{g}}$ and $\chi^2_{\rm{q}}<\chi^2_{\rm{s}}$, coupled with compactness criterion $R_{\rm{eff}}<0{\farcs}036$.
    \item r$150<2$ and $R_{\rm{eff}}<0{\farcs}036$ for magnitude brighter than 27.5 in F277W. We impose $\chi^2_{\rm{q}}<\chi^2_{\rm{s}}$ to remove the stars.
    \item X-ray detection with Chandra.
\end{itemize}

\begin{figure}[h!]
\includegraphics[width=1\columnwidth]{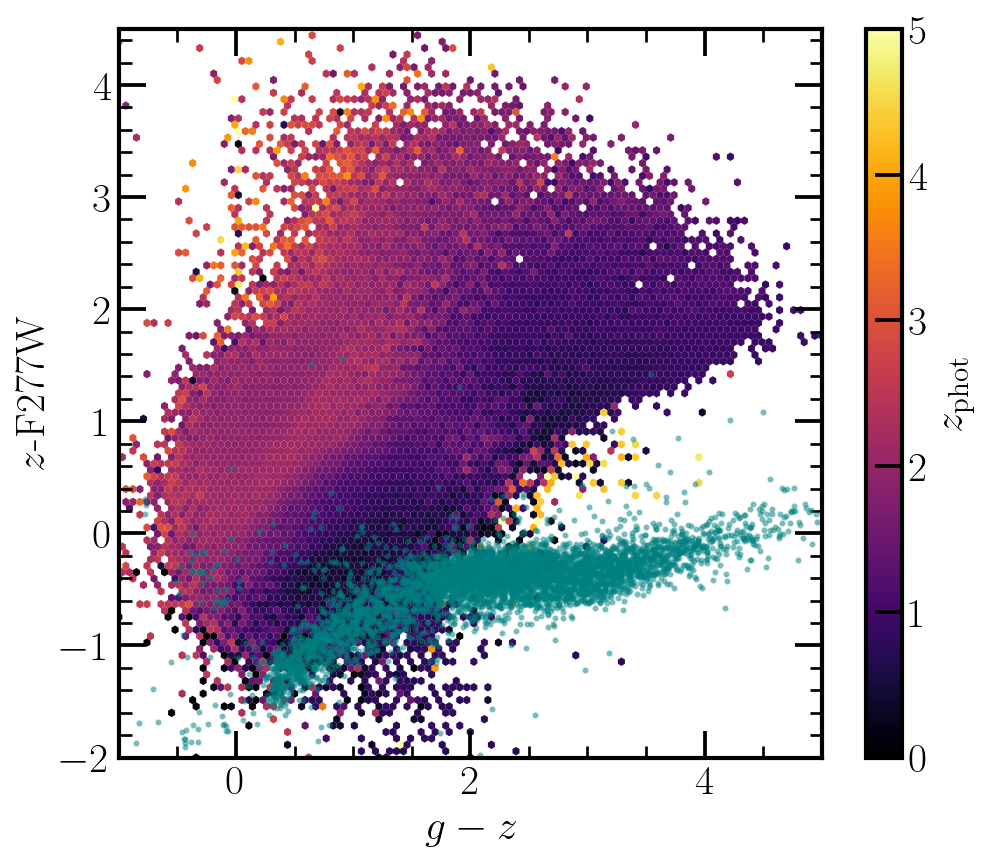}
\caption{Star and galaxy separation in the $g-z$ and $z-{\rm F444W}$ color space. The stars, selected using the criteria described in \ref{sec:star-galxy-separation}, are shown in teal dots. We only show S/N$>3$ sources outside of HSC star masks.
}
\label{fig:star-gal-separation}
\end{figure}

Sources are classified in the catalog with the keyword \texttt{type} with 0 for galaxies, 1 for stars and 2 for AGN. In Fig.~\ref{fig:star-gal-separation} we show the $g-z-F444W$ color space where stars are marked with teal points, while galaxies are color coded by their \zphot. Most of the sources classified as stars fall on the expected stellar locus of the $g-z-F444W$ color space.

We note these criteria are not exhaustive. There may be certain subtypes of QSOs/AGN that are contributing flux in the UV-MIR wavelengths, and thus the AGN Flag captures a fraction of the total AGN population. Though we note, this fraction likely represents the most luminous and relatively un-obscured QSO/AGN populations, which are the most problematic subtypes that can contaminate robust determination of galaxy properties via the multi-wavelength photometry provided in this catalogue.

\section{Physical properties} \label{sec:physical-prop}

\subsection{\lephare} \label{sec:physical-prop-lephare}

With the \lephare{} configuration adopted for COSMOS-Web, the physical parameters are derived simultaneously with the photometric redshift, as explained in Sect.~\ref{sec:photoz-methods}. We produce a PDF$_{\mstar}$ associated to the stellar mass by summing the probabilities P$_i$ at each stellar mass bin. By construction, these PDF$_{\mstar}$ include properly the photo-$z$ uncertainties. However, in the case of multi-peaked PDF$_z$, the stellar mass point estimate derived from PDF$_{\mstar}$ would not necessarily correspond to the selected redshift point estimate. 

To deal with this, we recomputed the physical parameters in a second run by setting the redshift to the point estimate established in the first run, i.e.,  the median of the PDF$_z$ by default. For homogeneity, we adopt this procedure for all sources of the catalog. In this case, the uncertainties associated with these physical parameters do not propagate redshift uncertainties.

\begin{figure}[t!]
\includegraphics[width=1\columnwidth]{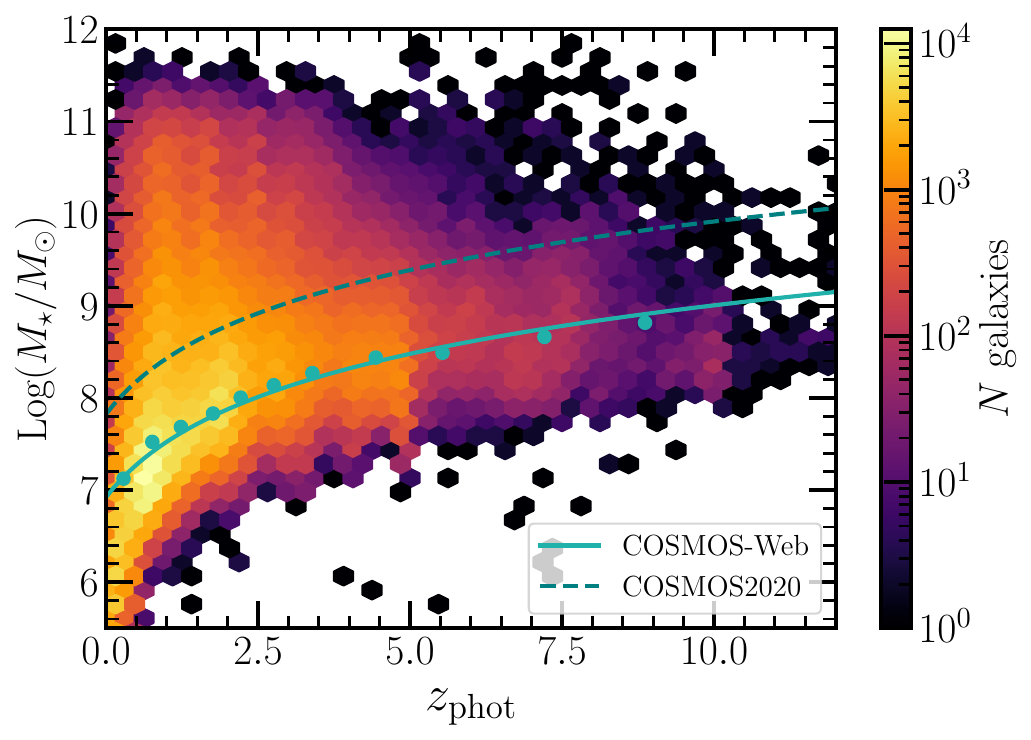}
\caption{Galaxy stellar mass vs. redshift diagram for the complete sample of the COSMOS-Web catalog. The circles show the stellar mass completeness limit computed derived by rescaling the stellar masses to the limiting magnitude of the survey and taking the $90$th percentile of this distribution \citep[following][]{pozzetti_zcosmos_2010}, while the solid curve shows the best-fit $(1+z)$ polynomial. The dashed line shows the completeness in the COSMOS2020 catalog that is about 1 dex shallower in $M_{\star}$.}
\label{fig:photoz-smass}
\end{figure}

In Fig.~\ref{fig:photoz-smass} we show the stellar mass vs. redshift distribution for the complete galaxy sample (c.f. \S \ref{sec:star-galxy-separation}). This shows a smooth distribution without significant striping or clustering in specific regions in the $M_{\star}-z$ space. We also show the stellar mass completeness computed with the \cite{pozzetti_zcosmos_2010} method. Briefly, we derive a rescaled stellar mass ($M_{\rm resc}$) by scaling the F444W magnitude to the magnitude limit of the survey ($m_{\rm lim}$)
\begin{equation} \label{eq:mass-completeness}
    \text{log}_{10}(M_{\text{resc}}) = \text{log}_{10} (M_{\star}) + 0.4 ( m_{\text{F444W}} - m_{\rm lim} ).
\end{equation}
Then, we define the limiting stellar mass, $M_{\star}^{\rm lim}$, as the 90th percentile of the $M_{\rm resc}$ distribution. The limiting magnitude is obtained by comparing the COSMOS-Web F444W magnitude number counts with those in the deeper catalog in the PRIMER footprint, requiring $f_{\rm compl.} = N_{\rm CWeb}/N_{\rm PRIMER} \approx 80\%$, resulting in $m_{\rm lim} = 27.5$ \citep{Shuntov2024}. We note that this is different from the fixed-aperture completeness limit discussed in \S\ref{sec:hotcold-complete-contam} because it refers to the total, model magnitude. The limiting stellar mass in several redshift bins is shown in the teal circles in Fig.~\ref{fig:photoz-smass} and the best-fit polynomial function in $(1+z)$ is shown in the solid line. For comparison, we also show the COSMOS2020 mass completeness function in the dashed line. Thanks to the deep $1-5 \, \si{\micron}$ detection in COSMOS-Web, we achieve a mass completeness improvement by about 1 dex compared to COSMOS2020. This makes COSMOS-Web an excellent catalog to carry out statistical population studies of galaxy evolution out $z\sim10$, with complete samples down to $M_{\star}\sim10^{9} \, \si{\Msun}$ at the highest redshifts, and down to $M_{\star}\sim10^{7} \, \si{\Msun}$ in the nearby Universe.

\begin{figure}[t!]
\centering
\includegraphics[width=0.9\columnwidth]{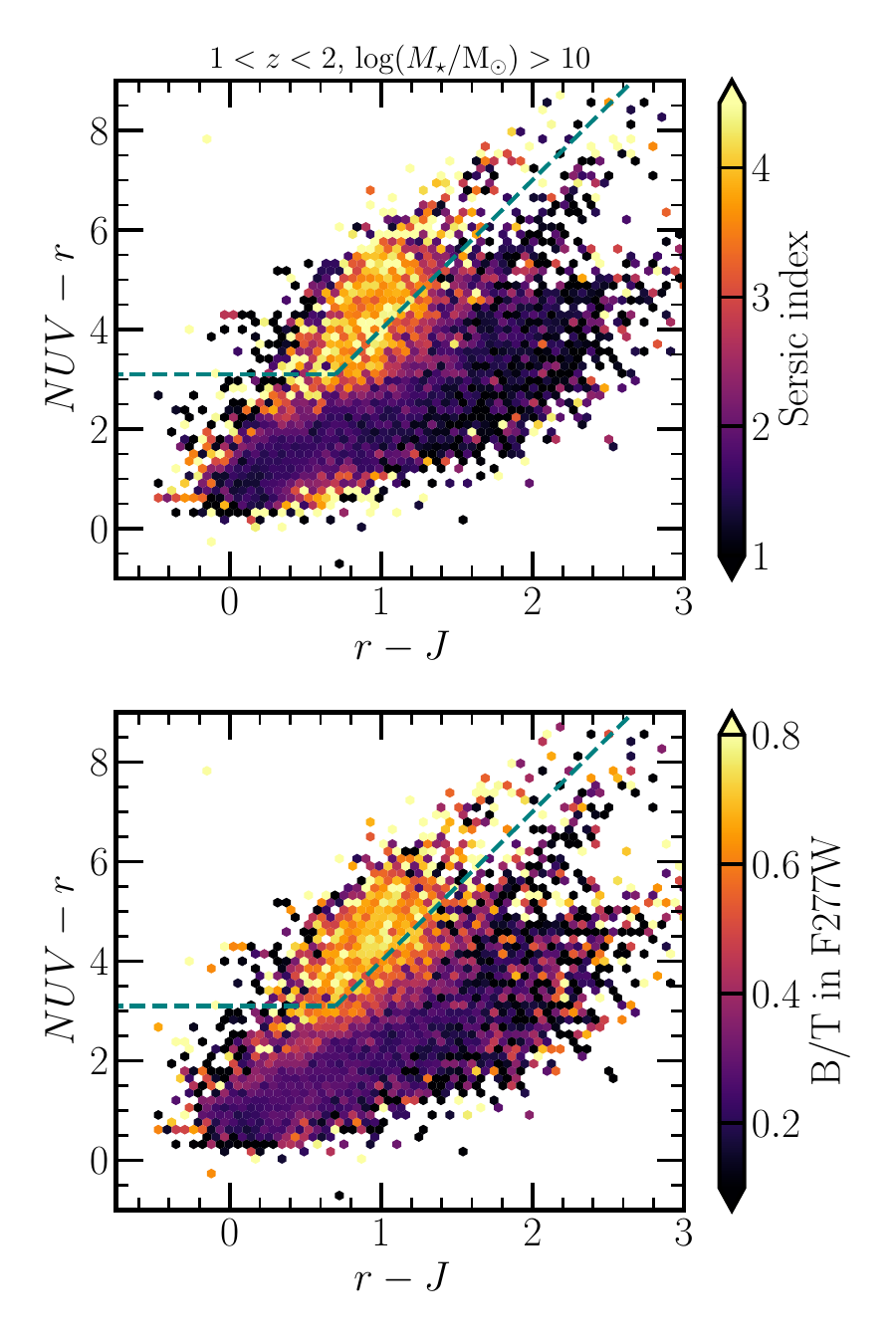}
\caption{Rest-frame $NUV-r-J$ color distribution for $1<z<2$ and $\log(M_{\star}/{\rm M}_{\odot})>10$ glaxies, color coded by their Sérsic index (top) and $B/T$ ratio (bottom). The dashed line marks the selection criteria for quiescent and star-forming galaxies \citep{ilbert_mass_2013}. Compact (high Sérsic index and $B/T$ ratio) predominantly lie in the quiescent region.}
\label{fig:morpho-color}
\end{figure}

Finally, we demonstrate the morphological information available in the catalog by exploring the correlation between star formation activity and morphology. In Fig.~\ref{fig:morpho-color} we show the $NUV-r-J$ color distribution for $1<z<2$ and $\log(M_{\star}/\si{\Msun})>10$ galaxies, color coded by their Sérsic index (top panel) and $B/T$ ratio (bottom panel). This shows that $NUV-r-J$ quiescent galaxies (marked by the dashed line) are predominantly compact, with high Sérsic index ($\gtrsim 3.5$) and $B/T$ ratio ($\gtrsim 0.6$). The region between the two populations, the so-called `green valley' is populated by intermediate values of $n_{\rm S}$ and $B/T$, while the star-forming region in predominantly disk-dominated. This is in good agreement with trends established at similar and lower redshifts by past works \citep[e.g.,][]{Daddi2005, Trujillo2007, Barro2013, Barro2017, 2016A&A...589A..35S}. Our catalogs allow for more detailed exploration of these morphological transformations that accompany galaxy evolution across large redshift ranges.

\subsection{\cigale } \label{sec:physical-prop-cigale }

In addition to \lephare{}, we derive a second set of physical properties independently through SED modeling using the \cigale\ code \citep{Boquien19}. This tool allows detailed analysis of galaxies' stellar populations and SFHs by fitting their observed photometry across a broad range of wavelengths. In particular, we focus on using \cigale's non-parametric SFH model, as this approach is highly suited to the investigation of galaxy evolution over cosmic time \citep{Arango-Toro2023,Arango-Toro2024, Ciesla2024b,Duan2024,Wuan2024}.

\cigale\ utilizes a Bayesian-like methodology to model the SED of galaxies, accounting for stellar and nebular emissions, dust attenuation, and, when necessary, contributions from active galactic nuclei (AGN). The core of this approach is reconstructing the galaxy's SFH through the \texttt{sfhNlevels} module \citep{Ciesla2023,Ciesla2024a}, which divides the SFH into bins defined by constant SFRs. This non-parametric approach avoids biases related to specific functional forms, such as exponentially declining models, and offers a flexible framework for accurately capturing complex star formation processes. The non-parametric SFHs are structured using time bins where the SFR can vary independently, and the \texttt{sfhNlevels} module incorporates a continuity-burst prior \citep{Leja19,Tacchella22}, suppressing sudden changes in the SFR while allowing for episodic bursts of star formation typical in high-redshift galaxies. This adaptability is crucial for accurately representing the diverse SFHs observed in the COSMOS-Web sample, which includes galaxies undergoing rapid changes due to interactions, mergers, and dynamic processes \citep{scoville_cosmic_2007,kartaltepe15b,davidzon_cosmos2015_2017}. 

\cigale\ estimates a comprehensive suite of physical parameters for each galaxy by comparing its observed photometry to model predictions. These parameters, derived through a Bayesian-like analysis, include the median values and associated uncertainties for stellar masses, star formation rates (both observed and averaged over the last 100 Myr), dust attenuation estimation, and stellar population metallicity. Additionally, we have provided detailed insights into the SFHs, such as the ages since the formation of the first stellar particle or the age at which 50\% of the total stellar mass was formed, the integrated SFH, and the SFR values in discrete time bins corresponding to different epochs in the galaxy's past. We provide as well the direction and normalization of the migration vector, which characterizes the displacement of a galaxy in the stellar mass vs. SFR plane over the last 250 Myr, further constraining the recent evolution of star formation activity \citep[see][for more details]{Arango-Toro2024}.

\begin{figure}[t!]
\includegraphics[width=1\columnwidth]{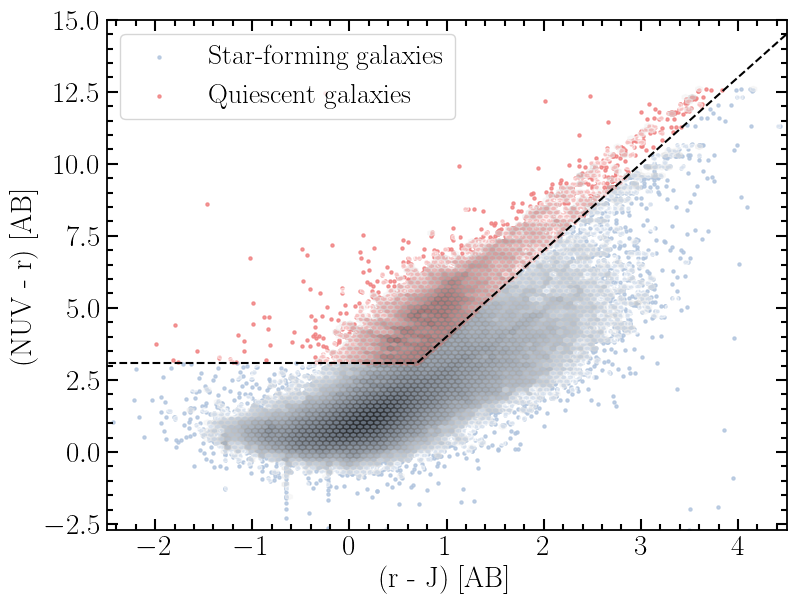}
\caption{Rest-frame $NUV-r-J$ color-color diagram showing the distribution of galaxies at redshifts between $1<z<3$ of the COSMOS-Web sample. Quiescent galaxies are highlighted in red, while star-forming galaxies are represented in blue. The black dashed line shows the selection criteria \citep{ilbert_mass_2013} for both populations. \label{fig:NRK}}
\end{figure}

\begin{figure}[t!]
\includegraphics[width=1\columnwidth]{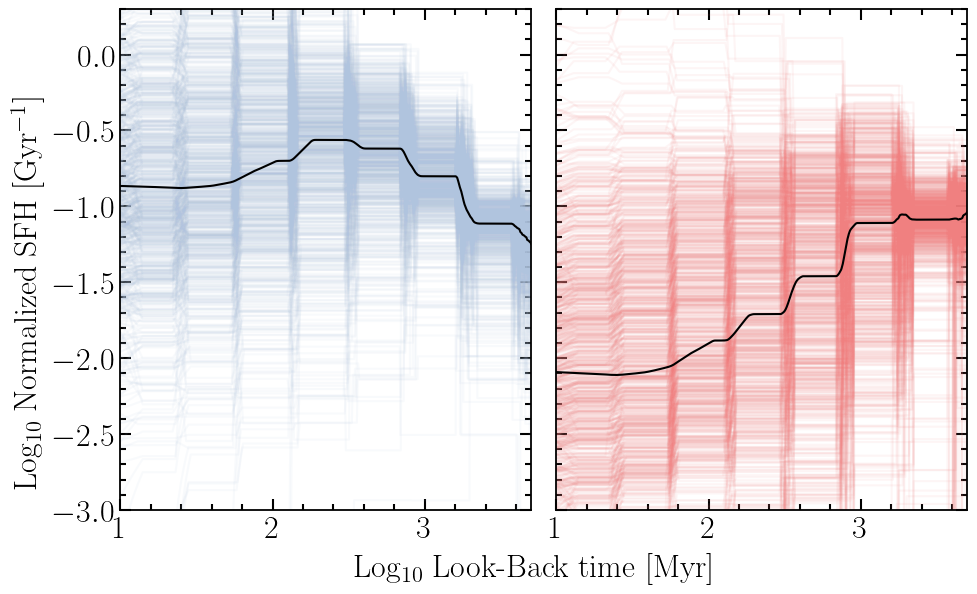}
\caption{Star formation histories for star-forming (left) and quiescent (right) galaxies selected with the $NUVrJ$ diagram. The colored, thin, lines show the SFHs of individual galaxies, while the solid black lines show the median SFH for the whole population.
\label{fig:sf_qs-sfhs}}
\end{figure}

The accuracy of the SED fitting method is assessed using both mock catalogs and simulations. Mock catalogs, based on the actual COSMOS-Web data, are generated by adding noise to the best-fit model fluxes and re-running the fitting procedure. The results from these mock analyses demonstrate that the stellar masses and SFRs derived using \cigale\ are consistent with the true values within small biases (typically <0.14 dex for stellar mass and <0.18 dex for SFR). 

Further validation is carried out by comparing the SED-derived physical parameters to those obtained from the Horizon-AGN cosmological simulation \citep{Dubois2014}. The comparison shows excellent agreement, particularly for key parameters such as stellar mass and SFR, with small biases and dispersions. This further confirms the robustness of the \cigale{} method in recovering physical parameters from SED fitting. These validations as well as a scientific application of the \cigale-derived physical properties in the context of the migration of galaxies over the SFR-$M_\star$ plane, are detailed in \citet[][Section 4.2]{Arango-Toro2024}.

Figure \ref{fig:NRK} presents the \lephare{} $NUV-r-J$ color-color diagram for the COSMOS-Web sample, for galaxies at $1<z<3$. Fig.~\ref{fig:sf_qs-sfhs} shows the \cigale{} non-parametric SFHs for star-forming and quiescent galaxies given the selection criteria over the $NUV-r-J$ diagram \citep{ilbert_mass_2013} . We show the normalized SFHs of individual galaxies in thin colored lines and the median SFH for the whole population in solid black lines. As expected, the star-forming population shows rising and elevated SFRs, where the median SFH rises by $\sim0.7$ dex in the first couple of Myr and then mildly decreases by $\sim 0.3$ dex since. The quiescent population shows SFHs that decline by $\sim1$ dex in about 3.5 Gyr. There is some scatter about the median SFH for both populations, with a few individual galaxies showing rising/declining SFHs even though they are selected as quiescent/star-forming from $NUV-r-J$. This could be due to the limitations of $NUV-r-J$ in efficiently distinguishing dust-reddened and star-forming from truly quiescent galaxies.
Nonetheless, Fig~\ref{fig:sf_qs-sfhs} demonstrates the consistency of the SFHs derived from \cigale{} for star-forming and quiescent galaxies selected using $NUV-r-J$ colors derived from \lephare{} and the synergy of the physical information added by both SED fitting methods. 

Finally, we compared the stellar mass measurements from \lephare{} and \cigale\ to investigate how the different assumptions in SFH and dust modeling would affect the resulting stellar masses. This comparison is shown in more detail in \citet[][Appendix F]{Shuntov2024}. In general,they agree well, albeit with a persistent bias towards higher $M_{\star}$ inferred by \cigale{} that is dependent on both redshift and mass. The difference is around 0.1 dex at $z<0.5$ and $\sim 0.2-0.3$ dex at higher redshifts. Concerning the mass, the difference is higher at the low-mass end and decreases with mass out to $z\sim 5.5$. At $z>5.5$ this trend reverses and \cigale results in higher masses with increasing \lephare{} mass by about $0.1-0.3$ dex. This trend is consistently observed when comparing stellar masses derived from non-parametric and parametric SFH \citep[e.g.,][]{Leja2020}, but could be due to other assumptions in both codes such as the set of attenuation curves.

\section{Conclusions} \label{sec:conclusions}

In this paper we presented the COSMOS2025 catalogue, the COSMOS-Web galaxy catalog of photometry, morphology, and physical parameters for over 700,000 galaxies in the central area of COSMOS. The foundation of this catalog is deep COSMOS-Web \JWST\ imaging that provides a $\sim 28$ mag NIR source detection from a positive-truncated $\sqrt{\chi^2}$ detection image made from a combination of all four NIRCam bands. We carried out source detection using a hot and cold technique and photometric extraction using two independent approaches: aperture photometry on the space-based ACS/F814W, NIRCam and MIRI/F770W bands and total photometry from profile-fitting using \SEpp\ on all 37 photometric bands. These include the UltraVISTA DR6 and the HSC PDR3 data which are both deeper and more homogenous than previous releases. 

Our profile-fitting technique provides morphological measurements for every source, both using a single Sérsic together with a combination of bulge and disk light profiles. To maximize scientific applications, we also provide independent morphological measurements based on Sérsic models from \texttt{Galight} as well as types derived from a machine-learning classification. These independent methods show good consistency in the distribution of the morphological properties and in direct comparisons of the same quantities. 

We carried out 34-band SED fitting on the total photometry to obtain photo-$z$ and physical parameters from \lephare{}. The photo-$z$ are highly accurate with $\sigma_{\rm MAD} = 0.012$ for $m_{\rm F444W} < 28$ and stay within $\sigma_{\rm MAD} \lesssim 0.03$ as a function of magnitude, color, and galaxy type. We also compared with a spec-$z$ sample of high-redshift $6.5\lesssim z \lesssim 10$ galaxies identified in COSMOS-3D, CAPERS and transient programs. Photo-$z$ performance remains excellent with $\sigma_{\rm MAD} = 0.038$ and $\eta = 7.9 \%$. There is a bias of about $-0.3$ towards higher photo-$z$ which is likely due to the lack of these objects in our calibration sample. Thanks to the deep NIRCam detection, our catalog improves the stellar mass completeness over COSMOS2020 by about 1 dex, and is $\sim 80 \%$ complete for ${\rm log} (M_{\star}/\si{\Msun}) \sim 9$ at $z\sim9$. We demonstrated the correlation between star-formation activity and morphology by showing that compact galaxies (high Sérsic index and $B/T$ ratio) predominantely occupy the quiescent region of the $NUV-r-J$ rest-frame color diagram. We obtained additional physical parameters from non-parametric star formation history modelling and SED fitting from \cigale{} by fixing the photo-$z$ to the solution from \lephare{}. 

In our COSMOS2020 paper, we imagined what form COSMOS2025 might take: a new catalogue comprising ultra-deep imaging from JWST combined with the final data releases from ground-based telescopes including UltraVISTA and HSC-SSP surveys, together with the rich heritage of existing multiwavelength COSMOS data. The catalogue described here is precisely that. This paper builds on the methodological advances of COSMOS2020 (robust light profile fitting for hundreds of thousands of galaxies) and combines them with a rigorous implementation of a two-threshold detection strategy to deliver a catalogue of over 700,000
galaxies with photometric measurements in 37 bands.  Many papers have used these catalogues in a range of impactful studies \citep[e.g.,][]{Casey2024, Franco2024, Akins2024, Gentile2024, Shuntov2024, Arango-Toro2024, Paquereau2025, MHC2025, Kaminsky2025, Toni2025, Yang2025, Nightingale2025}. With these catalogues and images now publicly available, many unexplored scientific avenues are now open to the community. 

COSMOS remains a foundation stone of extragalactic astronomy, and many more observations are either underway or in planning which give us today a glimpse of what COSMOS2030 might look like. COSMOS-3D (ID \#5893, PI: K. Kakiichi) is currently carrying out the largest NIRCam/WFSS survey covering $0.33$ deg$^2$, including 500 arcmin$^2$ in parallel with MIRI. This will also add imaging in two additional NIRCam (F200W and F356W) and MIRI (F1000W and F2100W) bands, as well as WFSS spectra for hundreds of sources at $z>7$. The HST Multi-Cycle Treasury program CLUTCH (ID \#17802, PI: J. Kartaltepe) will image $0.5$ deg$^2$, overlapping the areas observed by COSMOS-Web and COSMOS-3D. CLUTCH will observe the rest-frame UV to optical emission for galaxies at $z<7.5$ utilizing WFC3-UVIS F225+F275W, ACS F435W+F606W+F814W, and WFC3-IR F098M. Over a wider area, the \textit{Euclid} mission has already observed COSMOS in both optical and infrared bands, and these data will be delivered as part of the DR2 release in 2026. On slightly longer timescales, both the Roman Space telescope and the Rubin telescope will observe COSMOS.  The methods described in this paper provide a clear path to create future COSMOS catalogues which will optimally exploit these unique data. 

\section*{Data availability} \label{sec:data-availability}

We release the complete COSMOS-Web catalog, as well as the \JWST/NIRCam+MIRI imaging described in Franco et al.~(\textit{in prep.}) and Harish et al.~(\textit{in prep.}) at \url{https://cosmos2025.iap.fr/}. A \texttt{README} is provided with detailed descriptions of the catalog columns as well as example of simple use cases.
We provide an interactive map view of the COSMOS-Web field, constructed using FitsMap \citep{Hausen2022}, at \url{https://cosmos2025.iap.fr/fitsmap/}. This includes our COSMOS-Web catalog as an overlay, and SED plots for all 784k sources in the catalog.

\begin{acknowledgements}

Support for this work was provided by NASA through grant JWST-GO-01727 awarded by the Space Telescope Science Institute, which is operated by the Association of Universities for Research in Astronomy, Inc., under NASA contract NAS 5-26555.

The Cosmic Dawn Center (DAWN) is funded by the Danish National Research Foundation under grant DNRF140. 
This work was made possible by utilizing the CANDIDE cluster at the Institut d’Astrophysique de Paris, which was funded through grants from the PNCG, CNES, DIM-ACAV, and the Cosmic Dawn Center and maintained by S. Rouberol. French COSMOS team members are partly supported by the Centre National d’Etudes Spatiales (CNES). We acknowledge the funding of the French Agence Nationale de la Recherche for the project iMAGE (grant ANR-22-CE31-0007). 
This work has received funding from the Swiss State Secretariat for Education, Research and Innovation (SERI) under contract number MB22.00072.
This project has received funding from the European Union’s Horizon 2020 research and innovation programme under the Marie Skłodowska-Curie grant agreement No 101148925.
The data products presented herein were retrieved from the Dawn JWST Archive (DJA). DJA is an initiative of the Cosmic Dawn Center (DAWN), which is funded by the Danish National Research Foundation under grant DNRF140
B.T. acknowledges support from the European Research Council (ERC) under the European Union's Horizon 2020 research and innovation program (grant agreement number 950533). 
This research was supported by the Excellence Cluster ORIGINS which is funded by the Deutsche Forschungsgemeinschaft (DFG, German Research Foundation) under Germany's Excellence Strategy - EXC 2094 - 390783311.
The Hyper Suprime-Cam (HSC) collaboration includes the astronomical communities of Japan and Taiwan, and Princeton University. The HSC instrumentation and software were developed by the National Astronomical Observatory of Japan (NAOJ), the Kavli Institute for the Physics and Mathematics of the Universe (Kavli IPMU), the University of Tokyo, the High Energy Accelerator Research Organization (KEK), the Academia Sinica Institute for Astronomy and Astrophysics in Taiwan (ASIAA), and Princeton University. Funding was contributed by the FIRST program from the Japanese Cabinet Office, the Ministry of Education, Culture, Sports, Science and Technology (MEXT), the Japan Society for the Promotion of Science (JSPS), Japan Science and Technology Agency (JST), the Toray Science Foundation, NAOJ, Kavli IPMU, KEK, ASIAA, and Princeton University. 
DS carried out this research at the Jet Propulsion Laboratory, California Institute of Technology, under a contract with the National Aeronautics and Space Administration (80NM0018D0004).

This paper makes use of software developed for Vera C. Rubin Observatory. We thank the Rubin Observatory for making their code available as free software at http://pipelines.lsst.io/.

This paper is based on data collected at the Subaru Telescope and retrieved from the HSC data archive system, which is operated by the Subaru Telescope and Astronomy Data Center (ADC) at NAOJ. Data analysis was in part carried out with the cooperation of the Center for Computational Astrophysics (CfCA), NAOJ. We are honored and grateful for the opportunity of observing the Universe from Maunakea, which has cultural, historical and natural significance in Hawaii.

\end{acknowledgements}

\vspace{5mm}

\bibliographystyle{aa}
\bibliography{library}

\begin{appendix}

\section{Point-spread functions} \label{sec:appendix-psf}
In Sect.~\ref{sec:PSF-reconstruct-homogenize}

\begin{figure}[ht!]
\centering
\includegraphics[width=0.99\columnwidth]{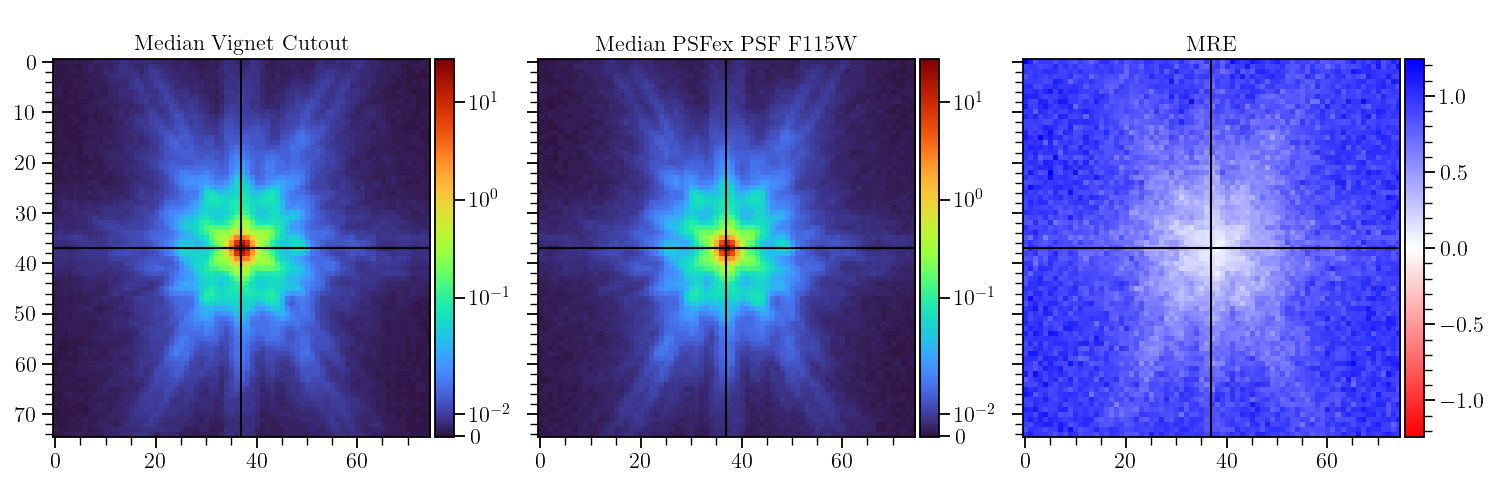}
\includegraphics[width=0.99\columnwidth]{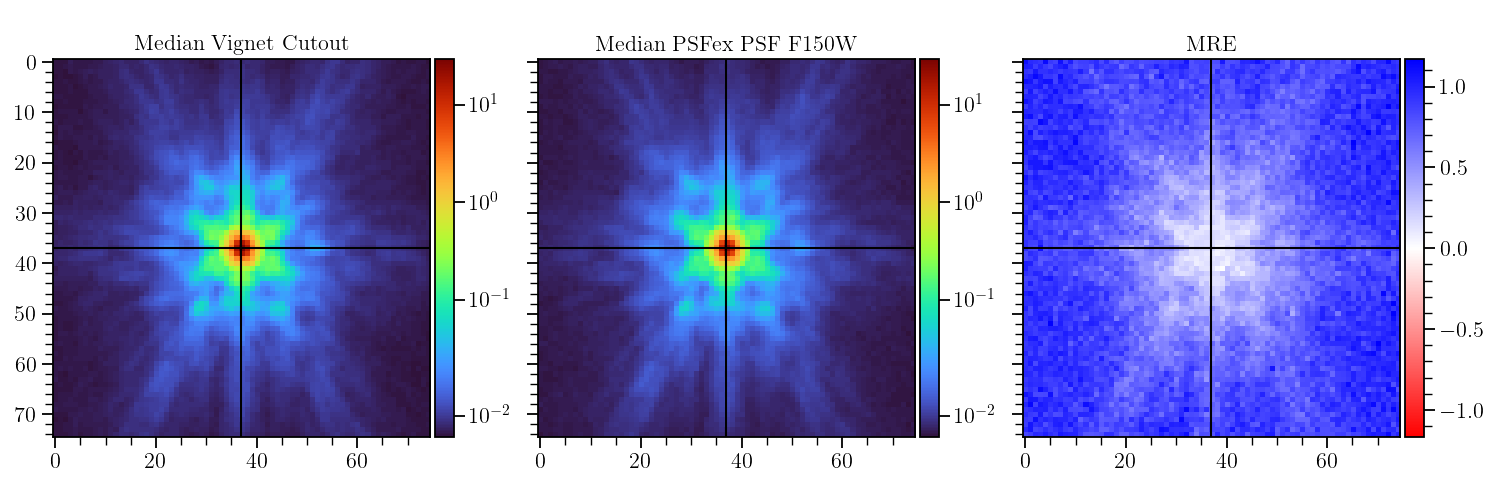}
\includegraphics[width=0.99\columnwidth]{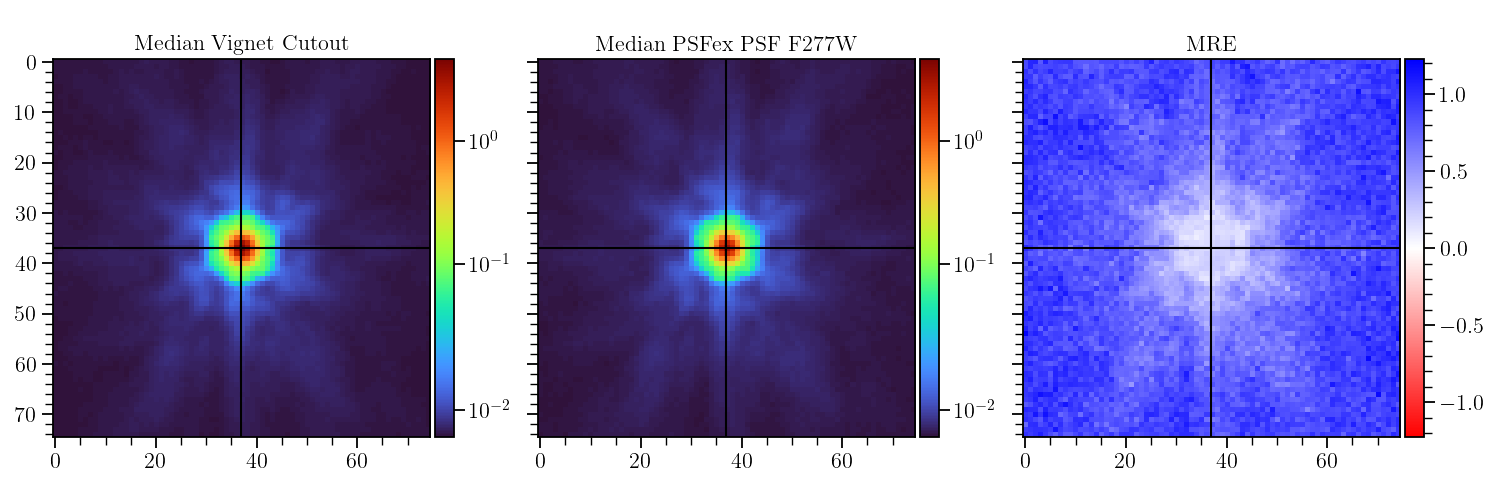}
\includegraphics[width=0.99\columnwidth]{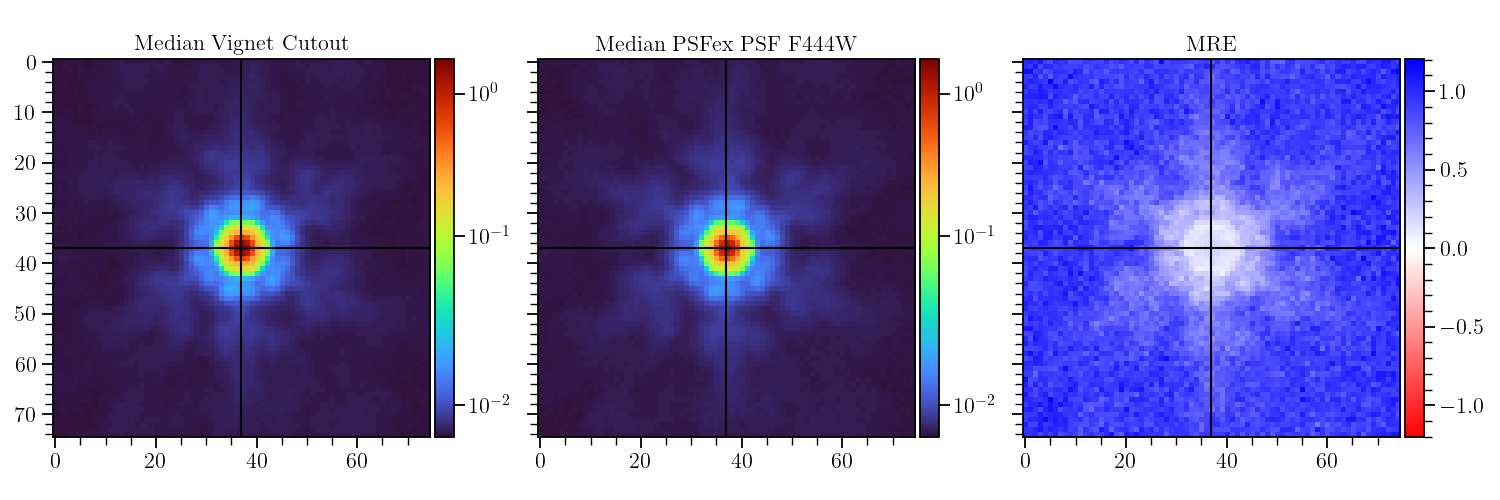}
\caption{\textit{Left panel:} the median of the input star catalog vignets. \textit{middle panel:} the median of the PSF models. \textit{Right panel:} the average relative error between star and PSF model.}
\label{fig:psf-MRE}
\end{figure}

\section{Source model priors in \SEpp} \label{sec:appendix-SEclparams}

\begin{figure}[h!]
\centering
\includegraphics[width=0.98\columnwidth]{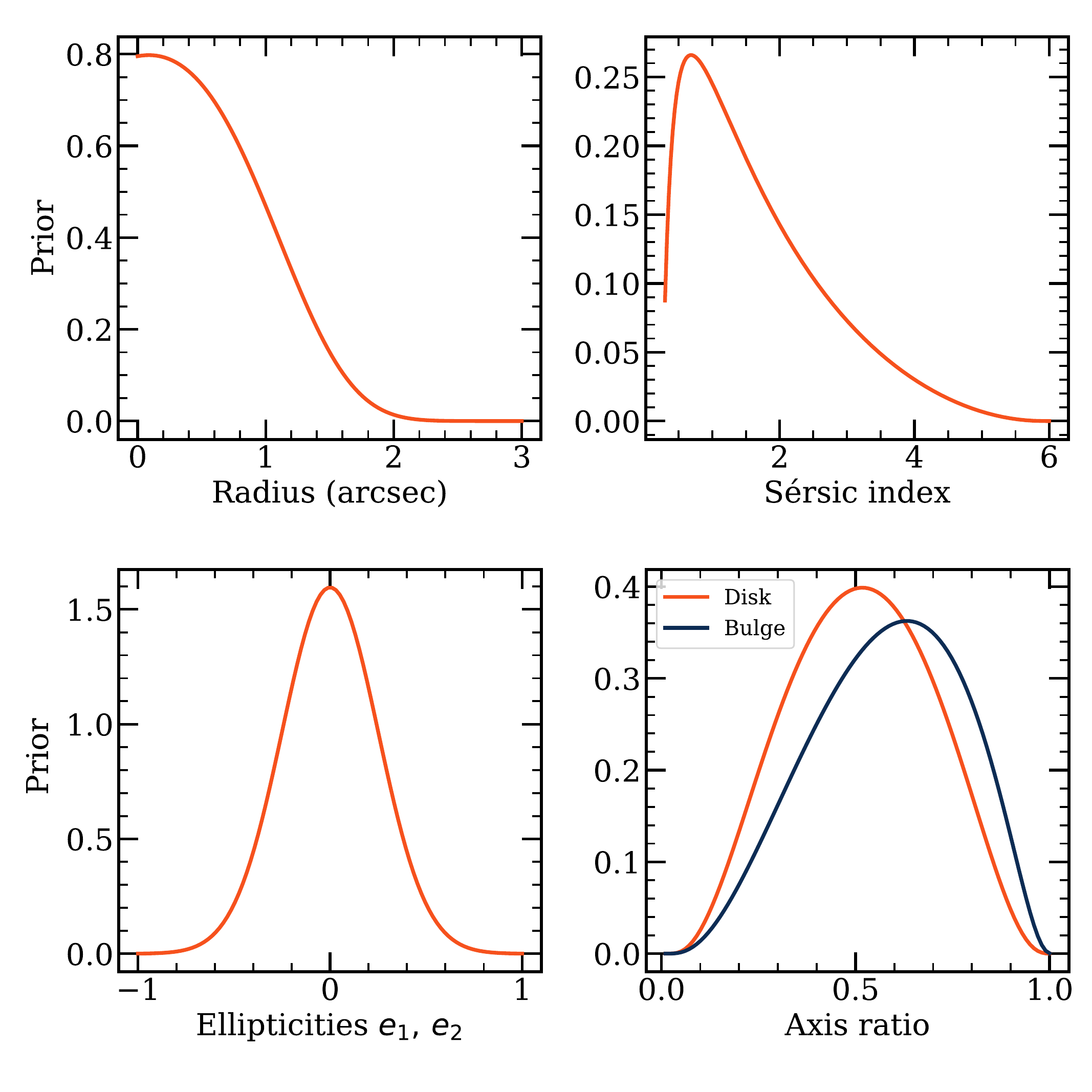}
\caption{Adopted priors in \SEpp}
\label{fig:priors}
\end{figure}

In Fig.~\ref{fig:priors} we show the priors adopted for the \texttt{SE++} Sérsic and Bulge+Disk models (\S \ref{sec:modeling-run}). The first three panels correspond to the priors on the effective radius (for both models), Sérsic index and ellipticities for the Sérsic model, while the fourth panel shows the priors on the axis ratio that we adopt for the bulge and disk fits.

\end{appendix}

\end{document}